\documentclass[11pt]{article}
\pdfoutput=1

\usepackage{braket}
\usepackage{color}
\usepackage{mhchem}
\usepackage{xcolor}
\usepackage{colortbl}
\usepackage{hhline}
\usepackage{cite}
\usepackage[colorlinks, linkcolor=red, anchorcolor=black,citecolor=green]{hyperref}
\usepackage[toc,page]{appendix}
\usepackage{mathtools}
\usepackage{amsfonts}
\usepackage{bbold}
\usepackage{textcomp}
\usepackage[DIV13]{typearea}
\usepackage{amsmath, amsthm, amssymb, mathtools,empheq,latexsym,dsfont}
\usepackage{bbm}
\usepackage{amsmath,amssymb,latexsym,dsfont}
\usepackage{slashed, simplewick}
\usepackage[utf8]{inputenc}
\usepackage{graphicx}
\usepackage{graphicx,placeins}
\usepackage{makeidx}
\usepackage[font=small,labelfont=bf]{caption}
\usepackage{nicefrac}
\usepackage{subfigure}
\usepackage{array, bigdelim,multirow,multicol}
\usepackage{mathrsfs}
\usepackage{adjustbox}
\usepackage{booktabs}
\usepackage{indentfirst}

\usepackage{booktabs}
\usepackage{tabulary}
\usepackage{longtable}
\usepackage{fancybox}
\usepackage{graphicx}
\usepackage{bm}
\usepackage{bbm}
\usepackage{tikz}
\usepackage{diagbox}
\usepackage{enumitem}
\usepackage[integrals]{wasysym}
\usepackage{pdflscape}

\graphicspath{{immagini/}}

\definecolor{Gray}{gray}{0.92}
\definecolor{highlight}{RGB}{255,255,0}

\newcommand{\ignore}[1]{}

\renewcommand{\arg}{{\rm Arg}}

\newcommand{\be}{\begin{equation}}
	\newcommand{\ee}{\end{equation}}
\newcommand{\bea}{\begin{equation}}
	\newcommand{\eea}{\end{equation}}

\definecolor{lightred}{rgb}{1,0.4,0.4}
\definecolor{lightgreen}{rgb}{0.4,1,0.4}
\definecolor{LightCyan}{rgb}{0.88,1,1}

\newcounter{Thm}[section]
\renewcommand{\theThm}{\arabic{section}.\arabic{Thm}}

\newcounter{nodecount}

\tikzstyle{every picture}+=[remember picture,baseline]
\tikzstyle{every node}+=[inner sep=0pt,anchor=base,
text depth=.25ex,outer sep=1.5pt]
\tikzstyle{every path}+=[thick, rounded corners]
\tikzset{
	plabel/.style={inner sep=2pt}
}

\makeatletter
\@addtoreset{equation}{section}
\makeatother

\begin{document}
	\title{\begin{center}
            {\Large\bf Lepton mixing from the $\Delta(96)$ Modular Littlest Seesaw}
	\end{center}}
	\date{}
	\author{Hai-Zhi Hao$^{a,b,c,d}$\footnote{E-mail: {\tt
			202521457@stumail.nwu.edu.cn}},  \
Li-Na Yan$^{a,b,c,d}$\footnote{E-mail: {\tt
			202321634@stumail.nwu.edu.cn}},  \
Xiang-Gan Liu$^{e,f}$\footnote{E-mail: {\tt
			xianggal@uci.edu}},  \
	Cai-Chang Li$^{a,b,c,d}$\footnote{E-mail: {\tt ccli@nwu.edu.cn}} \ \\*[20pt]
	\centerline{\begin{minipage}{\linewidth}
			\begin{center}
				$^a${\it\small School of Physics, Northwest University, Xi'an 710127, China}\\[2mm]
				$^b${\it\small Shaanxi Key Laboratory for Theoretical Physics Frontiers, Xi'an 710127, China}\\[2mm]
				$^c${\it\small NSFC-SPTP Peng Huanwu Center for Fundamental Theory, Xi'an 710127, China}\\[2mm]
                $^d${\it\small Fundamental Discipline Research Center for  Quantum Science and technology of Shaanxi Province, Xi'an 710127, China}\\[2mm]
                $^e${\it\small Department of Physics and Astronomy, University of California, Irvine, CA 92697-4575 USA}\\[2mm]
                $^f${\it\small Instituto de F\'isica, Universidad Nacional Aut\'onoma de M\'exico, Cd.\ de M\'exico C.P.\ 04510, M\'exico}
			\end{center}
	\end{minipage}} \\[10mm]}
	\maketitle
	
	\thispagestyle{empty}
	
	\centerline{\large\bf Abstract}
\begin{quote}
\indent
We perform the first comprehensive and model independent study of Modular Littlest Seesaw models based on the finite modular group $\Delta(96)$. We construct the  vector-valued modular forms (VVMFs) for all irreducible representations of 
modular $\Delta(96)$, classify the inequivalent symmetry-preserving fixed 
points, and derive the corresponding alignments of the low-weight and next-to-lowest-weight triplet VVMFs. These results allow an exhaustive scan 
over the residual symmetries in the charged lepton, atmospheric neutrino, and 
solar neutrino sectors. We identify 35 phenomenologically viable and inequivalent 
breaking patterns, including 21 with normal ordering and 14 with inverted 
ordering. The resulting Dirac neutrino mass matrices go beyond the conventional 
CSD$(n)$ structure, yielding new fixed PMNS columns and novel correlations 
among the lepton mixing parameters beyond the TM$_1$ paradigm. The viable models are highly predictive, 
giving narrow ranges for neutrino masses, mixing parameters and CP phases, and 
can be stringently tested by upcoming experiments such as JUNO, DUNE and T2HK.

\end{quote}

 \thispagestyle{empty}
 \vfill

 \clearpage

 {\hypersetup{linkcolor=black}
 \tableofcontents
 }

\section {Introduction}

The discovery of neutrino oscillations~\cite{McDonald:2016ixn,Kajita:2016cak} has established that neutrinos are massive and exhibit flavor mixing, providing clear evidence for physics beyond the Standard Model (SM). Over the past two decades, neutrino oscillation experiments have measured the mass squared differences and leptonic mixing angles with remarkable precision~\cite{ParticleDataGroup:2026aaa}. Explaining the tiny neutrino masses together with the observed pattern of lepton flavor mixing remains one of the central challenges in particle physics, motivating the construction of predictive extensions of the SM.

Among the proposed mechanisms for neutrino mass generation, the type-I seesaw is particularly attractive because it naturally explains the smallness of neutrino masses through the introduction of heavy right-handed neutrinos (RHNs)~\cite{Minkowski:1977sc}. However, the conventional seesaw model with 3RHNs contains too many free parameters to provide definite predictions. A more predictive framework is the minimal seesaw model with only 2RHNs~\cite{King:1999mb,Frampton:2002qc,Xing:2020ald}, which originated from the idea of sequential dominance (SD)~\cite{King:1998jw,King:1999cm,King:1999mb,King:2002nf}. Its predictive power is further enhanced by constrained sequential dominance (CSD)~\cite{King:2005bj}, where the Dirac neutrino mass matrix contains a single texture zero and its two columns satisfy specific alignment conditions.

A particularly successful realization is the CSD($n$) framework~\cite{King:2005bj,Antusch:2011ic,King:2013iva,King:2015dvf,King:2016yvg,Ballett:2016yod,King:2018fqh,King:2013xba,King:2013hoa,Bjorkeroth:2014vha}, in which the two columns of the Dirac neutrino mass matrix are aligned along $(0,1,-1)^T$ and $(1,n,2-n)^T$ respectively, where the parameter $n$ was originally taken to be a positive integer but can in general be any real number. For a fixed value of the real parameter $n$, all neutrino masses and lepton mixing parameters are determined by only three real input parameters, predicting the TM$_1$ mixing pattern~\cite{Xing:2006ms,Lam:2006wm}. The best-known example is the Littlest Seesaw (LS) model, corresponding to CSD($3$)~\cite{King:2013iva,King:2015dvf,King:2016yvg,Ballett:2016yod,King:2018fqh}, while viable variants based on CSD($4$)~\cite{King:2013xba,King:2013hoa}, CSD($-1/2$)~\cite{Chen:2019oey}, the Golden Littlest Seesaw~\cite{Ding:2017hdv}, and the tri-direct CP approach~\cite{Ding:2018fyz,Ding:2018tuj,Chen:2019oey,Yan:2025itm} have also been proposed. These constructions are typically realized within discrete flavor symmetries such as $S_4$ or $A_5$, where the required vacuum alignments arise from flavon fields preserving different residual symmetries. Such models, however, generally require numerous flavon fields together with auxiliary symmetries, making the scalar sector rather complicated.

Modular symmetry provides an elegant alternative to conventional flavor symmetries by eliminating the need for flavon fields~\cite{Feruglio:2017spp}; see Refs.~\cite{Kobayashi:2023zzc,Ding:2023htn} for recent reviews. In this framework, flavor symmetry is broken solely by the vacuum expectation value (VEV) of a single modulus field $\tau$, while Yukawa couplings are promoted to modular forms with definite modular weights and transformation properties. The systematic construction of modular multiplets is provided by vector-valued modular forms (VVMFs), whose structure is governed by modular linear differential equations (MLDE)~\cite{Liu:2021gwa}. From a top-down perspective, modular flavor symmetries naturally emerge from target-space dualities in string compactifications~\cite{Lauer:1989ax,Ferrara:1989bc,Kobayashi:2018rad,Baur:2019kwi,Nilles:2020kgo}. In this context, modular fixed points represent self-dual loci in moduli space, often with enhanced residual symmetry, and hence provide well-motivated candidates for modulus VEVs~\cite{Font:1990nt,Acharya:1995ag,Dent:2001ut,Ishiguro:2020tmo,Novichkov:2022wvg}. Combining modular symmetry with the Littlest Seesaw framework leads to highly predictive Modular Littlest Seesaw models~\cite{Ding:2019gof,deMedeirosVarzielas:2022fbw}, in which the flavor structure is determined by modular forms evaluated at symmetry preserving (fixed) points in moduli space.

Existing Modular Littlest Seesaw models are based mainly on modular $S_4$\cite{Ding:2019gof,Ding:2021zbg,deMedeirosVarzielas:2022fbw,deMedeirosVarzielas:2023ujt}, together with more recent extensions to $S_4^{\prime}$ and $A_5$~\cite{Shang:2026qkh}. They typically require three independent moduli, $\tau_{\ell}$, $\tau_{\text{atm}}$ and $\tau_{\text{sol}}$, whose VEVs are fixed at fixed points~\cite{Ding:2019gof,Novichkov:2018yse,Novichkov:2018ovf,Novichkov:2018nkm,deMedeirosVarzielas:2022ihu,deMedeirosVarzielas:2019cyj,King:2019vhv,deMedeirosVarzielas:2020kji}. Such a setup can be realized either through multiple modular symmetries~\cite{deMedeirosVarzielas:2022fbw,deMedeirosVarzielas:2023ujt} or orbifold compactifications with three factorisable tori~\cite{deAnda:2023udh}. Nevertheless, a systematic exploration of Modular Littlest Seesaw models based on larger finite modular groups is still lacking.

The group $\Delta(96)$, corresponding to the $n=4$ member of the $\Delta(6n^2)$ series, has been extensively explored as a non-Abelian flavor symmetry~\cite{deAdelhartToorop:2011re,Ding:2012xx,King:2012in,Ding:2014ssa,Bernigaud:2020wvn,Gautam:2020bnx,Alvarado:2022lzx,Yan:2025itm}. By contrast, its role as a finite modular group has received little attention. In particular, although $\Delta(96)$ is one of the four finite modular groups of order 96 arising from $\mathrm{SL}(2,\mathds{Z})$, a systematic investigation of model building based on modular $\Delta(96)$ has not previously been carried out.\footnote{A recent study has demonstrated that modular $\Delta(96)$ can naturally generate large fermion mass hierarchies with very few parameters~\cite{CentellesChulia:2026bkr}.} Beyond its rich group theoretical structure, modular $\Delta(96)$ possesses complex triplet representations and a sextet representation, which are absent in the modular groups $A_4$, $S_4$, and $A_5$. These additional representations considerably enlarge the possibilities for flavor model building, CP violation, and grand unified theories~\cite{King:2012in}. Motivated by these features, we present the first comprehensive and model independent study of Modular Littlest Seesaw models based on modular $\Delta(96)$. As a first step, we determine all inequivalent modulus fixed points and construct the complete set of VVMFs of modular $\Delta(96)$.

Using these results, we derive all independent vacuum alignments of lowest-weight and next-to-lowest-weight triplet VVMFs at the fixed points and perform an exhaustive scan over every possible assignment of residual symmetries in the charged lepton, atmospheric neutrino and solar neutrino sectors. We identify 35 inequivalent phenomenologically viable Modular Littlest Seesaw models, including 21 normal ordering (NO) and 14 inverted ordering (IO) cases. Remarkably, the resulting Dirac neutrino mass matrices exhibit structures considerably more general than those of the conventional CSD($n$) framework, leading to new correlations among the leptonic mixing parameters beyond the TM$_1$ paradigm. In contrast to residual-symmetry constructions with possible continuous alignment parameters, the fixed-point VVMF alignments form a finite set of algebraic directions fixed by modular covariance. This algebraic rigidity is a key source of predictivity in the present modular construction. A comprehensive numerical analysis shows that the viable models are highly predictive, yielding narrow ranges for neutrino masses, lepton mixing parameters, and CP-violating phases. Most of these predictions can be stringently tested by forthcoming neutrino oscillation experiments, including JUNO~\cite{JUNO:2022mxj,JUNO:2025gmd}, DUNE~\cite{DUNE:2020ypp}, and T2HK~\cite{Hyper-Kamiokande:2018ofw}.

The remainder of this paper is organized as follows. In section~\ref{sec:VVMFs}, we review the mathematical formalism of modular symmetry and VVMFs. In section~\ref{secVVMFs_Delta96}, we construct the relevant VVMFs and derive their vacuum alignments at the fixed points for modular group $\Delta(96)$. In section~\ref{sec:MLS}, we present the framework of Modular Littlest Seesaw. An exhaustive scan over all possible Modular Littlest Seesaw constructions is performed, the phenomenologically viable models based on $\Delta(96)$, the corresponding  numerical analysis and phenomenological results are presented in section~\ref{sec:model_building}. We conclude in section~\ref{sec:conclusion}. Appendix~\ref{sec:Delta96_group_theory} contains the group theory of $\Delta(96)$. In appendix~\ref{sec:qes_VVMFs}, we provide the explicit $q$-expressions of lowest- and next-to-lowest-weight VVMFs for all three-dimensional and six-dimensional representations of modular $\Delta(96)$. Appendix~\ref{sec:fixed_column} presents the analytical expressions for the fixed PMNS columns of the 27 viable patterns beyond the TM$_1$ mixing scheme, and appendix~\ref{sec:sum_rules} collects the explicit sum rules for selected six patterns.

\section {\label{sec:VVMFs}Modular symmetry and VVMFs}

The special linear group $\Gamma\equiv \mathrm{SL}(2, \mathds{Z})$, also called the (full) modular group, consists of $2\times2$ matrices with integer entries and determinant 1. It is generated by two elements $S$ and $T$ subject to the relations:
\begin{equation}
\Gamma\equiv \mathrm{SL}(2,\mathds{Z})=\left\langle S,T\mid S^{4}=(ST)^{3}=1,~~S^{2}T=TS^{2}\right\rangle\,,
\end{equation}
where the generators can be represented by the matrices
\begin{equation}
S=\begin{pmatrix}0&1\\-1&0
\end{pmatrix}
,\qquad 
T=\begin{pmatrix}1&1\\0&1
\end{pmatrix}\,.
\end{equation}
The group $\Gamma$ acts on the upper half-plane $\mathcal{H} = \left\{ \tau  \in \mathds{C} \left| \mathop{\rm Im} \left( \tau  \right) > 0 \right. \right\}$ via linear fractional transformations. Under the action of the element $\gamma\in\Gamma$, the modulus $\tau$  transforms as
\begin{equation} 
	\tau\xmapsto{~\gamma~} \dfrac{a\tau+b}{c\tau+d},\qquad 
		\gamma =
		\begin{pmatrix}
			a&b\\
			c&d
		\end{pmatrix}
		\in \Gamma\,. 
\end{equation}
Because $\pm \mathds{1}_{2}$ (with $\mathds{1}_{2}$ the $2\times2$ identity) act trivially on $\tau$, the faithful action on $\tau$ is that of the projective group 
\begin{equation}
\overline{\Gamma} = \mathrm{PSL}(2,\mathds{Z}) \cong \Gamma / {\{\pm \mathds{1}_{2}\}}=\left\langle S,T\mid S^{2}=(ST)^{3}=1\right\rangle\,.
\end{equation}

There exists a class of vector-valued holomorphic functions $Y(\tau)=(Y_1(\tau),\dots,Y_d(\tau))^{T}$ defined on the upper half-plane $\mathcal{H}$ extended to include the cusp $\tau\to i\infty$. These functions transform in a nontrivial manner under modular transformations $\gamma\in \Gamma$ as 
\begin{equation}
  \label{eq:defVVMF}
  Y(\tau)\xmapsto{~\gamma~} Y(\gamma\tau)=J_k(\gamma, \tau) \rho(\gamma)\,Y(\tau)\,,
\end{equation}
where $k\in\mathds{Z}$ denotes the modular weight of $Y(\tau)$, $J_k(\gamma, \tau)=(c\tau+d)^{k}$ is the so-called automorphy factor and  $\rho(\gamma)$ provides a unitary irreducible representation of $\Gamma$ (or $\overline{\Gamma}$) with finite image. 
This representation induces a finite modular group $\mathcal{G}_{f}\cong \Gamma/\text{ker}(\rho)$ (or $\mathcal{G}_{f}\cong \overline{\Gamma}/\text{ker}(\rho)$), obtained as the quotient of  $\Gamma$ (or $\overline{\Gamma}$) by the finite-index normal subgroup $\text{ker}(\rho)$, i.e., the kernel of the representation $\rho$~\cite{Liu:2021gwa}. Functions satisfying the transformation law in Eq.~\eqref{eq:defVVMF} are called VVMFs or modular multiplets and play a fundamental role in modular flavor symmetry frameworks.

In this work, we focus on VVMFs transforming as the 1-, 2- and 3-$d$ irreducible representations\footnote{For $d\leq 3$, the MLDE method gives general closed form expressions for the lowest-weight VVMFs in terms of generalized hypergeometric functions. For $d > 3$, the MLDE module framework remains applicable, but such universal closed-form solutions are generally unavailable; instead, the $q$-expansions are typically obtained recursively. A four-dimensional example can be found in Ref.~\cite{Ding:2023ydy}.}. For these low-dimensional representations ($d\leq3$), the explicit expressions of lowest-weight VVMFs in the $T$-diagonal basis have been derived in Ref.~\cite{Liu:2021gwa} by solving MLDE. 
For the 1-$d$ irreducible representation $\rho_{\bm{1_{p}}}$, with $\rho_{\bm{1_{p}}}(S)=i^{p}$ and $\rho_{\bm{1}_{p}}(T)=e^{p\pi i/6}$ ($p=0,1,\ldots,11$), the lowest-weight is $k_{0}=p$, and the corresponding VVMF is
\begin{equation}\label{eq:1d_VVMFs}
 Y^{(p)}_{\bm{1_{p}}}=\eta^{2p}(\tau)\,,
\end{equation}
where $\eta(\tau)$ denotes the Dedekind eta function, defined as
\begin{equation}
\eta(\tau)=q^{1/24}\prod_{n=1}^\infty \left(1-q^n \right),\qquad
        q\equiv e^{2 \pi i\tau}\,.
\end{equation}
For the 2- and 3-dimensional irreducible representations, we adopt a basis in which the representation matrices of $T$ are diagonal,
\begin{equation}
\rho_{\bm{2}}(T) = \text{diag}(e^{2\pi i r_1}, e^{2\pi i r_2}), \qquad \rho_{\bm{3}}(T) = \text{diag}(e^{2\pi i r_1}, e^{2\pi i r_2} , e^{2\pi i r_3}), \qquad 0 \le r_j < 1\,. 
\end{equation}
In this basis, the lowest-weights are given by 
\begin{equation}\label{eq:min_2d3d_k0}
k_{0}=\left\{\begin{array}{lll}
6(r_1 + r_2) - 1  &  \text{for} & \rho_{\bm{2}}\,, \\
 4(r_1+r_2+r_3)-2  & \text{for}  &   \rho_{\bm{3}}\,.
\end{array}\right.
\end{equation}

The explicit expressions for the lowest-weight doublet $Y^{(k_{0})}_{\bm{2}}(\tau)$ and triplet $Y^{(k_{0})}_{\bm{3}}(\tau)$ can be written as~\cite{Liu:2021gwa}

{\small\begin{eqnarray}
\nonumber \hskip-0.2in Y^{(k_{0})}_{\bm{2}}(\tau)&=&\begin{pmatrix}
	\eta^{12(r_1+r_2)-2}\;(\frac{K}{1728})^{\frac{6(r_1-r_2)+1}{12}}\;
	{}_2F_{\!1}\!\Bigl(\frac{6(r_1-r_2)+1}{12},\frac{6(r_1-r_2)+5}{12};\,r_1-r_2+1;K\Bigr)\\[4pt]
	C\;\eta^{12(r_1+r_2)-2}\;(\frac{K}{1728})^{\frac{6(r_2-r_1)+1}{12}}\;
	{}_2F_{\!1}\!\Bigl(\frac{6(r_2-r_1)+1}{12},\frac{6(r_2-r_1)+5}{12};\,r_2-r_1+1;K\Bigr)
\end{pmatrix}, \\
\label{eq:2d3d_VVMFs} \hskip-0.2in Y^{(k_{0})}_{\bm{3}}(\tau)&=&\begin{pmatrix}
	\eta^{8(r_1+r_2+r_3)-4}(\frac{K}{1728})^{a_1}\;
	{}_3F_{\!2}\!\Bigl(a_1,a_1+\frac{1}{3},a_1+\frac{2}{3};
	r_1-r_2+1,r_1-r_3+1;K\Bigr)\\[4pt]
	C_1\,\eta^{8(r_1+r_2+r_3)-4}(\frac{K}{1728})^{a_2}\;
	{}_3F_{\!2}\!\Bigl(a_2,a_2+\frac{1}{3},a_2+\frac{2}{3};
	r_2-r_1+1,r_2-r_3+1;K\Bigr)\\[4pt]
	C_2\,\eta^{8(r_1+r_2+r_3)-4}(\frac{K}{1728})^{a_3}\;
	{}_3F_{\!2}\!\Bigl(a_3,a_3+\frac{1}{3},a_3+\frac{2}{3};
	r_3-r_1+1,r_3-r_2+1;K\Bigr)
\end{pmatrix}, 
\end{eqnarray}}
with 
\begin{equation}
 a_1=\frac{4r_1-2r_2-2r_3+1}{6},\quad  a_2=\frac{4r_2-2r_1-2r_3+1}{6}, \quad  a_3=\frac{4r_3-2r_1-2r_2+1}{6}\,.
\end{equation}
The function $K(\tau)$ and  the generalized hypergeometric series ${}_nF_{\!n-1}$ appearing in Eq.~\eqref{eq:2d3d_VVMFs} are defined as \begin{equation}\label{eq:Ktau_HypSeries}
K(\tau)=1-\frac{E^{2}_{6}(\tau)}{E^{3}_{4}(\tau)}, \qquad {}_nF_{\!n-1}(\alpha_1,\dots,\alpha_n;\beta_1,\dots,\beta_{n-1};z)=\sum_{m\ge0}
\frac{\prod_{j=1}^{n}(\alpha_j)_m}{\prod_{k=1}^{n-1}(\beta_k)_m}\,\frac{z^m}{m!}, 
\end{equation}
where $m\in \mathds{N}$, $\alpha_{i},\beta_{i}\in\mathds{C}$ and $(x)_{m}$ is the Pochhammer symbol defined by
\begin{equation}
(x)_{m} = \begin{cases}
1 \,, ~~~~~~~~~~~~~~~~~~~~~~~~~~~~~~\qquad m=0 \,, \\
x(x+1)\dots(x+m-1) \,,\ \qquad m\geq 1\,.
\end{cases}
\end{equation}
The functions $E_{4}(\tau)$ and $E_{6}(\tau)$ in Eq.~\eqref{eq:Ktau_HypSeries} are the Eisenstein series of weights 4 and 6:
\begin{equation}
E_4(\tau) = 1 + 240\sum_{n=1}^\infty \sigma_3(n)q^n, \qquad
E_6(\tau) = 1 - 504\sum_{n=1}^\infty \sigma_5(n)q^n\,,
\end{equation}
where $\sigma_k(n) = \sum_{x|n} x^k$ is the sum of the $k$th power of the divisors of $n$. The constants $C$ (for the doublet) and $C_1$, $C_2$ (for the triplet) depend on the explicit form of the representation matrices $\rho_{\bm{2}}(S)$ and $\rho_{\bm{3}}(S)$, respectively. They are fixed by requiring that the VVMFs satisfy the defining transformation property in Eq.~\eqref{eq:defVVMF}.

In the modular invariant framework~\cite{Feruglio:2017spp}, one introduces a complex modulus $\tau$ together with a set of chiral supermultiplets $\Phi_{I}$ whose behavior under the modular group $\Gamma$ (or $\overline{\Gamma}$) is prescribed as
\begin{equation}
\Phi_I\xmapsto{~\gamma~} (c\tau+d)^{-k_I}\rho_I(\gamma)\Phi_I\,, \qquad \gamma=\begin{pmatrix}
a & b \\ c & d
\end{pmatrix} \in \Gamma\,,
\end{equation}
where $k_I$ denotes the modular weight of $\Phi_I$, while $\rho_I$ specifies a unitary irreducible representation of the finite modular flavor symmetry $\mathcal{G}_f$, obtained as a quotient of the modular group $\Gamma$ (or $\overline{\Gamma}$) by a suitable normal subgroup.

In this framework, the K\"ahler potential is usually assumed to take the minimal modular invariant form~\cite{Feruglio:2017spp}
\begin{equation}
\label{eq:minKahler}
\mathcal{K}(\Phi_I,\bar{\Phi}_I; \tau,\bar{\tau}) =-\Lambda^2 \log(-i\tau+i\bar\tau)+ \sum_I (-i\tau+i\bar\tau)^{-k_I} |\Phi_I|^2\,,
\end{equation}
where $\Lambda$ represents the characteristic mass scale of the theory, typically identified with the ultraviolet cutoff. Once the modulus $\tau$ acquires a non-zero vacuum expectation value, this K\"ahler potential yields canonical kinetic terms for both the matter superfields and the modulus field.

The superpotential $\mathcal{W}(\Phi_I,\tau)$ is expanded in powers of the matter supermultiplets
\begin{equation}
\mathcal{W}(\Phi_I,\tau) =\sum_n Y_{I_1...I_n}(\tau)~ \Phi_{I_1} \cdots \Phi_{I_n}\,.
\end{equation}
Modular symmetry requires the coefficient functions $Y_{I_1...I_n}(\tau)$ to be VVMFs. Their transformation law under $\Gamma$ is
\begin{equation}
Y_{I_1...I_n}(\tau)\xmapsto{~\gamma~}Y_{I_1...I_n}(\gamma\tau)=(c\tau+d)^{k_Y}\rho_{Y}(\gamma)Y_{I_1...I_n}(\tau)\,,
\end{equation}
where the modular weight and representation assignment satisfy
\begin{equation}
 k_Y=k_{I_1}+\cdots+k_{I_n}\,, \qquad    \rho_Y\otimes\rho_{I_1}\otimes\cdots\otimes\rho_{I_n}\ni\bm{1}\,,
\end{equation}
with $\bm{1}$ being trivial singlet of $\mathcal{G}_{f}$. 

\section{\label{secVVMFs_Delta96}VVMFs and fixed points of the finite modular group $\Delta(96)$ }

All VVMFs associated with the  $\Delta(96)$  group, collectively denoted by  $\mathcal{M}(\Delta(96))$, decompose into modules associated with the irreducible representations of $\Delta(96)$, so that the full space admits the direct sum decomposition
\begin{equation}
\mathcal{M}(\Delta(96))=\mathcal{M}(\bm{1_{0}})\oplus\mathcal{M}(\bm{1_{1}})\oplus\mathcal{M}(\bm{2})\oplus \mathcal{M}(\bm{3_{0}})\oplus \mathcal{M}(\bm{3_{1}}) \oplus \mathcal{M}(\bm{\bar{3}_{0}})\oplus \mathcal{M}(\bm{\bar{3}_{1}})\oplus \mathcal{M}(\bm{\hat{3}_{0}})\oplus \mathcal{M}(\bm{\hat{3}_{1}})\oplus\mathcal{M}(\bm{6})\,.
\end{equation}
According to the general theory of VVMFs~\cite{Liu:2021gwa}, for each irreducible representation $\rho$ of dimension $d$, the space $\mathcal{M}(\rho) = \bigoplus^{\infty}_{k=0}\mathcal{M}_k(\rho)$ is a free graded module of rank $d$ over the ring of scalar modular forms $\mathcal{M}(\bm{1})=\mathds{C}[E_4,E_6]$, whose structure is completely characterized by the weight profile $(k_0,k_1,\dots,k_{d-1})$ of a chosen basis\footnote{For low dimensions, the explicit weight profiles are $d=1:(k_0)$,\; $d=2:(k_0,\,k_0+2)$ and $d=3 :(k_0,\,k_0+2,\,k_0+4)$, with the lowest-weight $k_0$ given by Eq.~\eqref{eq:1d_VVMFs} and Eq.~\eqref{eq:min_2d3d_k0}. Additionally, the weight profile of the 6-$d$ case is $d=6 :(k_0,\,k_0+2,\,k_0+2,\,k_0+4,\,k_0+4,\,k_0+6)$, and the lowest-weight $k_0=2$.}, where $k_0$ denotes the minimal weight. The dimensions of the graded components $\mathcal{M}_k(\rho)$ are encoded in the Hilbert–Poincaré series~\cite{marks2009classification}:
\begin{equation}
P(\mathcal{M}(\rho))(t) \;:=\; \sum_{k\ge k_0} t^k \dim\mathcal{M}_k(\rho)
\;=\; \frac{t^{k_0}+t^{k_1}+\cdots+t^{k_{d-1}}}{(1-t^4)(1-t^6)} .
\end{equation}
Expanding the right‑hand side as a power series in $t$ directly yields $\dim\mathcal{M}_k(\rho)$, i.e., the number of linearly independent VVMFs of weight $k$ in the representation $\rho$.

\begin{table}[t!]
\begin{center}
\renewcommand{\tabcolsep}{4.0mm}
\renewcommand{\arraystretch}{1.1}
\begin{tabular}{|c|c|c|c|c|c|c|c|c|c|c|c|}\hline\hline
\diagbox{$\bm{r}$}{$k$}&$0$&$2$&$4$&$6$&$8$&$10$&$12$&$14$&$16$&$18$&$20$\\
			\hline
$\bm{1_{0}}$&\cellcolor{yellow!70}$1$&&$1$&$1$&$1$&$1$&$2$&$1$&$2$&$2$&$2$\\  \hline 
$\bm{1_{1}}$&&&&\cellcolor{yellow!70}$1$&&$1$&$1$&$1$&$1$&$2$&$1$\\ \hline 
$\bm{2}$&&\cellcolor{yellow!70}$1$&$1$&$1$&$2$&$2$&$2$&$3$&$3$&$3$&$4$\\ \hline
$\bm{3_{0}}$&&&&\cellcolor{yellow!70}$1$&$1$&$2$&$2$&$3$&$3$&$4$&$4$\\ \hline
$\bm{3_{1}}$&&&\cellcolor{yellow!70}$1$&$1$&$2$&$2$&$3$&$3$&$4$&$4$&$5$\\ \hline
$\bm{\bar{3}_{0}}$&&\cellcolor{yellow!70}$1$&$1$&$2$&$2$&$3$&$3$&$4$&$4$&$5$&$5$\\
\hline
$\bm{\bar{3}_{1}}$&&&\cellcolor{yellow!70}$1$&$1$&$2$&$2$&$3$&$3$&$4$&$4$&$5$\\ \hline
$\bm{\hat{3}_{0}}$&&\cellcolor{yellow!70}$1$&$1$&$2$&$2$&$3$&$3$&$4$&$4$&$5$&$5$\\
\hline
$\bm{\hat{3}_{1}}$&&&\cellcolor{yellow!70}$1$&$1$&$2$&$2$&$3$&$3$&$4$&$4$&$5$\\
\hline
$\bm{6}$&&\cellcolor{yellow!70}$1$&$2$&$3$&$4$&$5$&$6$&$7$&$8$&$9$&$10$\\ \hline\hline
\end{tabular}
\caption{\label{tab:VVMF dimensions}Dimensions of VVMF spaces $\mathcal{M}_k(\rho)$ for the relevant irreducible representations, where $k$ denotes the modular weight and $\bm{r}$ labels the representation of $\Delta(96)$. Blank entries indicate $\dim\mathcal{M}_k(\rho)=0$, namely the absence of a 
corresponding VVMF at that weight and representation; such vanishing entries 
are also commonly referred to as modular zeros~\cite{Liu:2025lym}.}
\end{center}
\end{table}

The group theory of the finite modular group $\Delta(96)$ is summarized in appendix~\ref{sec:Delta96_group_theory}. The representation matrices of the generators $S$ and $T$ for all irreducible representations are given in table~\ref{tab:Delta96_Reps}. In the chosen basis, $T$ is diagonal in all irreducible representations. Note that each of these representations can be regarded as a finite image representation of the modular group $\overline{\Gamma}$, and therefore admits an associated space of VVMFs.  Applying the formalism outlined above, we have determined, for each irreducible representation, the dimension $\dim \mathcal{M}_k(\rho)$ of the corresponding VVMF space for weights up to $k \leq 20$. The results are presented in table~\ref{tab:VVMF dimensions}.

In the construction of the Modular Littlest Seesaw models studied in this paper, we employ only triplet VVMFs. 
Guided by the principle of minimality and predictivity, we restrict our analysis to VVMFs of weights for which a unique triplet VVMF exists in each of the six representations $\bm{3_{m}}$, $\bm{\bar{3}_{m}}$ and $\bm{\hat{3}_{m}}$ ($m=0,1$). As shown in table~\ref{tab:VVMF dimensions}, these VVMFs include both the lowest-weight and next-to-lowest-weight triplet VVMFs for all six triplet representations. 
The next-to-lowest-weight VVMFs are obtained by combining the corresponding lowest-weight triplet VVMFs with the lowest-weight doublet VVMF in the representation $\bm{2}$. Their explicit expressions are presented below.

Using the representation matrices of the generators $S$ and $T$ for the two-dimensional irreducible representation $\bm{2}$ of $\Delta(96)$ in table~\ref{tab:Delta96_Reps}, together with the general expression in Eq.~\eqref{eq:2d3d_VVMFs}, we obtain the corresponding lowest-weight ($k_0=2$) doublet VVMF:
\begin{equation}
	Y_{\bm{2}}^{(2)}(\tau)=
	\begin{pmatrix}
		\eta^{4}(\tau)\Bigl(\dfrac{K(\tau)}{1728}\Bigr)^{-\frac{1}{6}}
		{}_2F_{\!1}\!\Bigl(-\frac{1}{6},\frac{1}{6};\,\frac{1}{2};K(\tau)\Bigr)\\[4pt]
		8\sqrt{3}\,\eta^{4}(\tau)\Bigl(\dfrac{K(\tau)}{1728}\Bigr)^{\frac{1}{3}}
		{}_2F_{\!1}\!\Bigl(\frac{1}{3},\frac{2}{3};\,\frac{3}{2};K(\tau)\Bigr)
	\end{pmatrix}\,,
\end{equation}
where the functions $K(\tau)$ and ${}_2F_{\!1}$ are defined in Eq.~\eqref{eq:Ktau_HypSeries}.

The finite modular group $\Delta(96)$ possesses six three-dimensional irreducible representations, listed in table~\ref{tab:Delta96_Reps}. Applying the above construction to each representation yields the corresponding lowest-weight triplet VVMFs. The two lowest-weight ($k_{0} = 2$) triplet VVMFs are given by:
\begin{eqnarray}
\nonumber 	Y_{\bm{\bar{3}_{0}}}^{(2)}(\tau) &=&
	\begin{pmatrix}
		\eta^{4}(\tau)\Bigl(\dfrac{K(\tau)}{1728}\Bigr)^{\frac{11}{24}}
		{}_3F_{\!2}\!\Bigl(\frac{11}{24},\frac{19}{24},\frac{9}{8};\,\frac{11}{8},\frac{3}{2};\,K(\tau)\Bigr)\\[4pt]
		\dfrac{\sqrt{2}}{2}\,\eta^{4}(\tau)\Bigl(\dfrac{K(\tau)}{1728}\Bigr)^{\frac{1}{12}}
		{}_3F_{\!2}\!\Bigl(\frac{1}{12},\frac{5}{12},\frac{3}{4};\,\frac{5}{8},\frac{9}{8};\,K(\tau)\Bigr)\\[4pt]
		-\dfrac{1}{2}\,\eta^{4}(\tau)\Bigl(\dfrac{K(\tau)}{1728}\Bigr)^{-\frac{1}{24}}
		{}_3F_{\!2}\!\Bigl(-\frac{1}{24},\frac{7}{24},\frac{5}{8};\,\frac{1}{2},\frac{7}{8};\,K(\tau)\Bigr)
	\end{pmatrix},\\
	Y_{\bm{\hat{3}_{0}}}^{(2)}(\tau) &=&
	\begin{pmatrix}
		\eta^{4}(\tau)\Bigl(\dfrac{K(\tau)}{1728}\Bigr)^{-\frac{1}{6}}
		{}_3F_{\!2}\!\Bigl(-\frac{1}{6},\frac{1}{6},\frac{1}{2};\,\frac{3}{4},\frac{1}{4};\,K(\tau)\Bigr)\\[4pt]
		-4\sqrt{2}\,\eta^{4}(\tau)\Bigl(\dfrac{K(\tau)}{1728}\Bigr)^{\frac{1}{12}}
		{}_3F_{\!2}\!\Bigl(\frac{1}{12},\frac{5}{12},\frac{3}{4};\,\frac{5}{4},\frac{1}{2};\,K(\tau)\Bigr)\\[4pt]
		-16\sqrt{2}\,\eta^{4}(\tau)\Bigl(\dfrac{K(\tau)}{1728}\Bigr)^{\frac{7}{12}}
		{}_3F_{\!2}\!\Bigl(\frac{7}{12},\frac{11}{12},\frac{5}{4};\,\frac{7}{4},\frac{3}{2};\,K(\tau)\Bigr)
	\end{pmatrix}\,.
\end{eqnarray}

The  three triplet  VVMFs with lowest-weight  $k_{0}=4$ are taken to be:
\begin{eqnarray}
\nonumber 	Y_{\bm{3_{1}}}^{(4)}(\tau) &=&
	\begin{pmatrix}
		\eta^{8}(\tau)\Bigl(\dfrac{K(\tau)}{1728}\Bigr)^{\frac{13}{24}}
		{}_3F_{\!2}\!\Bigl(\frac{13}{24},\frac{7}{8},\frac{29}{24};\,\frac{13}{8},\frac{3}{2};\,K(\tau)\Bigr)\\[4pt]
		-\dfrac{\sqrt{2}}{8}\,\eta^{8}(\tau)\Bigl(\dfrac{K(\tau)}{1728}\Bigr)^{-\frac{1}{12}}
		{}_3F_{\!2}\!\Bigl(-\frac{1}{12},\frac{1}{4},\frac{7}{12};\,\frac{3}{8},\frac{7}{8};\,K(\tau)\Bigr)\\[4pt]
		\dfrac{1}{2}\,\eta^{8}(\tau)\Bigl(\dfrac{K(\tau)}{1728}\Bigr)^{\frac{1}{24}}
		{}_3F_{\!2}\!\Bigl(\frac{1}{24},\frac{3}{8},\frac{17}{24};\,\frac{1}{2},\frac{9}{8};\,K(\tau)\Bigr)
	\end{pmatrix}\,,\\
\nonumber 	Y_{\bm{\bar{3}_{1}}}^{(4)}(\tau) &=&
	\begin{pmatrix}
		\eta^{8}(\tau)\Bigl(\dfrac{K(\tau)}{1728}\Bigr)^{-\frac{5}{24}}
		{}_3F_{\!2}\!\Bigl(-\frac{5}{24},\frac{1}{8},\frac{11}{24};\,\frac{3}{8},\frac{1}{2};\,K(\tau)\Bigr)\\[4pt]
		-16\sqrt{2}\,\eta^{8}(\tau)\Bigl(\dfrac{K(\tau)}{1728}\Bigr)^{\frac{5}{12}}
		{}_3F_{\!2}\!\Bigl(\frac{5}{12},\frac{3}{4},\frac{13}{12};\,\frac{13}{8},\frac{9}{8};\,K(\tau)\Bigr)\\[4pt]
		-10\,\eta^{8}(\tau)\Bigl(\dfrac{K(\tau)}{1728}\Bigr)^{\frac{7}{24}}
		{}_3F_{\!2}\!\Bigl(\frac{7}{24},\frac{5}{8},\frac{23}{24};\,\frac{3}{2},\frac{7}{8};\,K(\tau)\Bigr)
	\end{pmatrix}\,,\\
	Y_{\bm{\hat{3}_{1}}}^{(4)}(\tau) &=&
	\begin{pmatrix}
		\eta^{8}(\tau)\Bigl(\dfrac{K(\tau)}{1728}\Bigr)^{\frac{1}{6}}
		{}_3F_{\!2}\!\Bigl(\frac{1}{6},\frac{1}{2},\frac{5}{6};\,\frac{3}{4},\frac{5}{4};\,K(\tau)\Bigr)\\[4pt]
		\sqrt{2}\,\eta^{8}(\tau)\Bigl(\dfrac{K(\tau)}{1728}\Bigr)^{\frac{5}{12}}
		{}_3F_{\!2}\!\Bigl(\frac{5}{12},\frac{3}{4},\frac{13}{12};\,\frac{5}{4},\frac{3}{2};\,K(\tau)\Bigr)\\[4pt]
		-\dfrac{\sqrt{2}}{4}\,\eta^{8}(\tau)\Bigl(\dfrac{K(\tau)}{1728}\Bigr)^{-\frac{1}{12}}
		{}_3F_{\!2}\!\Bigl(-\frac{1}{12},\frac{1}{4},\frac{7}{12};\,\frac{3}{4},\frac{1}{2};\,K(\tau)\Bigr)
	\end{pmatrix}\,.
\end{eqnarray}

The explicit expression of the triplet VVMF of lowest-weight $k_{0}=6$ is 
\begin{equation}
	Y_{\bm{3_{0}}}^{(6)}(\tau) =
	\begin{pmatrix}
		\eta^{12}(\tau)\Bigl(\dfrac{K(\tau)}{1728}\Bigr)^{-\frac{1}{8}}
		{}_3F_{\!2}\!\Bigl(-\frac{1}{8},\frac{5}{24},\frac{13}{24};\,\frac{5}{8},\frac{1}{2};\,K(\tau)\Bigr)\\[4pt]
		-4\sqrt{2}\,\eta^{12}(\tau)\Bigl(\dfrac{K(\tau)}{1728}\Bigr)^{\frac{1}{4}}
		{}_3F_{\!2}\!\Bigl(\frac{1}{4},\frac{7}{12},\frac{11}{12};\,\frac{11}{8},\frac{7}{8};\,K(\tau)\Bigr)\\[4pt]
		-6\,\eta^{12}(\tau)\Bigl(\dfrac{K(\tau)}{1728}\Bigr)^{\frac{3}{8}}
		{}_3F_{\!2}\!\Bigl(\frac{3}{8},\frac{17}{24},\frac{25}{24};\,\frac{3}{2},\frac{9}{8};\,K(\tau)\Bigr)
	\end{pmatrix}\,.
\end{equation}

Appendix~\ref{sec:qes_VVMFs} collects the $q$-expansions of the lowest-weight doublet and triplet VVMFs, together with the tensor-product decompositions used to construct the corresponding next-to-lowest-weight VVMFs. It also provides the $q$-expansions of the next-to-lowest-weight triplet VVMFs for all six three-dimensional irreducible representations of $\Delta(96)$. For completeness, the explicit construction of the six-dimensional VVMFs is included as well.

\subsection{\label{sec:VVMFs_VEVs}VVMFs at fixed points of the finite modular group $\Delta(96)$}

A point $\tau_{\gamma}\in\mathcal{H}^{*}$ is called a fixed point of a modular transformation $\gamma\in\overline{\Gamma}$ if it is invariant under the action of $\gamma$, namely $\gamma\tau_{\gamma}=\tau_{\gamma}$. Here $\mathcal{H}^{*}=\mathcal{H}\cup\mathds{Q}\cup\{i\infty\}$ denotes the compactified upper half-plane, including the cusp fixed by $T$. The set of all elements of $\overline{\Gamma}$ that leave $\tau_{\gamma}$ invariant forms the corresponding stabilizer subgroup
\begin{equation}
\Gamma_{\gamma}=\{\tilde\gamma\in\overline{\Gamma}\mid \tilde\gamma\tau_{\gamma}=\tau_{\gamma}\}\,.
\end{equation}

In ordinary single-modulus modular models, all sectors share the same modulus. A modular transformation then conjugates all residual symmetries simultaneously by the same finite modular element, and therefore leaves the physical predictions unchanged. The modulus can consequently be restricted to the fundamental domain
\begin{equation}
\label{eq:fundamental_domain}
\mathcal{D}=\left\{\tau\in\mathcal{H}\,\Big|\,|\tau|\geq 1,\ -\frac12\leq \operatorname{Re}\tau\leq 0\right\}\cup\left\{\tau\in\mathcal{H}\,\Big|\,|\tau|>1,\ 0<\operatorname{Re}\tau<\frac12\right\}\cup\{i\infty\}\,.
\end{equation}
In the construction of Modular Littlest Seesaw models, however, three independent moduli may be stabilized at different fixed points. In general, these three fixed points cannot be brought simultaneously into the fundamental domain by the same modular transformation. We therefore first classify the inequivalent fixed-point representatives in $\mathcal{H}^{*}$.

All fixed points in $\mathcal{H}^{*}$ are related by modular transformations to one of the three fixed points in the standard fundamental domain $\mathcal{D}$\footnote{The point $\tau_{TS}=\frac12+i\frac{\sqrt3}{2}$ is also a fixed point, but it is related to $\tau_{ST}$ by a modular transformation, for example $T\tau_{ST}=\tau_{TS}$. Since the fundamental domain defined in Eq.~\eqref{eq:fundamental_domain} includes only the left boundary, $\tau_{ST}$ is chosen as the unique representative of this equivalence class.}~\cite{Ding:2019gof},
\begin{equation}
\label{eq:FD_fixed_points}
\tau_S=i,\qquad \tau_{ST}=\omega,\qquad \tau_T=i\infty\,,
\end{equation}
where we denote $\omega\equiv e^{2\pi i/3}$. Their stabilizer subgroups are
\begin{equation}
\label{eq:FD_stabilizers}
\Gamma_S=Z_2^S=\{1,S\},\qquad \Gamma_{ST}=Z_3^{ST}=\{1,ST,(ST)^2\},\qquad \Gamma_T=Z_{\infty}^{T}=\{1,T,T^{-1},\cdots\}\,.
\end{equation}
Thus the infinitely many fixed points of the full modular group are partitioned into three inequivalent $\overline{\Gamma}$-orbits. Let $\tau_{\gamma_0}\in\{\tau_S,\tau_{ST},\tau_T\}$ be one of these representatives, with stabilizer subgroup $\Gamma_{\gamma_0}$ generated by $\gamma_0$. Any fixed point in the same orbit can be written as $\tau_{\gamma'}=\gamma\tau_{\gamma_0}$, and its stabilizer  subgroup is the conjugate subgroup $\Gamma_{\gamma'}=\gamma\Gamma_{\gamma_0}\gamma^{-1}$.

For Modular Littlest Seesaw model building, the relevant objects are not the fixed points themselves, but the vacuum alignments of the VVMFs evaluated at them. Let $\rho$ be the finite-image representation carried by a VVMF, let $K\equiv\ker(\rho)$, and let $\mathcal{G}_f\cong\overline{\Gamma}/K$ be the corresponding finite modular group. Using the transformation law in Eq.~\eqref{eq:defVVMF}, for $\tau_{\gamma'}=\gamma\tau_{\gamma_0}$ one obtains
\begin{equation}\label{eq:fixed_point_condition}
Y^{(k)}(\tau_{\gamma'})=Y^{(k)}(\gamma\tau_{\gamma_0})=J_k(\gamma,\tau_{\gamma_0})\rho(g)Y^{(k)}(\tau_{\gamma_0})\,,\qquad g=\gamma K\in\mathcal{G}_f\,.
\end{equation}
At the seed fixed point itself,
\begin{equation}
Y^{(k)}(\tau_{\gamma_0})=Y^{(k)}(\gamma_0\tau_{\gamma_0})=J_k(\gamma_0,\tau_{\gamma_0})\rho(\gamma_0)Y^{(k)}(\tau_{\gamma_0})\,.
\end{equation}
Therefore $Y^{(k)}(\tau_{\gamma_0})$ is an eigenvector of $\rho(\gamma_0)$ with eigenvalue $J_k^{-1}(\gamma_0,\tau_{\gamma_0})$. The overall factor $J_k(\gamma,\tau_{\gamma_0})$ in Eq.~\eqref{eq:fixed_point_condition} can be absorbed into the corresponding Yukawa coupling, so the physical alignment is determined by $\rho(g)Y^{(k)}(\tau_{\gamma_0})$. Two types of redundancy arise in the construction of VVMF vacuum alignments at the fixed points. First, right multiplication of a modular transformation $\gamma$ by an element of the stabilizer subgroup $\Gamma_{\gamma_0}$ leaves the seed fixed point unchanged, so different transformations related in this way generate the same fixed point. Second, left multiplication of $\gamma$ by an element of the kernel $K$ produces an equivalent point on the modular curve associated with $K$ and changes the VVMF only by an overall automorphy factor, leaving the physical vacuum alignment unchanged.

Hence the inequivalent vacuum alignments in the orbit of $\tau_{\gamma_0}$ are labelled by the double cosets
\begin{equation}\label{eq:double_cosets}
K\backslash\overline{\Gamma}/\Gamma_{\gamma_0}\cong \mathcal{G}_f/\mathcal{G}_{\gamma_0}\,,
\end{equation}
where $\mathcal{G}_{\gamma_0}\equiv \pi(\Gamma_{\gamma_0})\subset\mathcal{G}_f$ and $\pi:\overline{\Gamma}\to\mathcal{G}_f$ is the natural projection. Choosing left coset representatives $\{g_i\}$ of $\mathcal{G}_f/\mathcal{G}_{\gamma_0}$, all inequivalent alignments are generated by
\begin{equation}
\label{eq:coset_alignments}
Y^{(k)}(g_i\tau_{\gamma_0})\propto \rho(g_i)Y^{(k)}(\tau_{\gamma_0})\,,\qquad \mathcal{G}_f=\bigcup_i g_i\mathcal{G}_{\gamma_0}\,.
\end{equation}
In practice, it is therefore sufficient to determine the VVMF vacuum alignments at the three fixed points in the fundamental domain. All remaining inequivalent alignments are obtained by acting with the representation matrices of the corresponding finite-group coset representatives.

Applying this construction to the finite modular group $\Delta(96)$, the images of the three stabilizer subgroups $\Gamma_S$, $\Gamma_{ST}$ and $\Gamma_T$ are the cyclic subgroups $Z_2^S$, $Z_3^{ST}$ and $Z_8^T$, respectively. Since $|\Delta(96)|=96$, the numbers of inequivalent representatives in the three orbits are
\begin{equation}
\label{eq:Delta96_fixed_point_numbers}
N_S=\frac{|\Delta(96)|}{|Z_2^S|}=\frac{96}{2}=48\,,\quad N_{ST}=\frac{|\Delta(96)|}{|Z_3^{ST}|}=\frac{96}{3}=32\,,\quad N_T=\frac{|\Delta(96)|}{|Z_8^T|}=\frac{96}{8}=12\,.
\end{equation}
Altogether, this gives $48+32+12=92$ inequivalent fixed-point representatives associated with the modular $\Delta(96)$ symmetry, as listed in table~\ref{tab:fixed_points}. The representatives originating from $\tau_T$ are cusps of the modular curve, while those originating from $\tau_S$ and $\tau_{ST}$ are interior fixed points. Every residual symmetry preserved at such a fixed point is conjugate to one of the three cyclic subgroups $Z_2^S$, $Z_3^{ST}$ and $Z_8^T$.

\begin{figure}[t!]
\centering
\makebox[\linewidth][c]{\includegraphics[width=1.1\linewidth]{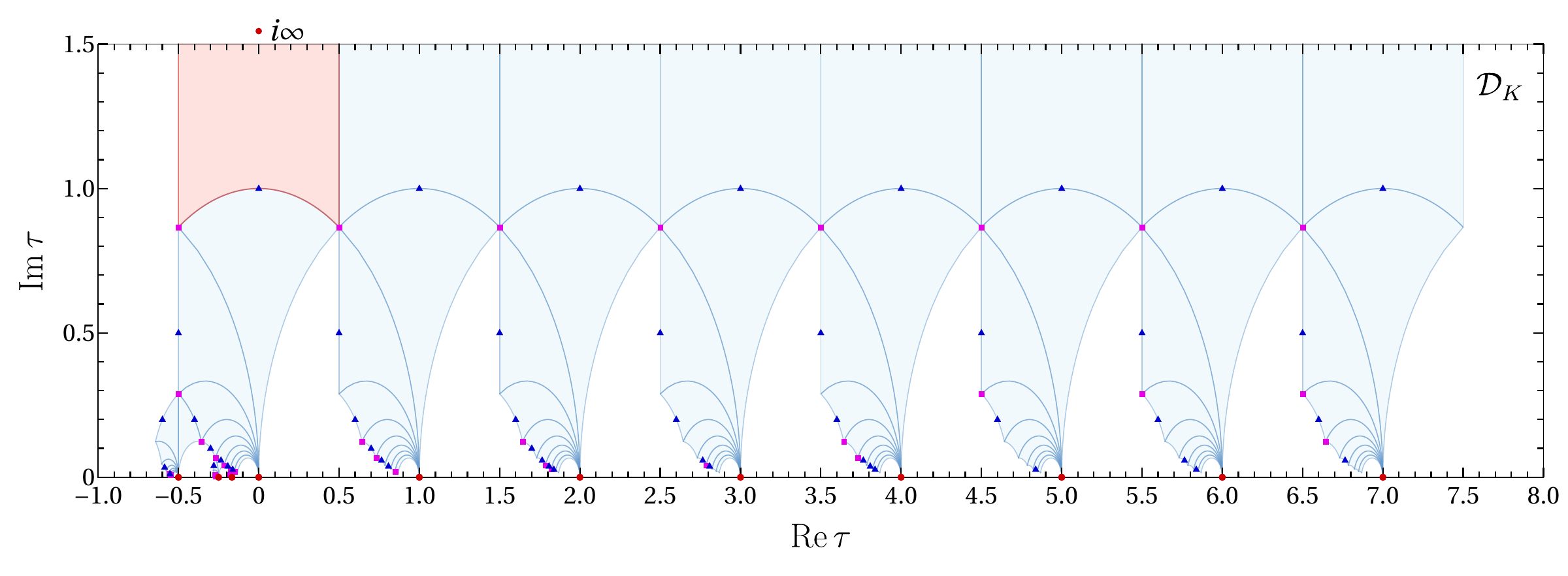}}
\caption{\label{fig:Ker_Delta96_fundamental_domain}
A compactified fundamental domain $\mathcal{D}_{K}$ of $K=\ker(\Delta(96))$ and the images of the $92$ inequivalent fixed points. The tiles are generated from the fundamental domain $\mathcal{D}$ of $\overline{\Gamma}$ by the determinant-one matrix representatives in Eq.~\eqref{eq:Delta_96_CC}. Pink squares denote the images of $\tau_{ST}=\omega$, blue triangles denote the images of $\tau_S=i$, and red circles denote the cusp images of $\tau_T=i\infty$.
}
\end{figure}

It is also useful to give a geometric realization of the above fixed-point
representatives. Let $K=\ker(\Delta(96))$ be the finite-index normal
subgroup of the modular group whose quotient gives the finite modular group
$\Delta(96)$. A fundamental domain for $K$ can be obtained by unfolding the
standard fundamental domain $\mathcal{D}$ of $\Gamma$. We choose the determinant-one
matrix representatives $\mathcal{R}=\{\gamma_i\}$ of the $\Delta(96)$ elements given in
Eq.~\eqref{eq:Delta_96_CC}, such that
\begin{equation}
	\overline{\Gamma}=\bigsqcup_{\gamma_i\in\mathcal{R}} K\gamma_i .
\end{equation}
The corresponding $K$-fundamental domain is then
\begin{equation}
	\label{eq:Ker_Delta96_domain}
	\mathcal{D}_{K}\equiv\bigcup_{\gamma_i\in\mathcal{R}}\gamma_i\mathcal{D}.
\end{equation}
Then $\mathcal{H}^{*}=\overline{\Gamma}\mathcal{D}=\bigcup_{\gamma_i\in\mathcal{R}}K\gamma_i\mathcal{D} =K\mathcal{D}_{K}$. The boundary convention of $\mathcal{D}_{K}$ is inherited from that of $\mathcal{D}$. Thus $\mathcal{D}_{K}$
should be viewed as a representative fundamental region for the modular curve
$K \backslash\mathcal{H}^{*}$, with the usual boundary identifications induced by
the action of $K$. Since the compactified domain $\mathcal{D}$ contains the cusp
$i\infty$, the compactified region $\mathcal{D}_{K}$ also contains the cusp images
$\gamma_i(i\infty)$. The representatives in Eq.~\eqref{eq:Delta_96_CC} are chosen so that the
corresponding tiles form a connected region: if two representatives differ by
right multiplication with $S$ or $T^{\pm1}$, the closures of the two tiles share
a boundary edge. Therefore the union displayed in
figure~\ref{fig:Ker_Delta96_fundamental_domain} gives a connected compactified
fundamental domain for $K$. Since $K$ is normal in $\Gamma$, the modular action descends to the quotient
curve $K\backslash\mathcal{H}^{*}$, whose deck transformation group is
$\Gamma/K\simeq\Delta(96)$. The $92$ inequivalent fixed-point representatives
obtained above can therefore be chosen in $\mathcal{D}_K$. They are the images of the three seed fixed points $\tau_S=i$, $\tau_{ST}=\omega$ and $\tau_T=i\infty$, after the double-coset redundancies in Eq.~\eqref{eq:double_cosets} have been removed. In this
way, figure~\ref{fig:Ker_Delta96_fundamental_domain} displays the fixed points which generate the inequivalent VVMF vacuum alignments.

For the lowest- and next-to-lowest-weight VVMFs in the six three-dimensional irreducible representations of $\Delta(96)$, the vacuum alignments at all inequivalent fixed points can be constructed systematically. Since the complete list is rather lengthy, table~\ref{tab:3D_VEVs_FDfixed} presents only the generating alignments at the three fundamental fixed points. The alignments at all remaining inequivalent fixed points are obtained by acting on these generating alignments with the representation matrices of the corresponding left coset representatives $g_i$ listed in table~\ref{tab:fixed_points}. Altogether, this procedure generates 792 non-vanishing vacuum alignments\footnote{Different fixed points may lead to identical alignment directions.}. These residual-symmetry-induced alignments severely constrain the allowed Yukawa structures and, consequently, lead to highly predictive lepton mass matrices and flavor mixing.

\begin{table}[t!]
	\begin{center}
		\renewcommand{\tabcolsep}{0.95mm}
		\renewcommand{\arraystretch}{1.15}
		\begin{tabular}{|c|c|c||c|c|c||c|c|c|}
			\hline\hline
$\mathcal{G}_{\gamma^{\prime}}$&$g_i$&$\tau_{\gamma^{\prime}}$
&$\mathcal{G}_{\gamma^{\prime}}$&$g_i$&$\tau_{\gamma^{\prime}}$
&$\mathcal{G}_{\gamma^{\prime}}$&$g_i$&$\tau_{\gamma^{\prime}}$ \\
			\hline
			
		    \multirow{4}{*}{$Z_2^{S}$}&\cellcolor{yellow!70}$1$&\cellcolor{yellow!70}$i$
			&\multirow{4}{*}{$Z_2^{T^2ST^6}$}&$T^2$&$2+i$
			&\multirow{4}{*}{$Z_2^{T^6ST^2}$}&$T^6$&$6+i$\\
			\cline{2-3}\cline{5-6}\cline{8-9}
			&$T^6ST^6$&$\frac{216}{37}+\frac{i}{37}$
			&&$T^6ST^4$&$\frac{98}{17}+\frac{i}{17}$
			&&$ST^2$&$-\frac{2}{5}+\frac{i}{5}$\\
			\cline{2-3}\cline{5-6}\cline{8-9}
			&$T^2ST^2$&$\frac{8}{5}+\frac{i}{5}$
			&&$ST^6$&$-\frac{6}{37}+\frac{i}{37}$
			&&$T^4ST^6$&$\frac{142}{37}+\frac{i}{37}$\\
			\cline{2-3}\cline{5-6}\cline{8-9}
			&$T^4ST^4$&$\frac{64}{17}+\frac{i}{17}$
			&&$T^4ST^2$&$\frac{18}{5}+\frac{i}{5}$
			&&$T^2ST^4$&$\frac{30}{17}+\frac{i}{17}$\\
			\hline
			
			\multirow{4}{*}{$Z_2^{T^4ST^4}$}&$T^4$&$4+i$
			&\multirow{4}{*}{$Z_2^{ST^6ST^3}$}&$T^2ST^3$&$\frac{17}{10}+\frac{i}{10}$
			&\multirow{4}{*}{$Z_2^{ST^2ST^5}$}&$ST^5$&$-\frac{5}{26}+\frac{i}{26}$\\
			\cline{2-3}\cline{5-6}\cline{8-9}
			&$ST^4$&$-\frac{4}{17}+\frac{i}{17}$
			&&$ST^3$&$-\frac{3}{10}+\frac{i}{10}$
			&&$TST^3$&$\frac{7}{10}+\frac{i}{10}$\\
			\cline{2-3}\cline{5-6}\cline{8-9}
			&$T^6ST^2$&$\frac{28}{5}+\frac{i}{5}$
			&&$TST^5$&$\frac{21}{26}+\frac{i}{26}$
			&&$T^4ST^5$&$\frac{99}{26}+\frac{i}{26}$\\
			\cline{2-3}\cline{5-6}\cline{8-9}
			&$T^2ST^6$&$\frac{68}{37}+\frac{i}{37}$
			&&$T^3ST^5$&$\frac{73}{26}+\frac{i}{26}$
			&&$T^2ST^5$&$\frac{47}{26}+\frac{i}{26}$\\
			\hline
			
			\multirow{4}{*}{$Z_2^{T^3ST^5}$}&$T^3$&$3+i$
			&\multirow{4}{*}{$Z_2^{T^5ST^3}$}&$T^5$&$5+i$
			&\multirow{4}{*}{$Z_2^{TST^7}$}&$T$&$1+i$\\
			\cline{2-3}\cline{5-6}\cline{8-9}
			&$ST^2ST^3$&$-\frac{17}{29}+\frac{i}{29}$
			&&$T^7ST^2$&$\frac{33}{5}+\frac{i}{5}$
			&&$ST^2ST$&$-\frac{3}{5}+\frac{i}{5}$\\
			\cline{2-3}\cline{5-6}\cline{8-9}
			&$T^7ST^4$&$\frac{115}{17}+\frac{i}{17}$
			&&$ST^2ST^5$&$-\frac{47}{85}+\frac{i}{85}$
			&&$T^3ST^2$&$\frac{13}{5}+\frac{i}{5}$\\
			\cline{2-3}\cline{5-6}\cline{8-9}
			&$T^5ST^2$&$\frac{23}{5}+\frac{i}{5}$
			&&$TST^4$&$\frac{13}{17}+\frac{i}{17}$
			&&$ST^4ST$&$-\frac{7}{25}+\frac{i}{25}$\\
			\hline
			
			\multirow{4}{*}{$Z_2^{ST^2ST}$}&$ST$&$-\frac{1}{2}+\frac{i}{2}$
			&\multirow{4}{*}{$Z_2^{T^7ST}$}&$TST^2$&$\frac{3}{5}+\frac{i}{5}$
			&\multirow{4}{*}{$Z_2^{ST^6ST^7}$}&$TST$&$\frac{1}{2}+\frac{i}{2}$\\
			\cline{2-3}\cline{5-6}\cline{8-9}
			&$T^6ST$&$\frac{11}{2}+\frac{i}{2}$
			&&$T^5ST^6$&$\frac{179}{37}+\frac{i}{37}$
			&&$T^7ST$&$\frac{13}{2}+\frac{i}{2}$\\
			\cline{2-3}\cline{5-6}\cline{8-9}
			&$T^2ST$&$\frac{3}{2}+\frac{i}{2}$
			&&$T^3ST^4$&$\frac{47}{17}+\frac{i}{17}$
			&&$T^5ST$&$\frac{9}{2}+\frac{i}{2}$\\
			\cline{2-3}\cline{5-6}\cline{8-9}
			&$T^4ST$&$\frac{7}{2}+\frac{i}{2}$
			&&$T^7$&$7+i$
			&&$T^3ST$&$\frac{5}{2}+\frac{i}{2}$\\
			\hline\hline
			
			\multirow{2}{*}{$Z_3^{ST}$}&\cellcolor{yellow!70}$1$&\cellcolor{yellow!70}$\omega$
			&\multirow{2}{*}{$Z_3^{T^6ST^3}$}&$T^6$&$\frac{11}{2}+\frac{i\sqrt{3}}{2}$
			&\multirow{2}{*}{$Z_3^{T^4ST^5}$}&$T^4$&$\frac{7}{2}+\frac{i\sqrt{3}}{2}$\\
			\cline{2-3}\cline{5-6}\cline{8-9}
			&$T^2ST^6$&$\frac{113}{62}+\frac{i\sqrt{3}}{62}$
			&&$ST^6$&$-\frac{11}{62}+\frac{i\sqrt{3}}{62}$
			&&$TST^3$&$\frac{9}{14}+\frac{i\sqrt{3}}{14}$\\
			\hline
			
			\multirow{2}{*}{$Z_3^{T^2ST^7}$}&$T^7ST^3$&$\frac{93}{14}+\frac{i\sqrt{3}}{14}$
			&\multirow{2}{*}{$Z_3^{T^3ST^6}$}&$T^3$&$\frac{5}{2}+\frac{i\sqrt{3}}{2}$
			&\multirow{2}{*}{$Z_3^{T^5ST^4}$}&$T^5$&$\frac{9}{2}+\frac{i\sqrt{3}}{2}$\\
			\cline{2-3}\cline{5-6}\cline{8-9}
			&$T^2$&$\frac{3}{2}+\frac{i\sqrt{3}}{2}$
			&&$ST^3$&$-\frac{5}{14}+\frac{i\sqrt{3}}{14}$
			&&$T^2ST^3$&$\frac{23}{14}+\frac{i\sqrt{3}}{14}$\\
			\hline
			
			\multirow{2}{*}{$Z_3^{TS}$}&$S$&$\frac{1}{2}+\frac{i\sqrt{3}}{2}$
			&\multirow{2}{*}{$Z_3^{T^6ST}$}&$T^7$&$\frac{13}{2}+\frac{i\sqrt{3}}{2}$
			&\multirow{2}{*}{$Z_3^{TST^4}$}&$T^7ST^2$&$\frac{13}{2}+\frac{i\sqrt{3}}{6}$\\
			\cline{2-3}\cline{5-6}\cline{8-9}
			&$ST^2ST^6$&$-\frac{113}{206}+\frac{i\sqrt{3}}{206}$
			&&$T^4ST^3$&$\frac{51}{14}+\frac{i\sqrt{3}}{14}$
			&&$ST^4$&$-\frac{7}{26}+\frac{i\sqrt{3}}{26}$\\
			\hline
			
			\multirow{2}{*}{$Z_3^{T^2ST^3}$}&$ST^2$&$-\frac{1}{2}+\frac{i\sqrt{3}}{6}$
			&\multirow{2}{*}{$Z_3^{T^3ST^2}$}&$ST^7$&$-\frac{13}{86}+\frac{i\sqrt{3}}{86}$
			&\multirow{2}{*}{$Z_3^{T^4ST}$}&$ST^5$&$-\frac{3}{14}+\frac{i\sqrt{3}}{42}$\\
			\cline{2-3}\cline{5-6}\cline{8-9}
			&$TST^4$&$\frac{19}{26}+\frac{i\sqrt{3}}{26}$
			&&$ST^4ST^3$&$-\frac{51}{186}+\frac{i\sqrt{3}}{186}$
			&&$TST^7$&$\frac{73}{86}+\frac{i\sqrt{3}}{86}$\\
			\hline
			
			\multirow{2}{*}{$Z_3^{T^7ST^6}$}&$T^5ST^2$&$\frac{9}{2}+\frac{i\sqrt{3}}{6}$
			&\multirow{2}{*}{$Z_3^{ST^5}$}&$ST^4ST^4$&$-\frac{97}{362}+\frac{i\sqrt{3}}{362}$
			&\multirow{2}{*}{$Z_3^{T^5S}$}&$ST^6ST^2$&$-\frac{33}{182}+\frac{i\sqrt{3}}{182}$\\
			\cline{2-3}\cline{5-6}\cline{8-9}
			&$T^3ST^5$&$\frac{117}{42}+\frac{i\sqrt{3}}{14}$
			&&$T^6ST^2$&$\frac{11}{2}+\frac{i\sqrt{3}}{6}$
			&&$T^4ST^4$&$\frac{97}{26}+\frac{i\sqrt{3}}{26}$\\
			\hline
			
			\multirow{2}{*}{$Z_3^{T^6ST^7}$}&$ST^2ST^5$&$-\frac{75}{134}+\frac{i\sqrt{3}}{134}$
			&&&&&&\\
			\cline{2-3}
			&$T^2ST^5$&$\frac{25}{14}+\frac{i\sqrt{3}}{42}$
			&&&&&&\\
			\hline\hline
			
			\multirow{2}{*}{$Z_8^{T}$}&\cellcolor{yellow!70}$1$&\cellcolor{yellow!70}$i\infty$
			&\multirow{2}{*}{$Z_8^{ST^2}$}&$T^7ST^7$&$7$
			&\multirow{2}{*}{$Z_8^{ST^6}$}&$TS$&$1$\\
			\cline{2-3}\cline{5-6}\cline{8-9}
			&$ST^4ST^4$&$-\frac{1}{4}$
			&&$T^3ST^5$&$3$
			&&$T^5ST^2$&$5$\\
			\hline
			
			\multirow{2}{*}{$Z_8^{ST^4ST^3}$}&$ST^6ST^2$&$-\frac{1}{6}$
			&\multirow{2}{*}{$Z_8^{TST}$}&$T^4ST^4$&$4$
			&\multirow{2}{*}{$Z_8^{TST^5}$}&$T^6ST^2$&$6$\\
			\cline{2-3}\cline{5-6}\cline{8-9}
			&$ST^2ST^6$&$-\frac{1}{2}$
			&&$S$&$0$
			&&$T^2ST^6$&$2$\\
			\hline\hline
		\end{tabular}
\caption{\label{tab:fixed_points}The 92 inequivalent fixed points associated with the finite modular group $\Delta(96)$, obtained from the left-coset decomposition of $\Delta(96)$. They are organized into three orbits originating from the fundamental fixed points $\tau_S$, $\tau_{ST}$, and $\tau_T$, containing 48, 32, and 12 points, respectively. The corresponding images of the stabilizer subgroups in $\Delta(96)$ are the cyclic groups with order 2, 3 and 8, respectively. The positions of all these fixed-point representatives in $\mathcal{H}^{*}$ are displayed in figure~\ref{fig:Ker_Delta96_fundamental_domain}.}
\end{center}
\end{table}

\begin{table}[t!]
	\begin{center}
		\renewcommand{\tabcolsep}{2.2mm}
		\renewcommand{\arraystretch}{1.3}
		\begin{tabular}{|c|c|c|c|}\hline\hline
			&$\tau_S=i$&$\tau_{ST}=\omega$&$\tau_T=i\infty$\\ \hline
			$Y_{\bm{3_{0}}}^{(6)}$&$\left(1+2^{3/4},-\sqrt{2},1-2^{3/4}\right)^{T}$&$\left(i \cot\frac{\pi}{8},-2\omega_{16}\cos\frac{\pi}{8},-1\right)^{T} $&\multirow{5}{*}{$
				\left(0,0,0\right)^{T}$}\\\cline{1-3}
			$Y_{\bm{3_{1}}}^{(4)}$&$\left(1-2^{1/4},2^{3/4},-1-2^{1/4}\right)^{T}$&$\left(i \left(\sqrt{2}-\sqrt{3}\right),2\sqrt{2}\omega_{16}^{9} \sin\frac{5\pi}{24},\cot\frac{\pi}{8}\right)^{T}$&\\
			\cline{1-3}
			$Y_{\bm{\bar{3}_{0}}}^{(2)}$&$\left(1-2^{1/4},-\sqrt{2},1+2^{1/4}\right)^{T}$&$\left(i \left(\sqrt{3}-\sqrt{2}\right),2\sqrt{2}\omega_{16}^{7} \sin\frac{5\pi}{24},\cot\frac{\pi}{8}\right)^{T}$&\\
			\cline{1-3}
			$Y_{\bm{\bar{3}_{1}}}^{(4)}$&$\left(1+2^{5/4},-\sqrt{2},1-2^{5/4}\right)^{T}$&$\left(i \left(\sqrt{2}+\sqrt{3}\right),2\sqrt{2}\omega_{16}^{7} \cos\frac{\pi}{24},-\cot\frac{\pi}{8}\right)^{T}$&\\
			\cline{1-3}
			$Y_{\bm{\hat{3}_{1}}}^{(4)}$&$\left(\sqrt{2},\tan\frac{\pi}{8},-\cot\frac{\pi}{8}\right)^{T}$&$\left(2\sqrt{2}\omega_{16}^{2} \cos\frac{\pi}{12},1,-i\cot\frac{\pi}{12}\right)^{T}$&\\
			\hline
			$Y_{\bm{\hat{3}_{0}}}^{(2)}$&$\left(2,-\cot\frac{\pi}{8},-\tan\frac{\pi}{8}\right)^{T}$&$\left(2\sqrt{2}\omega_{16}^{2} \sin\frac{\pi}{12},-1,i\tan\frac{\pi}{12}\right)^{T}$&$\left(1,0,0\right)^{T}$\\
			\hline\hline
			$Y_{\bm{3_{0}}}^{(8)}$&\multirow{4}{*}{$\left(1,\sqrt{2},1\right)^{T}$}&$\left(i \left(\sqrt{2}+\sqrt{3}\right),2\sqrt{2}\omega_{16} \cos\frac{\pi}{24},\cot\frac{\pi}{8}\right)^{T}$&\multirow{5}{*}{$\left(0,0,0\right)^{T}$}\\
			\cline{1-1}\cline{3-3}
			$Y_{\bm{3_{1}}}^{(6)}$&&$\left(1,2i\omega_{16}\sin\frac{\pi}{8},i \tan\frac{\pi}{8}\right)^{T}$&\\
			\cline{1-1}\cline{3-3}
			$Y_{\bm{\bar{3}_{0}}}^{(4)}$&&$\left(i \left(\sqrt{2}+\sqrt{3}\right),2\sqrt{2}\omega_{16}^{7} \cos\frac{\pi}{24},-\cot\frac{\pi}{8}\right)^{T}$&\\
			\cline{1-1}\cline{3-3}
			$Y_{\bm{\bar{3}_{1}}}^{(6)}$&&$\left(1,-2\omega_{16}^3\sin\frac{\pi}{8},-i\tan\frac{\pi}{8}\right)^{T}$&\\
			\cline{1-3}
			$Y_{\bm{\hat{3}_{1}}}^{(6)}$&\multirow{2}{*}{$\left(\sqrt{2},1,1\right)^{T}$}&$\left(1,-\omega_{16}^{6},\omega_{16}^2\right)^{T}$&\\
			\cline{1-1}\cline{3-4}
			$Y_{\bm{\hat{3}_{0}}}^{(4)}$&&$\left(2\sqrt{2}\omega_{16}^2 \cos\frac{\pi}{12},1,-i \cot\frac{\pi}{12}\right)^{T}$&$\left(1,0,0\right)^{T}$\\
			\hline\hline
\end{tabular}
\caption{\label{tab:3D_VEVs_FDfixed}Vacuum alignments at the fundamental domain fixed points for the lowest and next-to-lowest weight VVMFs in each three-dimensional irreducible representation of $\Delta(96)$, where $\omega_{16} = e^{i\pi/8}$. In our convention, an unphysical overall factor has been omitted from each triplet VEV at these fixed points; these VEVs are used in place of the true VEVs throughout the discussion.}
	\end{center}
\end{table}

\section{\label{sec:MLS} Modular Littlest Seesaw}

The Modular Littlest Seesaw provides a predictive framework for neutrino masses and lepton flavor mixing by combining the minimal type-I seesaw mechanism, involving two right-handed neutrinos, with modular flavor symmetry. Its primary goal is to account for the observed neutrino oscillation data within a highly constrained setup containing only a small number of free parameters. In this setup, the three left-handed lepton doublets $L$ are usually assigned to a triplet representation $\bm{3}$ of a finite modular group $\mathcal{G}_f$, while the right-handed charged leptons may transform as singlets, doublets, or triplets depending on the specific construction. The two right-handed neutrinos, denoted by  the ``atmospheric'' neutrino $N_{\text{atm}}^{c}$ with modular weight $k_{N^c_{\text{atm}}}$ and the ``solar'' neutrino $N_{\text{sol}}^{c}$ with modular weight $k_{N^c_{\text{sol}}}$, are distinguished by different modular weights and/or different representations of $\mathcal{G}_f$, which naturally forbids the mixed Majorana mass term $N^{c}_{\text{atm}}N^{c}_{\text{sol}}$. The Higgs fields $H_u$ and $H_d$ are generally taken to be invariant under the modular symmetry.

A characteristic feature of the Modular Littlest Seesaw scenario is the introduction of three independent moduli $\tau_{\ell}$, $\tau_{\text{atm}}$, and $\tau_{\text{sol}}$~\cite{Ding:2019gof}, whose vacuum expectation values determine the residual symmetries preserved in the charged lepton sector, atmospheric neutrino sector and solar neutrino sector, respectively. Such a structure can be realised either within theories involving multiple modular symmetries~\cite{deMedeirosVarzielas:2022fbw,deMedeirosVarzielas:2023ujt} or in orbifold constructions based on three factorisable tori~\cite{deAnda:2023udh}. 

In the Modular Littlest Seesaw framework, the lepton sector superpotential can be written as
\begin{eqnarray}
\nonumber \mathcal{W}&=&-Y^{\ell}_{ij}(\tau_{\ell})L_i E^{c}_jH_d -y_{\text{atm}}LY_{\text{atm}}(\tau_{\text{atm}})N^c_{\text{atm}}H_u-y_{\text{sol}}LY_{\text{sol}}(\tau_{\text{sol}})N^c_{\text{sol}}H_u \\
&&-\frac{1}{2}M_{\text{atm}}N^c_{\text{atm}}N^c_{\text{atm}}
	-\frac{1}{2}M_{\text{sol}}N^{c}_{\text{sol}}N^c_{\text{sol}}+\text{h.c.} \,,
\end{eqnarray}
where each term is modular invariant, and all possible contractions into singlet should be considered. The quantities $Y^\ell_{ij}(\tau_\ell)$, $Y_{\text{atm}}(\tau_{\text{atm}})$, and $Y_{\text{sol}}(\tau_{\text{sol}})$ are modular forms with transformation properties fixed by the underlying symmetry. In the neutrino sector, modular invariance requires the combinations $LY_{\text{atm}}$ and $LY_{\text{sol}}$ to contain singlet contractions. As a result, the modular forms $Y_{\text{atm}}$ and $Y_{\text{sol}}$ must transform as triplet representations of $\mathcal{G}_{f}$. Furthermore, invariance of the Majorana mass terms implies that $M_{\text{atm}}$ and $M_{\text{sol}}$ should be regarded as dimension-one constants multiplying singlet modular forms of weights $2k_{N^c_{\text{atm}}}$ and $2k_{N^c_{\text{sol}}}$, respectively. These constraints completely determine the flavor structure of the Dirac neutrino mass matrix once the modular forms and field assignments are specified. 

In the Modular Littlest Seesaw framework, the three moduli $\tau_{\ell}$, $\tau_{\text{atm}}$, and $\tau_{\text{sol}}$ are stabilized at modular fixed points, and the vacuum structure is dictated directly by the residual modular symmetries rather than by introducing additional flavon fields and auxiliary alignment mechanisms. Specifically, we set $\langle \tau_{\ell} \rangle$, $\langle \tau_{\text{atm}} \rangle$, and $\langle \tau_{\text{sol}} \rangle$ to three distinct modular fixed points. Denoting the corresponding stabilizer elements by $\gamma_{\ell}$, $\gamma_{\text{atm}}$ and $\gamma_{\text{sol}}$, one has
\begin{equation}
\gamma_{\ell}\langle\tau_{\ell}\rangle=\langle\tau_{\ell}\rangle, \qquad \gamma_{\text{atm}}\langle\tau_{\text{atm}}\rangle=\langle\tau_{\text{atm}}\rangle, \qquad \gamma_{\text{sol}}\langle\tau_{\text{sol}}\rangle=\langle\tau_{\text{sol}}\rangle\,,
\end{equation}
As a result, non-trivial residual modular subgroups generated by these stabilizer  elements are  preserved separately in the charged lepton, atmospheric neutrino, and solar neutrino sectors.

Once electroweak symmetry is broken, the Dirac neutrino mass matrix is generated from the modular forms evaluated at the fixed points:
\begin{equation}
M_D =\begin{pmatrix} \; y_{\text{atm}}\bm{v}_{\text{atm}} &
y_{\text{sol}}\bm{v}_{\text{sol}}  \end{pmatrix}v_{u} ,
\end{equation}
where the two column vectors $\bm{v}_{\text{atm}}$ and $\bm{v}_{\text{sol}}$ are defined as 
\begin{equation}
\bm{v}_{\text{atm}}=U_{a} Y_{\text{atm}}\left(\langle \tau_{\text{atm}} \rangle\right),\qquad \bm{v}_{\text{sol}}=U_{s} Y_{\text{sol}}\left(\langle \tau_{\text{sol}} \rangle\right)\,,
\end{equation}
where $U_a$ and $U_s$ are constant matrices determined by the CG coefficients appearing in the atmospheric and solar neutrino Yukawa contractions, respectively. As the right-handed neutrino mass matrix is taken to be diagonal, $M_{N} = \operatorname{diag}(M_{\text{atm}}, M_{\text{sol}})$. The seesaw relation then gives the following light neutrino mass matrix:
\begin{equation}\label{eq:light_nu_mass}
M_{\nu} = - M_{D} M_{N}^{-1} M_{D}^{T} = m_a \left( \bm{v}_{\text{atm}} \bm{v}_{\text{atm}}^T + r e^{i\eta} \bm{v}_{\text{sol}} \bm{v}_{\text{sol}}^T \right)\,,
\end{equation}
where the three input parameters $m_a$, $r$ and $\eta$ are defined as:
 \begin{equation}\label{eq:input_parameters}
 m_a = -\frac{y_{\text{atm}}^2 v_u^2}{M_{\text{atm}}}, \qquad    \eta = \arg\!\left( \frac{y_{\text{sol}}^2 M_{\text{atm}}}{y_{\text{atm}}^2 M_{\text{sol}}} \right), \qquad  r = \left| \frac{y_{\text{sol}}^2 M_{\text{atm}}}{y_{\text{atm}}^2 M_{\text{sol}}} \right|\,. 
 \end{equation}
Note that  the phase of the parameter $m_{a}$ in Eq.~\eqref{eq:light_nu_mass} is unphysical and it can be absorbed by lepton fields. Thus, it can be taken to be positive real. Therefore, the entire light neutrino mass matrix depends on two positive real parameters $m_a$ and $r$, together with one physical phase $\eta\in[0,2\pi)$.

Since the model contains only two right-handed neutrinos, the light neutrino mass matrix $M_\nu$ has rank two, and consequently one light neutrino is massless. One can readily verify that the normalized vector
\begin{equation}
\hat{\bm{v}}_{\text{fix}}=\frac{ \bm{v}_{\text{atm}} \times \bm{v}_{\text{sol}} }{| \bm{v}_{\text{atm}} \times \bm{v}_{\text{sol}} |} \,,
\end{equation}
 satisfies
\begin{equation}
M_\nu \hat{\bm{v}}_{\text{fix}}=(0,0,0)^T\, .
\end{equation}
Accordingly, $\hat{\bm{v}}_{\text{fix}}$ corresponds to the first column of the neutrino diagonalization matrix $U_\nu$ for NO, and to the third column for IO:
\begin{equation}
U_\nu^T M_\nu U_\nu =
\begin{cases} 
	\text{diag}(0,m_2,m_3) & \text{for NO} \\
	\text{diag}(m_1,m_2,0) & \text{for IO}
\end{cases}\, .
\end{equation}
The explicit form of $U_\nu$ in terms of the input parameters $r$ and $\eta$ can be found in Refs.~\cite{Ding:2018tuj,Yan:2025itm}.

In the charged lepton sector, the residual symmetry generated by the stabilizer element $\gamma_{\ell}$ imposes strong constraints on the charged lepton mass matrix, i.e.  the hermitian matrix $M_{\ell}^{\dagger}M_{\ell}$ satisfies
\begin{equation}
\rho^{\dagger}_{\bm{3}}\left(\gamma_{\ell}\right)  M_{\ell}^{\dagger}\left(\langle \tau_{\ell} \rangle\right) M_{\ell}\left(\langle \tau_{\ell} \rangle\right) \rho_{\bm{3}}\left(\gamma_{\ell}\right)=M_{\ell}^{\dagger}\left(\langle \tau_{\ell} \rangle\right) M_{\ell}\left(\langle \tau_{\ell} \rangle\right)\,,
\end{equation}
where $\rho_{\bm{3}}(\gamma_{\ell})$ denotes the triplet representation matrix of $\gamma_{\ell}$ acting on the left-handed lepton doublets $L$. Consequently, $M_{\ell}^{\dagger}M_{\ell}$ commutes with $\rho_{\bm{3}}(\gamma_{\ell})$, implying that the two matrices can be diagonalized simultaneously by the same unitary matrix $U_{\ell}$. Hence, the charged lepton diagonalization matrix is determined by the eigenvectors of $\rho_{\bm{3}}(\gamma_{\ell})$. Explicitly,
\begin{equation}
U_{\ell}^{\dagger} \, M_{\ell}^{\dagger}\left(\langle \tau_{\ell} \rangle\right) M_{\ell}\left(\langle \tau_{\ell} \rangle\right) \, U_{\ell} = \operatorname{diag}(m_e^2, m_\mu^2, m_\tau^2),\qquad
U_{\ell}^{\dagger} \, \rho_{\bm{3}}(\gamma_{\ell}) \, U_{\ell} = \hat{\rho}_{\bm{3}}(\gamma_{\ell}),
\end{equation}
where $\hat{\rho}_{\bm{3}}(\gamma_{\ell})$ is a diagonal phase matrix. When the eigenvalues of $\hat{\rho}_{\bm{3}}(\gamma_{\ell})$ are non-degenerate, $U_{\ell}$ is fixed up to an unphysical diagonal phase redefinition and a column permutation. The latter is used to enforce the conventional mass ordering $\text{diag}(m_e^2,m_{\mu}^2,m_{\tau}^2)$, and it may take one of the following six forms:
\begin{eqnarray}
\nonumber P_{123} = \begin{pmatrix} 1 & 0 & 0 \\ 0 & 1 & 0 \\ 0 & 0 & 1 \end{pmatrix},\qquad  
P_{231} = \begin{pmatrix} 0 & 1 & 0 \\ 0 & 0 & 1 \\ 1 & 0 & 0 \end{pmatrix},\qquad  
P_{312} = \begin{pmatrix} 0 & 0 & 1 \\ 1 & 0 & 0 \\ 0 & 1 & 0 \end{pmatrix},\\[4pt]
P_{132} = \begin{pmatrix} 1 & 0 & 0 \\ 0 & 0 & 1 \\ 0 & 1 & 0 \end{pmatrix},\qquad  
P_{213} = \begin{pmatrix} 0 & 1 & 0 \\ 1 & 0 & 0 \\ 0 & 0 & 1 \end{pmatrix},\qquad  
P_{321} = \begin{pmatrix} 0 & 0 & 1 \\ 0 & 1 & 0 \\ 1 & 0 & 0 \end{pmatrix}\,.
\end{eqnarray}

The lepton mixing matrix is obtained from the mismatch between the diagonalization matrices of the charged lepton and neutrino sectors,
\begin{equation}\label{eq:PMNS_MLSS}
U_\text{PMNS} = Q_{\ell} P_{\ell} U_{\ell}^{\dagger} U_{\nu},
\end{equation}
where the unphysical phase matrix $Q_{\ell}$ can be absorbed by the charged lepton fields, and $P_{\ell}$ denotes a permutation matrix. Once the modular forms and fixed points are specified, one column of $U_\text{PMNS}$ is fixed to be $P_{\ell} U_{\ell}^{\dagger}\hat{\bm{v}}_{\text{fix}}$, which yields definite predictions for the lepton mixing angles, neutrino mass spectrum and Dirac CP phase. These predictions can then be confronted with current neutrino oscillation data, making the Modular Littlest Seesaw an economical and predictive approach to lepton flavor physics. 

In the standard parametrisation~\cite{ParticleDataGroup:2026aaa}, the lepton mixing matrix is expressed as:
\begin{equation}\label{eq:PMNS_standard}
	U_\text{PMNS}=\begin{pmatrix}
		c_{12}c_{13}  &   s_{12}c_{13}   &   s_{13}e^{-i\delta_{CP}}  \\
		-s_{12}c_{23}-c_{12}s_{13}s_{23}e^{i\delta_{CP}}   &  c_{12}c_{23}-s_{12}s_{13}s_{23}e^{i\delta_{CP}}  &  c_{13}s_{23}  \\
		s_{12}s_{23}-c_{12}s_{13}c_{23}e^{i\delta_{CP}}   & -c_{12}s_{23}-s_{12}s_{13}c_{23}e^{i\delta_{CP}}  &  c_{13}c_{23}
	\end{pmatrix}\text{diag}(1,e^{i\frac{\beta}{2}},1)\,,
\end{equation}
where $c_{ij} \equiv \cos\theta_{ij}$, $s_{ij} \equiv \sin\theta_{ij}$, $\delta_{CP}$ is the Dirac CP phase and $\beta$ is the only physical Majorana phase. The mixing angles $\theta_{ij}$ are restricted to the first quadrant, $\theta_{ij} \in [0, \pi/2]$, while the Dirac and Majorana  CP-violating phases range over $[0, 2\pi)$. As regards the CP violation, two weak basis invariants $J_{\text{CP}}$~\cite{Jarlskog:1985ht} and $I_1$~\cite{Branco:1986gr,Nieves:1987pp,Nieves:2001fc,Jenkins:2007ip,Branco:2011zb} associated with the Dirac CP phase $\delta_{CP}$ and Majorana CP phase $\beta$ respectively, are given by
\begin{eqnarray}
\nonumber && J_{\text{CP}} = \text{Im}(U_\text{PMNS,11} U_\text{PMNS,33} U_\text{PMNS,13}^* U_\text{PMNS,31}^*) = \frac{1}{8} \sin 2\theta_{12} \sin 2\theta_{13} \sin 2\theta_{23} \cos \theta_{13} \sin \delta_{\text{CP}}\,, \\
&& I_1=\begin{cases}
	\text{Im}(U_\text{PMNS,12}^2 U_\text{PMNS,13}^{*2}) = \frac{1}{4} \sin^2 \theta_{12} \sin^2 2\theta_{13} \sin(\beta + 2\delta_{\text{CP}}) & \text{for NO}, \\
	\text{Im}(U_\text{PMNS,12}^2 U_\text{PMNS,11}^{*2}) = \frac{1}{4} \cos^4 \theta_{13} \sin^2 2\theta_{12} \sin \beta & \text{for IO}.
\end{cases}
\end{eqnarray}
Evaluating these invariants from our predicted $U_{\text{PMNS}}$ and matching them to standard parameterization forms allows us to unambiguously extract all physical CP phases. In particular, if the fixed column of $U_{\text{PMNS}}$ is taken to be $(u_{1},u_{2},u_{3})$, one obtains the following relations 
\begin{eqnarray}
\nonumber &&\begin{cases}
	\sin^{2}\theta_{12}=1-\frac{|u_{1}|^2}{\cos^2\theta_{13}}, \\
	\cos\delta_{CP}=\frac{(|u_{2}|^2-|u_{3}|^2)\cos^2\theta_{13}+\left[(1+|u_{1}|^2)\sin^2\theta_{13}-(1-|u_{1}|^2)\right]
\cos2\theta_{23}}{2|u_{1}|\sqrt{\cos^2\theta_{13}-|u_{1}|^2}\;\sin2\theta_{23}\sin\theta_{13}},
    \end{cases} \quad \text{for} \quad \text{NO} \,,\\
\label{eq:sum_rules}&&\begin{cases}
	\sin^{2}\theta_{13}=|u_{1}|^2, \\
	\sin^{2}\theta_{23}=\frac{|u_{2}|^2}{1-|u_{1}|^2},
\end{cases} \quad \text{for} \quad \text{IO}\,,
\end{eqnarray}
which may be tested by future high-precision measurements of $\theta_{12}$, $\theta_{23}$ and $\delta_{CP}$ at JUNO, DUNE and T2HK for a given fixed column.

Based on the preceding theoretical setup, we shall perform a comprehensive survey of all viable triplet alignments arising from modular fixed points within the finite modular group $\Delta(96)$. Each candidate alignment is tested against the latest global fit to neutrino oscillation data. This scan yields a substantial set of phenomenologically allowed fixed points and their associated modular form alignments. These selected configurations pinpoint the specific Modular Littlest Seesaw models whose predictions for neutrino masses, mixing angles, and CP-violating phases will be analyzed in detail in the sections that follow.

\section{\label{sec:model_building}$\Delta(96)$ Modular Littlest Seesaw models}

In this section, we systematically investigate the viable lepton mixing patterns generated by the modular $\Delta(96)$ symmetry within the Modular Littlest Seesaw framework. The relevant group theoretical properties of $\Delta(96)$, including the representation matrices and CG coefficients, are summarized in appendix~\ref{sec:Delta96_group_theory}. In this framework, the lepton mixing matrix is determined by the diagonalization matrix $U_\ell(\tau_{l})$ and the permutation matrix $P_\ell$  associated with the charged lepton sector, and the two Yukawa coupling vectors $Y_{\text{atm}}(\tau_{\text{atm}})$ and $Y_{\text{sol}}(\tau_{\text{sol}})$ that control the neutrino mass structure. The VEVs $\langle\tau_\ell\rangle$, $\langle\tau_{\text{atm}}\rangle$ and $\langle\tau_{\text{sol}}\rangle$ of the three independent moduli are assumed to be stabilised at fixed points preserving the residual symmetries generated by $\gamma_\ell$, $\gamma_{\text{atm}}$ and $\gamma_{\text{sol}}$, respectively. Consequently, $U_\ell$ diagonalises the representation matrix $\rho_{\bm{3}}(\gamma_\ell)$, where $\bm{3}$ denotes the representation of the left-handed lepton doublet $L$. In the neutrino sector, the corresponding Yukawa vectors $Y_{\text{atm}}(\tau_{\text{atm}})$ and $Y_{\text{sol}}(\tau_{\text{sol}})$ are identified with the values of the lowest- and next-to-lowest-weight triplet VVMFs of $\Delta(96)$ evaluated at the fixed points $\langle\tau_{\text{atm}}\rangle$ and $\langle\tau_{\text{sol}}\rangle$, respectively. For each viable configuration, the resulting lepton mixing matrix and neutrino masses are characterised by only three real parameters: the overall neutrino mass scale $m_a$, the ratio $r$ and the relative phase $\eta$, as defined in Eq.~\eqref{eq:input_parameters}.  

In the charged lepton sector, once the residual symmetry generated by $\gamma_\ell$ is specified, the matrix $U_\ell$ is further determined by the representation assignment of the left-handed lepton doublets $L$. We restrict our analysis to models in which the three generations of $L$ transform as three-dimensional irreducible representations $\rho_{\ell}$ of the modular group $\Delta(96)$, which contains six such representations, denoted by $\bm{3_{m}}$, $\bm{\bar{3}_{m}}$ and $\bm{\hat{3}_{m}}$ with $m=0,1$. The representations with $m=0$ and $m=1$ differ only by an overall sign in the action of the generators $S$ and $T$, and therefore yield equivalent predictions for $U_\ell$. Thus, it is sufficient to consider only $\bm{3_{0}}$, $\bm{\bar{3}_{0}}$, and $\bm{\hat{3}_{0}}$. Moreover, since $\bm{3_{0}}$ and $\bm{\bar{3}_{0}}$ are complex conjugate representations, their corresponding charged lepton diagonalization matrices are related by complex conjugation.

Given the triplet assignment of $L$, the representations of the Yukawa coupling vectors $Y_{\text{atm}}$ and $Y_{\text{sol}}$ are fixed by modular invariance. Since $N_{\rm atm}^{c}$ and $N_{\rm sol}^{c}$ are $\Delta(96)$ singlets and only $\bm{3_{m}}\otimes\bm{\bar{3}_{n}}$ and $\bm{\hat{3}_{m}}\otimes\bm{\hat{3}_{n}}$ contain a singlet, as shown in Eq.~\eqref{eq:kronecker_product}, $Y_{\text{atm}}$ and $Y_{\text{sol}}$ must transform as $\bm{\bar{3}_{m}}$, $\bm{3_{m}}$ and $\bm{\hat{3}_{m}}$ when $L$ is assigned to $\bm{3_{0}}$, $\bm{\bar{3}_{0}}$, and $\bm{\hat{3}_{0}}$, respectively.

We now classify the independent symmetry breaking patterns of the modular group $\Delta(96)$ within the Modular Littlest Seesaw framework. Each pattern is specified by a set of stabilizers $\{\gamma_{\ell}, \gamma_{\text{atm}}, \gamma_{\text{sol}}\}$. Two sets of stabilizers, $\{\gamma_{\ell}, \gamma_{\text{atm}}, \gamma_{\text{sol}}\}$ and $\{\gamma^{\prime}_{\ell}, \gamma^{\prime}_{\text{atm}}, \gamma^{\prime}_{\text{sol}}\}$, are physically equivalent if they are related by a common modular conjugation,
\begin{equation}\label{eq:equivalent_stabilizers}
	\gamma_{\ell}' = \gamma \gamma_{\ell} \gamma^{-1},\qquad \gamma_{\text{atm}}' = \gamma \gamma_{\text{atm}} \gamma^{-1},\qquad \gamma_{\text{sol}}' = \gamma \gamma_{\text{sol}} \gamma^{-1},
\end{equation}
for some modular transformation $\gamma$. In this case, the charged lepton and neutrino mass matrices transform as $M_{\ell}^{\dagger}M_{\ell} \rightarrow \rho_{L}(\gamma) M_{\ell}^{\dagger}M_{\ell} \rho_{L}^{\dagger}(\gamma)$ and $M_{\nu} \rightarrow \rho_{L}^{*}(\gamma) M_{\nu} \rho_{L}^{\dagger}(\gamma)$ which leaves the resulting PMNS matrix unchanged. Therefore, stabilizer sets related by Eq.~\eqref{eq:equivalent_stabilizers} lead to identical lepton mixing predictions and should be regarded as the same symmetry breaking pattern~\cite{Ding:2013bpa,Li:2014eia,Ding:2018tuj,Shang:2026qkh}. 

In the present multi-modulus framework the fixed points associated with different sectors cannot in general be chosen simultaneously within the fundamental domain. For a single modulus all lepton sectors share the same modulus. If $\tau_f=\gamma\tau_0$ with $\tau_0$ in the fundamental domain, the residual symmetries in all sectors are conjugated by the same modular transformation, which implies that the PMNS matrix remains unchanged. The fixed point may therefore be chosen within the fundamental domain without loss of generality. In our multi-modulus framework the three independent moduli $\tau_\ell$, $\tau_{\text{atm}}$ and $\tau_{\text{sol}}$ may be stabilized at different fixed points. Bringing each fixed point separately into the fundamental domain generally requires different modular transformations, which does not correspond to such a common conjugation and therefore does not in general leave the PMNS matrix invariant. We may therefore use the freedom of common conjugation to fix the charged lepton modulus $\langle\tau_\ell\rangle$ in the fundamental domain. The fixed points of $\tau_{\text{atm}}$ and $\tau_{\text{sol}}$ must then be allowed throughout the upper half-plane, with inequivalent representatives taken modulo $\ker(\rho)$ as illustrated in table~\ref{tab:fixed_points}.
\begin{table}[t!]
	\begin{center}
		\renewcommand{\tabcolsep}{2.5mm}
		\renewcommand{\arraystretch}{1.3}
		\begin{tabular}{|c|c|c|}\hline\hline
			$\rho_\ell$&$U_\ell$ for $G_\ell=Z_{3}^{ST}$&$U_\ell$ for $G_\ell=Z_{8}^{T}$\\ \hline
			& &  \\[-0.2in]
			$\bm{3_{0}}$&$\frac{1}{\sqrt{3}}
			\begin{pmatrix}
				-i\sqrt{2}\sin\frac{\pi}{24} &
				i\sqrt{2}\cos\frac{5\pi}{24} &
				-i\sqrt{2}\cos\frac{\pi}{8}
				\\
				-\omega_{16} & \omega_{16} & \omega_{16}
				\\
				\sqrt{2}\cos\frac{\pi}{24} &
				\sqrt{2}\sin\frac{5\pi}{24} &
				\sqrt{2}\sin\frac{\pi}{8}
			\end{pmatrix}$&  \\ [0.33in] \cline{1-2}
			& &  \\[-0.2in]
			$\bm{\bar{3}_{0}}$&$\frac{1}{\sqrt{3}}
			\begin{pmatrix}
				i\sqrt{2}\sin\frac{\pi}{24} &
				-i\sqrt{2}\cos\frac{5\pi}{24} &
				i\sqrt{2}\cos\frac{\pi}{8}
				\\
				\omega_{16}^7 & -\omega_{16}^7 & -\omega_{16}^7
				\\
				\sqrt{2}\cos\frac{\pi}{24} &
				\sqrt{2}\sin\frac{5\pi}{24} &
				\sqrt{2}\sin\frac{\pi}{8}
			\end{pmatrix}$& $\begin{pmatrix}
				1 & 0 & 0 \\
				0 & 1 & 0 \\
				0 & 0 & 1 \\
			\end{pmatrix}$\\ [0.33in]
			\cline{1-2}
			& &  \\[-0.22in]
			$\bm{\hat{3}_{0}}$&
			$\frac{1}{\sqrt{3}}
			\begin{pmatrix}
				-\omega_{16}^{-2} & \omega_{16}^{-2} & \omega_{16}^{-2}
				\\[2mm]
				i\sqrt{2}\sin\frac{\pi}{12} &
				i\sqrt{2}\cos\frac{\pi}{12} &
				-i\sqrt{2}\cos\frac{\pi}{4}
				\\[2mm]
				\sqrt{2}\cos\frac{\pi}{12} &
				\sqrt{2}\sin\frac{\pi}{12} &
				\sqrt{2}\cos\frac{\pi}{4}
			\end{pmatrix}$&\\ [0.31in]
			\hline\hline
		\end{tabular}
		\caption{\label{tab:Ul}Charged lepton diagonalisation matrices $U_\ell$ for the residual symmetries $G_\ell = Z_3^{ST}$ and $G_\ell = Z_8^{T}$, given up to column permutations and phase freedoms, where $\omega_{16} = e^{i\pi/8}$.}
	\end{center}
\end{table}

With $\langle\tau_\ell\rangle$ fixed in the fundamental domain, the residual symmetry in the charged lepton sector is restricted to one of the corresponding stabilizer subgroups, namely $Z_2^S$, $Z_3^{ST}$ or $Z_8^T$. To distinguish the three charged leptons, the residual symmetry must possess three non-degenerate eigenvalues in the triplet representation $\rho_{\ell}$ carried by $L$. This excludes $Z_2^S$, leaving only $Z_3^{ST}$ and $Z_8^T$ as viable possibilities. The corresponding charged lepton diagonalization matrices $U_\ell$ for the three chosen triplet representations are listed in table~\ref{tab:Ul}. The independent residual symmetry breaking patterns are then determined by the possible choices of $Y_{\mathrm{atm}}$ and $Y_{\mathrm{sol}}$ at the fixed points $\langle\tau_{\mathrm{atm}}\rangle$ and $\langle\tau_{\mathrm{sol}}\rangle$, respectively. The vacuum alignments of the lowest- and next-to-lowest-weight VVMFs for different triplet representations are presented in section~\ref{sec:VVMFs_VEVs}. Furthermore, interchanging the transformation properties of $N^c_{\mathrm{atm}}$ and $N^c_{\mathrm{sol}}$ only swaps the two columns of the Dirac neutrino mass matrix, leaving the neutrino masses and mixing parameters unchanged. Therefore, distinct breaking patterns need only be classified up to the exchange $Y_{\mathrm{atm}} \leftrightarrow Y_{\mathrm{sol}}$. 

For each assignment of lepton field representations and modular weights, residual symmetry breaking patterns of the finite modular group $\Delta(96)$ give rise to definite predictions for neutrino masses and lepton mixing. In the following, we systematically analyze all such breaking patterns and determine which are compatible with current neutrino oscillation data.

\subsection{Numerical analysis}

We now proceed to a systematic statistical analysis of the lepton mixing patterns arising from the modular $\Delta(96)$ symmetry within the Modular Littlest Seesaw framework. We find 89662 independent combinations of field assignments and symmetry breaking patterns described in the previous section. For each breaking pattern, we confront the resulting predictions for the three mixing angles  $\theta_{12}$, $\theta_{13}$, $\theta_{23}$, the Dirac CP phase $\delta_{CP}$,  the two mass squared differences $\Delta m^2_{21}$ and $\Delta m^2_{3\ell}$ (where $\Delta m^2_{3\ell}=m^{2}_{3}-m^{2}_{1}>0$ for NO and  $\Delta m^2_{3\ell}=m^{2}_{3}-m^{2}_{2}<0$ for IO) with the \texttt{NuFIT} 6.1 global-fit results with Super-Kamiokande atmospheric data~\cite{Esteban:2024eli}.  The best fit values together with the $1\sigma$ and $3\sigma$ allowed ranges for the mixing parameters and neutrino mass squared differences are summarised in table~\ref{tab:bf_13sigma_data}. 

\begin{table}[t!]
	\centering
	\begin{tabular}{|c|c|c||c|c|}\hline \hline
		\multirow{2}{*}{Observables}  &  	\multicolumn{2}{c||}{NO}   &      \multicolumn{2}{c|}{IO}       \\ \cline{2-3} \cline{4-5}
		
		& $\text{bf}\pm1\sigma$  & $3\sigma$ region & $\text{bf}\pm1\sigma$ & $3\sigma$ region   \\ \hline
		
		&   &  &  &    \\[-0.150in]
		
		$\sin^2\theta_{13}$ & $0.02248^{+0.00055}_{-0.00059}$ & $[0.02064,0.02418]$ & $0.02262^{+0.00057}_{-0.00056}$ &  $[0.02093,0.02441]$  \\ [0.050in]
		
		$\sin^2\theta_{12}$ & $0.3088^{+0.0067}_{-0.0066}$ & $[0.2893,0.3295]$ & $0.3088^{+0.0067}_{-0.0066}$ & $[0.2893,0.3295]$  \\ [0.050in]
		
		$\sin^2\theta_{23}$  & $0.470^{+0.017}_{-0.014}$  & $[0.435,0.584]$ & $0.550^{+0.013}_{-0.016}$  & $[0.439,0.584]$   \\ [0.050in]
		
		$\delta_{CP}/\pi$  & $1.178^{+0.144}_{-0.200}$ & $[0.694,2.028]$  & $1.522^{+0.122}_{-0.139}$ & $[1.128,1.861]$  \\ [0.050in]
		
		$\frac{\Delta m^2_{21}}{10^{-5}\text{eV}^2}$ & $7.537^{+0.094}_{-0.10}$ & $[7.236,7.823]$ & $7.537^{+0.094}_{-0.10}$ & $[7.236,7.822]$  \\ [0.050in]
		
		$\frac{\Delta m^2_{3\ell}}{10^{-3}\text{eV}^2}$ & $2.511^{+0.021}_{-0.020}$ & $[2.450,2.576]$ & $-2.483^{+0.020}_{-0.020}$ & $[-2.547,-2.421]$ \\ [0.050in]
		
		\hline \hline
		
	\end{tabular}
	\caption{\label{tab:bf_13sigma_data}
		The best fit values, $1\sigma$ and $3\sigma$ ranges for the mixing parameters and neutrino mass squared differences, where the experimental data and uncertainties for both the NO and IO neutrino mass spectra are sourced from \texttt{NuFIT} 6.1 with Super-Kamiokande atmospheric data~\cite{Esteban:2024eli}. It is important to note that $\Delta m^2_{3\ell}=\Delta m^2_{31}>0$ for NO and $\Delta m^2_{3\ell}=\Delta m^2_{32}<0$ for IO.}
\end{table}

To assess the compatibility of each symmetry breaking pattern with current neutrino oscillation data, we evaluate a global $\chi^{2}$ function for every possible case. The test statistic is constructed from the one-dimensional $\chi^{2}$ projections provided by \texttt{NuFIT},
\begin{equation}\label{eq:chi2_def}
\chi^{2}(\vec{o})=\sum_{i=1}^{6}\chi_{i}^{2}(o_i)\,, \qquad
\vec{o}=\left(\sin^2\theta_{12},\,\sin^2\theta_{13},\,\sin^2\theta_{23},\,\delta_{CP},\, \Delta m^2_{21},\, \Delta m^2_{3\ell}\right)\,,
\end{equation}
where $\chi_i^{2}(o_i)$ denotes the marginalized contribution associated with the observable $o_i$. Throughout this work, we adopt the conservative choice $\chi^{2}_{\min,\mathrm{NO}}=\chi^{2}_{\min,\mathrm{IO}}=0$. For a given breaking pattern, the observables in $\vec{o}$ are not independent but are constrained by model-specific sum rules determined by three continuous real parameters, $m_a$, $r$ and $\eta$. The best-fit values of these parameters, together with the permutation matrix $P_\ell$ appearing in the PMNS matrix of Eq.~\eqref{eq:PMNS_MLSS}, are obtained by minimizing the global $\chi^{2}$ function in Eq.~\eqref{eq:chi2_def}. The corresponding predictions for the neutrino masses and lepton mixing parameters are then extracted at the minimum.

\begin{table}[t!]
\begin{center}
\renewcommand{\tabcolsep}{1.mm}
\renewcommand{\arraystretch}{1.1}
\begin{tabular}{|c|c|c|c|c|c|c|c|}\hline\hline
			
Order&Fixed column&Case&$\rho_\ell$&$G_\ell$&$P_\ell$& $Y_{\text{atm}}$&$Y_{\text{sol}}$\\ \hline
\multirow{21}{*}{NO} & \multirow{8}{*}{($\sqrt{\frac{2}{3}},\frac{1}{\sqrt{6}},\frac{1}{\sqrt{6}}$)}&$\mathcal{N}_{1}$&\multirow{4}{*}{$\bm{3_{0}}$}&\multirow{9}{*}{$Z^{ST}_{3}$}&$P_{321}$& $\rho_{\bm{\bar{3}_{0}}}(T^{6})Y_{\bm{\bar{3}_{0}}}^{(2)} (\tau _S)$&$\rho_{\bm{\bar{3}_{0}}}(T^{2})Y_{\bm{\bar{3}_{0}}}^{(4)} (\tau _S)$\\  \cline{3-3}\cline{6-8}
&& $\mathcal{N}_{2}$&&&$P_{321}$&$\rho_{\bm{\bar{3}_{0}}}(ST^4ST)Y_{\bm{\bar{3}_{0}}}^{(2)} (\tau _S)$&$\rho_{\bm{\bar{3}_{0}}}(T^5)Y_{\bm{\bar{3}_{0}}}^{(4)} (\tau _S)$\\ \cline{3-3}\cline{6-8}
&&$\mathcal{N}_{3}$&&&$P_{312}$&$\rho_{\bm{\bar{3}_{0}}}(T^5)Y_{\bm{\bar{3}_{0}}}^{(4)} (\tau _S)$&$\rho_{\bm{\bar{3}_{1}}}(T^7)Y_{\bm{\bar{3}_{1}}}^{(6)} (\tau _{ST})$\\ \cline{3-3}\cline{6-8}
&& $\mathcal{N}_{4}$&&&$P_{312}$&$\rho_{\bm{\bar{3}_{0}}}(T^2)Y_{\bm{\bar{3}_{0}}}^{(4)} (\tau _S)$&$\rho_{\bm{\bar{3}_{1}}}(ST^6ST^2)Y_{\bm{\bar{3}_{1}}}^{(6)} (\tau _{ST})$\\ \cline{3-4}\cline{6-8}
&&$\mathcal{N}_{5}$&\multirow{2}{*}{$\bm{\bar{3}_{0}}$}&&$P_{321}$&$\rho_{\bm{3_{0}}}(T^{4})Y_{\bm{3_{0}}}^{(6)} (\tau _{ST})$&$\rho_{\bm{3_{1}}}(ST^{2})Y_{\bm{3_{1}}}^{(4)} (\tau _{S})$\\\cline{3-3}\cline{6-8}
&&$\mathcal{N}_{6}$&&&$P_{321}$&$\rho_{\bm{3_{0}}}(T^{4})Y_{\bm{3_{0}}}^{(6)} (\tau _{ST})$&$\rho_{\bm{3_{1}}}(T^{2}ST^4)Y_{\bm{3_{1}}}^{(4)} (\tau _{S})$\\\cline{3-4}\cline{6-8}
&&$\mathcal{N}_{7}$&\multirow{2}{*}{$\bm{\hat{3}_{0}}$}&&$P_{321}$&$\rho_{\bm{\hat{3}_{0}}}(T^6ST^6)Y_{\bm{\hat{3}_{0}}}^{(2)} (\tau _{S})$&$\rho_{\bm{\hat{3}_{0}}}(T^2)Y_{\bm{\hat{3}_{0}}}^{(4)} (\tau _{S})$\\\cline{3-3}\cline{6-8}
&&$\mathcal{N}_{8}$&&&$P_{321}$&$\rho_{\bm{\hat{3}_{0}}}(ST^5)Y_{\bm{\hat{3}_{0}}}^{(2)} (\tau _{S})$&$\rho_{\bm{\hat{3}_{0}}}(T^2ST^3)Y_{\bm{\hat{3}_{0}}}^{(4)} (\tau _{S})$\\\cline{2-4}\cline{5-8}
			
&\multirow{2}{*}{(0.827, 0.290, 0.481)}&$\mathcal{N}_{9}$&\multirow{5}{*}{$\bm{3_{0}}$}&\multirow{3}{*}{$Z^{T}_{8}$}&$P_{132}$&$Y_{\bm{\bar{3}_{0}}}^{(2)} (\tau _{ST})$&$\rho_{\bm{\bar{3}_{1}}}(TST^5)Y_{\bm{\bar{3}_{1}}}^{(4)} (\tau _S)$\\\cline{3-3}\cline{6-8}
&&$\mathcal{N}_{10}$&&&$P_{312}$&$\rho_{\bm{\bar{3}_{0}}}(T^4ST^4)Y_{\bm{\bar{3}_{0}}}^{(2)} (\tau _{ST})$&$\rho_{\bm{\bar{3}_{1}}}(ST)Y_{\bm{\bar{3}_{1}}}^{(4)} (\tau _S)$\\\cline{2-3}\cline{6-8}
			
& (0.818, 0.334, 0.468)&$\mathcal{N}_{11}$&&&$P_{312}$&$\rho_{\bm{\bar{3}_{0}}}(ST^4ST^4)Y_{\bm{\bar{3}_{0}}}^{(2)} (\tau _{ST})$&$\rho_{\bm{\bar{3}_{1}}}(T^4ST)Y_{\bm{\bar{3}_{1}}}^{(4)} (\tau _{S})$\\ \cline{2-3}\cline{5-8}
&\multirow{2}{*}{(0.820, 0.295, 0.490)}&$\mathcal{N}_{12}$&&\multirow{4}{*}{$Z^{ST}_{3}$}&$P_{231}$&$\rho_{\bm{\bar{3}_{1}}}(T^2ST^5)Y_{\bm{\bar{3}_{1}}}^{(4)} (\tau _{S})$&$\rho_{\bm{\bar{3}_{1}}}(T^6)Y_{\bm{\bar{3}_{1}}}^{(6)} (\tau _{ST})$\\\cline{3-3}\cline{6-8} 
&&$\mathcal{N}_{13}$&&&$P_{231}$&$\rho_{\bm{\bar{3}_{1}}}(T^2ST^5)Y_{\bm{\bar{3}_{1}}}^{(4)} (\tau _{S})$&$\rho_{\bm{\bar{3}_{1}}}(T^7ST^3)Y_{\bm{\bar{3}_{1}}}^{(6)} (\tau _{ST})$\\ \cline{2-4}\cline{6-8} 
			
&(0.813, 0.380, 0.442)&	$\mathcal{N}_{14}$&\multirow{8}{*}{$\bm{\bar{3}_{0}}$}&&$P_{231}$&$\rho_{\bm{3_{0}}}(T^2ST^5)Y_{\bm{3_{0}}}^{(6)} (\tau _{S})$&$\rho_{\bm{3_{0}}}(ST^2)Y_{\bm{3_{0}}}^{(8)} (\tau _{ST})$\\ \cline{2-3}\cline{6-8}
			
&(0.813, 0.442, 0.380)&$\mathcal{N}_{15}$&&&$P_{213}$&$\rho_{\bm{3_{0}}}(T^2ST^6)Y_{\bm{3_{0}}}^{(6)} (\tau _{S})$&$\rho_{\bm{3_{0}}}(S)Y_{\bm{3_{0}}}^{(8)} (\tau _{ST})$\\ \cline{2-3}\cline{5-8}
			
&\multirow{2}{*}{(0.818, 0.314, 0.482)}&$\mathcal{N}_{16}$&&\multirow{6}{*}{$Z^{T}_{8}$}&$P_{132}$&$\rho_{\bm{3_{0}}}(ST^4)Y_{\bm{3_{0}}}^{(6)} (\tau _{S})$&$\rho_{\bm{3_{0}}}(ST)Y_{\bm{3_{0}}}^{(6)} (\tau _{S})$\\ \cline{3-3}\cline{6-8}
&&$\mathcal{N}_{17}$&&&$P_{132}$&$\rho_{\bm{3_{0}}}(T^2ST^4)Y_{\bm{3_{0}}}^{(6)} (\tau _{S})$&$\rho_{\bm{3_{0}}}(T^3ST)Y_{\bm{3_{0}}}^{(6)} (\tau _{S})$\\ \cline{2-3}\cline{6-8}
			
&(0.823, 0.454, 0.341)&$\mathcal{N}_{18}$&&&$P_{123}$&$\rho_{\bm{3_{0}}}(ST^5)Y_{\bm{3_{0}}}^{(8)} (\tau _{S})$&$\rho_{\bm{3_{1}}}(T^4)Y_{\bm{3_{1}}}^{(4)} (\tau _{S})$\\ \cline{2-3}\cline{6-8}
			
&(0.823, 0.341, 0.454)&$\mathcal{N}_{19}$&&&$P_{312}$&$\rho_{\bm{3_{0}}}(TST)Y_{\bm{3_{0}}}^{(8)} (\tau _{S})$&$\rho_{\bm{3_{1}}}(T^4ST^4)Y_{\bm{3_{1}}}^{(4)} (\tau _{S})$\\\cline{2-3}\cline{6-8}			
			
&(0.832, 0.371, 0.413)&$\mathcal{N}_{20}$&&&$P_{312}$&$Y_{\bm{3_{0}}}^{(6)}( \tau _{ST})$&$\rho_{\bm{3_{1}}}(T^3ST^4)Y_{\bm{3_{1}}}^{(4)} (\tau _{S})$\\ \cline{2-3}\cline{6-8}
			
&(0.832, 0.413, 0.371)&$\mathcal{N}_{21}$&&&$P_{123}$&$\rho_{\bm{3_{0}}}(ST^2)Y_{\bm{3_{0}}}^{(6)} (\tau _{ST}) $&$\rho_{\bm{3_{1}}}(T^2)Y_{\bm{3_{1}}}^{(4)} (\tau _{S})$\\ \hline\hline

\multirow{14}{*}{IO}
			
&(0.154, 0.693, 0.704)&$\mathcal{I}_{1}$&\multirow{5}{*}{$\bm{3_{0}}$}&\multirow{11}{*}{$Z^{ST}_{3}$}&$P_{132}$&$\rho_{\bm{\bar{3}_{1}}}(T^2ST^2)Y_{\bm{\bar{3}_{1}}}^{(4)} (\tau _{S})$&$\rho_{\bm{\bar{3}_{1}}}(T^4ST^6)Y_{\bm{\bar{3}_{1}}}^{(4)} (\tau _{S})$\\ \cline{2-3}\cline{6-8}
			
&\multirow{2}{*}{(0.154, 0.704, 0.693)}&$\mathcal{I}_{2}$&&&$P_{213}$&$\rho_{\bm{\bar{3}_{1}}}(T^3)Y_{\bm{\bar{3}_{1}}}^{(4)} (\tau _{S})$&$\rho_{\bm{\bar{3}_{1}}}(ST^2ST)Y_{\bm{\bar{3}_{1}}}^{(4)} (\tau _{S})$\\\cline{3-3}\cline{6-8}
&&$\mathcal{I}_{3}$&&&$P_{213}$&$\rho_{\bm{\bar{3}_{1}}}(ST^2)Y_{\bm{\bar{3}_{1}}}^{(4)} (\tau _{S})$&$\rho_{\bm{\bar{3}_{1}}}(T^4)Y_{\bm{\bar{3}_{1}}}^{(4)} (\tau _{S})$\\ \cline{2-3}\cline{6-8}
			
&(0.149, 0.717, 0.681)&$\mathcal{I}_{4}$&&&$P_{132}$&$\rho_{\bm{\bar{3}_{1}}}(T^4ST^5)Y_{\bm{\bar{3}_{1}}}^{(4)} (\tau _{S})$&$\rho_{\bm{\bar{3}_{1}}}(T^7ST^4)Y_{\bm{\bar{3}_{1}}}^{(4)} (\tau _{S})$\\ \cline{2-3}\cline{6-8}
			
&(0.149, 0.681, 0.717)&$\mathcal{I}_{5}$&&&$P_{123}$&$\rho_{\bm{\bar{3}_{1}}}(T^4)Y_{\bm{\bar{3}_{1}}}^{(4)} (\tau _{S})$&$\rho_{\bm{\bar{3}_{1}}}(T^3)Y_{\bm{\bar{3}_{1}}}^{(4)} (\tau _{S})$\\ \cline{2-4}\cline{6-8}
			
&\multirow{2}{*}{(0.145, 0.687, 0.712)}&$\mathcal{I}_{6}$&\multirow{9}{*}{$\bm{\bar{3}_{0}}$}&&$P_{231}$&$\rho_{\bm{3_{0}}}(ST^5)Y_{\bm{3_{0}}}^{(8)} (\tau _{ST})$&$\rho_{\bm{3_{1}}}(T^3ST)Y_{\bm{3_{1}}}^{(4)} (\tau _{S})$\\ \cline{3-3}\cline{6-8}
&&$\mathcal{I}_{7}$&&&$P_{132}$&$\rho_{\bm{3_{0}}}(TST^7)Y_{\bm{3_{0}}}^{(8)} (\tau _{ST})$&$\rho_{\bm{3_{1}}}(TST)Y_{\bm{3_{1}}}^{(4)} (\tau _{S})$\\ \cline{2-3}\cline{6-8}
			
&(0.145, 0.712, 0.687)&$\mathcal{I}_{8}$&&&$P_{213}$&$\rho_{\bm{3_{0}}}(T^4)Y_{\bm{3_{0}}}^{(8)} (\tau _{ST})$&$\rho_{\bm{3_{1}}}(T^3ST^2)Y_{\bm{3_{1}}}^{(4)} (\tau _{S})$\\ \cline{2-3}\cline{6-8}
			
&\multirow{2}{*}{(0.153, 0.699, 0.699)}&$\mathcal{I}_{9}$&&&$P_{213}$&$\rho_{\bm{3_{0}}}(T^2)Y_{\bm{3_{0}}}^{(8)} (\tau _{ST})$&$\rho_{\bm{3_{1}}}(T^5ST^2)Y_{\bm{3_{1}}}^{(4)} (\tau _{S})$\\\cline{3-3}\cline{6-8}
&&$\mathcal{I}_{10}$&&&$P_{132}$&$\rho_{\bm{3_{0}}}(T^5)Y_{\bm{3_{0}}}^{(8)} (\tau _{ST})$&$\rho_{\bm{3_{1}}}(ST)Y_{\bm{3_{1}}}^{(4)} (\tau _{S})$\\\cline{2-3}\cline{6-8}
			
&(0.145, 0.712, 0.687)&$\mathcal{I}_{11}$&&&$P_{213}$&$\rho_{\bm{3_{0}}}(T^4)Y_{\bm{3_{0}}}^{(8)} (\tau _{ST})$&$\rho_{\bm{3_{1}}}(T^4ST^6)Y_{\bm{3_{1}}}^{(4)} (\tau _{S})$\\ \cline{2-3}\cline{5-8}
			
&(0.145, 0.681, 0.717)&$\mathcal{I}_{12}$&&\multirow{3}{*}{$Z^{T}_{8}$}&$P_{312}$&$\rho_{\bm{3_{0}}}(T^4)Y_{\bm{3_{0}}}^{(8)} (\tau _{ST})$&$\rho_{\bm{3_{1}}}(T^4ST^2)Y_{\bm{3_{1}}}^{(4)} (\tau _{S})$\\ \cline{2-3}\cline{6-8}
			
&(0.151, 0.659, 0.737)&$\mathcal{I}_{13}$&&&$P_{132}$&$\rho_{\bm{3_{0}}}(T^4ST^2)Y_{\bm{3_{0}}}^{(6)} (\tau _{S})$&$\rho_{\bm{3_{0}}}(T^5)Y_{\bm{3_{0}}}^{(8)} (\tau _{ST})$\\ \cline{2-3}\cline{6-8}
			
& (0.151, 0.737, 0.659)&$\mathcal{I}_{14}$&&&$P_{321}$&$\rho_{\bm{3_{0}}}(T^4ST^2)Y_{\bm{3_{0}}}^{(6)} (\tau _{S})$&$\rho_{\bm{3_{0}}}(ST^4ST^4)Y_{\bm{3_{0}}}^{(8)} (\tau _{ST})$\\ \hline\hline
\end{tabular}
\caption{\label{tab:viable_BPs}The permutation matrices $P_\ell$,  the representation $\rho_{\ell}$ of $L$, the residual symmetry $G_\ell$ of the charged lepton sector, the modular forms $Y_{\text{atm}}$ and $Y_{\text{sol}}$ at the fixed points $\langle\tau_{\text{atm}}\rangle$ and $\langle\tau_{\text{sol}}\rangle$, and the corresponding fixed columns of the PMNS matrix for the 35 viable breaking patterns. The analytical expressions for the fixed columns of the latter 27 viable patterns are given in appendix~\ref{sec:fixed_column}. }
	\end{center}
\end{table}

\begin{table}[t!]
\begin{center}
\renewcommand{\tabcolsep}{1.6mm}
\renewcommand{\arraystretch}{1.1}
\begin{tabular}{|c||c|c|c||c|c|c|c|c|c|c|c|c|}\hline\hline
			
NO&	$\eta/\pi$&	$\frac{m_{a}}{\text{meV}}$&$r$&$\chi^2_{\text{min}}$&$\sin^{2}\theta_{13}$&$\sin^{2}\theta_{12}$&$\sin^{2}\theta_{23}$&$\delta_{CP}/\pi$&$\beta/\pi$ &$\frac{m_{2}}{\text{meV}}$&$\frac{m_{3}}{\text{meV}}$ & $\frac{m_{ee}}{\text{meV}}$\\ \hline
			
$\mathcal{N}_1$	&1.53	&3.52	&3.10	&5.73	&0.02247	&0.3180	&0.546	&1.56	&1.24	&8.68	&50.1	&3.32\\ \hline
$\mathcal{N}_2$	&0.469	&3.53	&3.08	&5.15	&0.02266	&0.3179	&0.457	&1.44	&0.757	&8.66	&50.2	&3.33\\ \hline
$\mathcal{N}_3$	&1.85	&12.3	&1.24	&7.30	&0.02205	&0.3183	&0.450	&1.43	&1.63	&8.70	&50.1	&2.96\\ \hline
$\mathcal{N}_4$	&1.66	&12.2	&1.25	&5.76	&0.02248	&0.3180	&0.555	&1.58	&0.362	&8.69	&50.1	&2.98\\ \hline
$\mathcal{N}_5$	&0.636	&6.11	&1.16	&5.77	&0.02255	&0.3180	&0.545	&1.56	&0.402	&8.68	&50.1	&3.01\\ \hline
$\mathcal{N}_6$	&0.863	&6.10	&1.17	&5.10	&0.02237	&0.3181	&0.457	&1.44	&1.60	&8.67	&50.2	&2.98\\ \hline
$\mathcal{N}_7$	&0.756	&3.49	&4.47	&6.64	&0.02293	&0.3177	&0.452	&1.43	&0.463	&8.66	&50.2	&2.32\\ \hline
$\mathcal{N}_8$	&1.24	&3.47	&4.52	&5.70	&0.02256	&0.3179	&0.554	&1.57	&1.53	&8.68	&50.2	&2.31\\ \hline
$\mathcal{N}_9$	&0.489	&5.09	&0.218	&5.69	&0.02237	&0.3001	&0.528	&0.913	&1.94	&8.68	&50.1	&3.47\\ \hline
$\mathcal{N}_{10}$	&1.34	&5.10	&0.218	&4.42	&0.02235	&0.3001	&0.528	&1.09	&0.0558	&8.68	&50.2	&3.46\\ \hline
$\mathcal{N}_{11}$	&1.41	&5.33	&0.151	&4.58	&0.02230	&0.3151	&0.568	&1.34	&2.00	&8.69	&50.1	&2.30\\ \hline
$\mathcal{N}_{12}$	&0.113	&3.39	&2.11	&3.90	&0.02242	&0.3122	&0.540	&0.886	&0.618	&8.68	&50.1	&3.21\\ \hline
$\mathcal{N}_{13}$	&1.39	&3.39	&2.11	&1.70	&0.02249	&0.3121	&0.540	&1.12	&1.38	&8.68	&50.2	&3.21\\ \hline
$\mathcal{N}_{14}$	&1.07	&8.54	&0.266	&8.32	&0.02259	&0.3240	&0.542	&1.44	&0.588	&8.68	&50.2	&2.88\\ \hline
$\mathcal{N}_{15}$	&1.43	&8.55	&0.265	&8.55	&0.02243	&0.3241	&0.460	&1.56	&1.41	&8.68	&50.1	&2.87\\ \hline
$\mathcal{N}_{16}$	&1.29	&3.10	&1.81	&3.26	&0.02259	&0.3152	&0.492	&1.04	&0.0500	&8.68	&50.1	&3.74\\ \hline
$\mathcal{N}_{17}$	&0.714	&3.10	&1.81	&3.57	&0.02258	&0.3152	&0.492	&0.961	&1.95	&8.68	&50.2	&3.74\\ \hline
$\mathcal{N}_{18}$	&0.449	&4.23	&1.71	&5.91	&0.02213	&0.3067	&0.451	&1.65	&1.25	&8.69	&50.1	&2.69\\ \hline
$\mathcal{N}_{19}$	&1.54	&4.23	&1.70	&2.28	&0.02242	&0.3065	&0.552	&1.36	&0.741	&8.68	&50.1	&2.69\\ \hline
$\mathcal{N}_{20}$	&1.19	&3.29	&1.37	&8.21	&0.02235	&0.2925	&0.460	&1.37	&1.02	&8.67	&50.2	&3.38\\ \hline
$\mathcal{N}_{21}$	&0.306	&3.29	&1.37	&10.5	&0.02238	&0.2925	&0.541	&1.63	&0.976	&8.68	&50.1	&3.38\\ \hline\hline
			
IO&	$\eta/\pi$&	$\frac{m_{a}}{\text{meV}}$&$r$&$\chi^2_{\text{min}}$&$\sin^{2}\theta_{13}$&$\sin^{2}\theta_{12}$&$\sin^{2}\theta_{23}$&$\delta_{CP}/\pi$&$\beta/\pi$ &$\frac{m_{1}}{\text{meV}}$&$\frac{m_{2}}{\text{meV}}$ & $\frac{m_{ee}}{\text{meV}}$\\ \hline
			
$\mathcal{I}_1$	&1.86	&4.15	&0.988	&7.26	&0.02383	&0.3089	&0.493	&1.57	&1.93	&49.1	&49.8	&16.0\\ \hline
$\mathcal{I}_2$	&0.135	&4.15	&0.988	&7.89	&0.02383	&0.3086	&0.507	&1.43	&0.0735	&49.1	&49.8	&16.0\\ \hline
$\mathcal{I}_3$	&0.139	&4.14	&0.991	&7.85	&0.02383	&0.3087	&0.507	&1.60	&0.0762	&49.1	&49.8	&16.0\\ \hline
$\mathcal{I}_4$	&0.616	&3.84	&0.992	&6.70	&0.02228	&0.3103	&0.526	&1.24	&1.21	&49.1	&49.8	&13.8\\ \hline
$\mathcal{I}_5$	&1.38	&3.84	&0.992	&6.42	&0.02228	&0.3108	&0.474	&1.76	&0.787	&49.1	&49.8	&13.8\\ \hline
$\mathcal{I}_6$	&1.14	&2.63	&3.05	&10.8	&0.02108	&0.3087	&0.482	&1.36	&1.81	&49.1	&49.8	&15.2\\ \hline
$\mathcal{I}_7$	&1.68	&2.61	&3.09	&9.59	&0.02108	&0.3090	&0.482	&1.57	&1.82	&49.1	&49.8	&15.4\\ \hline
$\mathcal{I}_8$	&0.359	&2.63	&3.05	&10.8	&0.02108	&0.3091	&0.518	&1.64	&0.189	&49.1	&49.8	&14.6\\ \hline
$\mathcal{I}_9$	&1.74	&3.90	&3.06	&12.1	&0.02339	&0.3080	&0.500	&1.16	&0.488	&49.1	&49.8	&15.2\\ \hline
$\mathcal{I}_{10}$	&1.09	&3.90	&3.06	&12.6	&0.02339	&0.3079	&0.500	&1.84	&1.51	&49.1	&49.8	&15.9\\ \hline
$\mathcal{I}_{11}$	&1.15	&2.61	&3.09	&10.4	&0.02108	&0.3090	&0.518	&1.43	&0.184	&49.1	&49.8	&14.4\\ \hline
$\mathcal{I}_{12}$	&0.214	&2.35	&3.08	&27.2	&0.02098	&0.3295	&0.474	&1.14	&1.02	&49.1	&49.8	&16.8\\ \hline
$\mathcal{I}_{13}$	&0.309	&5.36	&0.415	&13.7	&0.02284	&0.3082	&0.444	&1.82	&1.63	&49.1	&49.8	&15.6\\ \hline
$\mathcal{I}_{14}$	&0.191	&5.36	&0.415	&6.91	&0.02284	&0.3081	&0.556	&1.18	&0.374	&49.1	&49.8	&15.6\\ \hline \hline
\end{tabular}
\caption{\label{tab:bf_value}Best fit values of the input parameters and predicted lepton observables for the 21 phenomenologically viable and experimentally favored breaking patterns in the NO case and the 14 viable breaking patterns in the IO case listed in table~\ref{tab:viable_BPs}.}
\end{center}
\end{table}
\FloatBarrier

For any given set of input parameters, the model predicts the neutrino masses, mixing parameters, and the corresponding $\chi^{2}$ value. By performing a comprehensive scan of the parameter space, the global minimum of $\chi^{2}$ can be determined. In our analysis, the input parameters $m_a$, $r$, and $\eta$ are randomly sampled over the ranges $[0,1\,\text{eV}]$, $[0,100]$, and $[0,2\pi]$, respectively. Through a systematic exploration of all 89662 independent breaking patterns, we identify 21 (14) phenomenologically viable breaking patterns for the NO (IO) neutrino mass spectrum. A breaking pattern is regarded as viable if, at its $\chi^{2}$ minimum, the predicted values of the three mixing angles $\theta_{12}$, $\theta_{13}$, $\theta_{23}$, the Dirac CP phase $\delta_{CP}$, and the mass squared differences $\Delta m^2_{21}$ and $\Delta m^2_{3\ell}$ all lie in the corresponding $3\sigma$ intervals summarized in table~\ref{tab:bf_13sigma_data}. The resulting 35 viable breaking patterns are summarized in table~\ref{tab:viable_BPs}, together with the associated permutation matrices $P_\ell$, the representation assignments of the left-handed lepton doublets $L$, the residual symmetry $G_\ell$ of the charged lepton sector, the modular forms $Y_{\text{atm}}$ and $Y_{\text{sol}}$ evaluated at the fixed points $\langle\tau_{\text{atm}}\rangle$ and $\langle\tau_{\text{sol}}\rangle$, and the corresponding fixed columns of the PMNS matrix. Specifically, among the 21 viable NO breaking patterns, 8 of them ($\mathcal{N}_{1}$ to $\mathcal{N}_{8}$) yield the well-known TM$_1$ mixing. In contrast, the fixed PMNS columns associated with the remaining NO patterns and all IO patterns are obtained here for the first time. For brevity, only the numerical results are listed in table~\ref{tab:viable_BPs}, and the corresponding analytical expressions are given in appendix~\ref{sec:fixed_column}.

For the 21 viable NO patterns and 14 viable IO patterns, the best-fit values of the input parameters, along with the resulting predictions for the lepton mixing angles, CP-violating phases, neutrino masses, and the effective Majorana mass $m_{ee}$ relevant for neutrinoless double-beta decay ($0\nu\beta\beta$), are presented in table~\ref{tab:bf_value}. Furthermore, the two breaking patterns $\mathcal{N}_{7}$ and $\mathcal{N}_{8}$ correspond to the so-called Modular Littlest Seesaw model CSD$(1+\sqrt{6})$~\cite{deMedeirosVarzielas:2022fbw} and CSD$(1-\sqrt{6})$~\cite{Ding:2019gof,Ding:2021zbg,deMedeirosVarzielas:2022fbw} based on the finite modular symmetry $S_{4}$. This connection is natural because the two three-dimensional representations $\bm{\hat{3}_{m}}$ are not faithful representations of $\Delta(96)$. They generate the $S_{4}$ group, so the results are expected to be consistent with $S_{4}$. 

Although both this work and Ref.~\cite{Yan:2025itm} are based on the finite group $\Delta(96)$ and minimal seesaw scenario, they realize flavor symmetry in different ways. Ref.~\cite{Yan:2025itm} adopts a tri-direct CP framework, where the lepton structures are determined by residual flavor and generalized CP symmetries, in which vacuum alignments are fixed by residual symmetries, and in some cases an additional continuous parameter $x$ appears, whose phase may be further constrained by CP consistency conditions. In contrast, the present work constructs a Modular Littlest Seesaw based on VVMFs evaluated at symmetry preserving fixed points. Although residual stabilizer symmetries may allow degenerate eigenspaces, the full modular covariance uniquely fixes the VVMFs at fixed points (up to normalization), eliminating any additional continuous freedom. 
The resulting alignments therefore belong to a finite set of exact algebraic directions rather than to a continuously deformable vacuum ansatz. This algebraic rigidity makes the modular construction more constrained, and the resulting vacua are in general not contained as special cases of those in Ref.~\cite{Yan:2025itm}. The two approaches can thus be regarded as complementary realizations of the $\Delta(96)$ flavor symmetry.

\begin{figure}[t!]
	\centering
	\begin{tabular}{c}
		\includegraphics[width=0.92\linewidth]{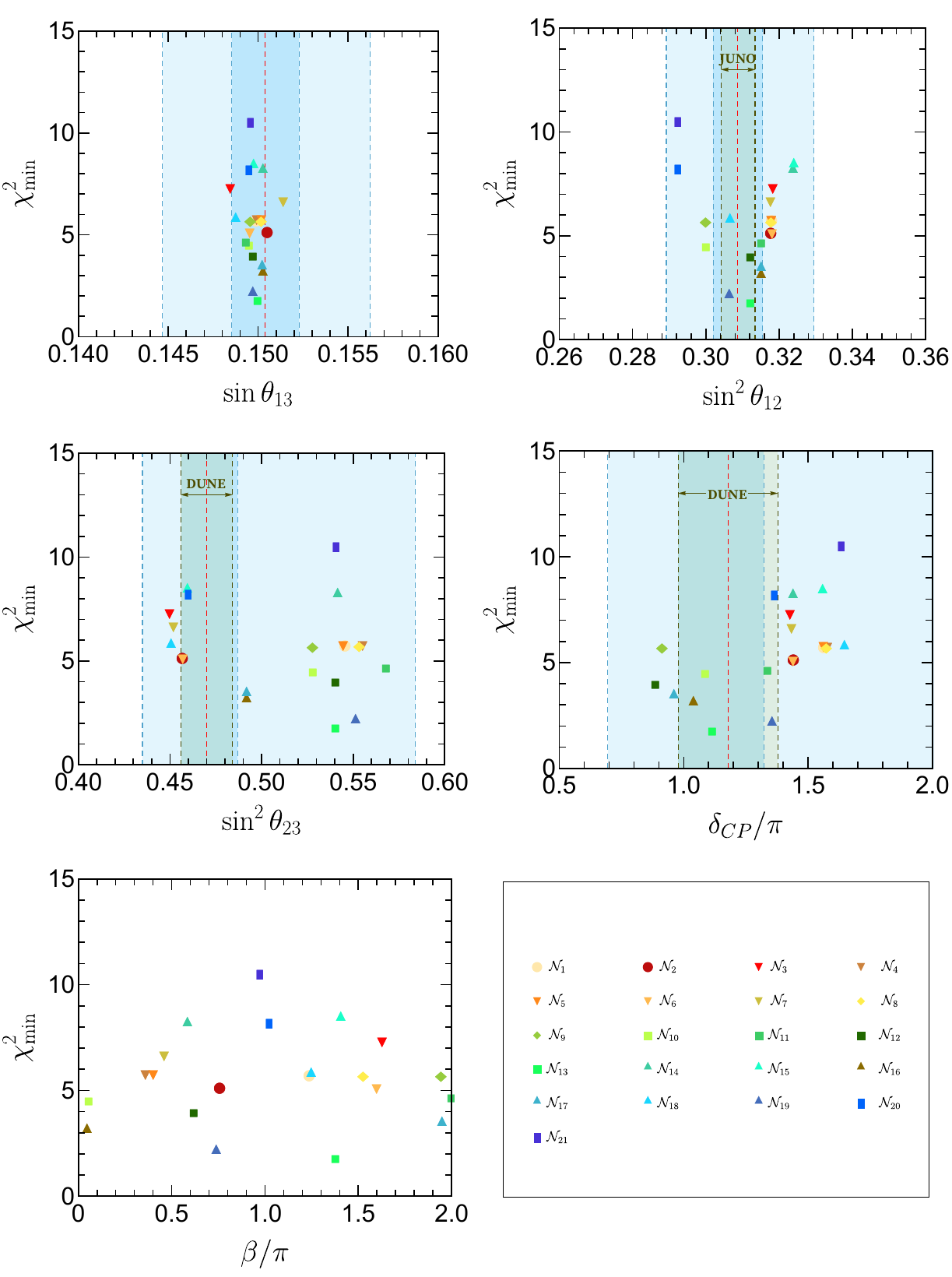}
	\end{tabular}
	\caption{\label{fig:bf_mixing_NO} The best fit results for 21 viable breaking patterns with NO neutrino mass spectrum, showing $\chi^2$ minima, lepton mixing angles, and CP violation phases. Red dashed lines indicate best fit values, light blue bands show the $1\sigma$ and $3\sigma$ ranges from \texttt{NuFIT} 6.1 with Super-Kamiokande atmospheric data~\cite{Esteban:2024eli}. The pale green band shows the predicted $3\sigma$ range for $\sin^{2}\theta_{12}$ after 6 years of JUNO data~\cite{JUNO:2022mxj,JUNO:2025gmd}. The faint green areas indicate the resolution (in degrees) for $\sin^{2}\theta_{23}$ and $\delta_{CP}$ after 15 years of DUNE operation~\cite{DUNE:2020ypp}. }
\end{figure}

\begin{figure}[t!]
	\centering
	\begin{tabular}{c}
		\includegraphics[width=0.92\linewidth]{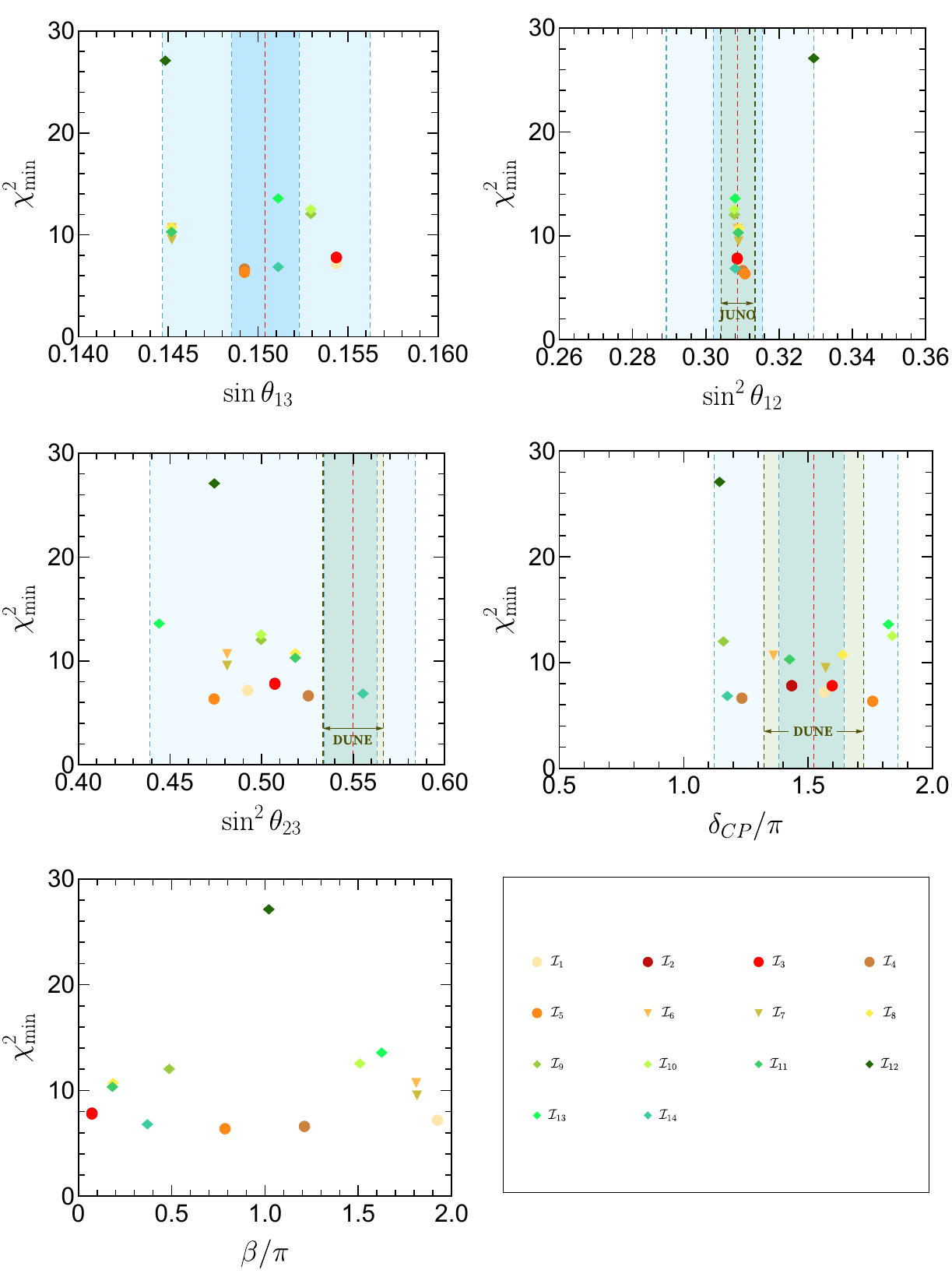}
	\end{tabular}
	\caption{\label{fig:bf_mixing_IO} The best fit $\chi^2$ for all 14 viable breaking patterns with IO neutrino mass spectrum, covering the three lepton mixing angles and CP violation phases.}
\end{figure}

To visually complement the quantitative summary presented in table~\ref{tab:bf_value}, we display in figures~\ref{fig:bf_mixing_NO} and~\ref{fig:bf_mixing_IO} the best fit predictions of all viable symmetry breaking patterns alongside the currently allowed $3\sigma$ regions from \texttt{NuFIT} and the future sensitivity bands of JUNO, DUNE and T2HK.  In the figures, red dashed lines and light blue bands denote the central values, the $1\sigma$ and $3\sigma$ ranges from the \texttt{NuFIT} global fit. From these figures, we observe that nearly all viable models predict mixing angles and the Dirac CP phase slightly outside the experimentally allowed $1\sigma$ ranges~\cite{Esteban:2024eli}. Future high-precision measurements from neutrino oscillation experiments and cosmological surveys will be crucial for discriminating among these Modular Littlest Seesaw models. JUNO is expected to significantly refine $\theta_{12}$, with the pale green band in $\sin^2\theta_{12}$ panels representing the projected $3\sigma$ range after six years~\cite{JUNO:2022mxj,JUNO:2025gmd}:
\begin{equation}
\label{eq:s12sq_JUNO_6years}
0.3042 \leq \sin^{2}\theta_{12} \leq 0.3134 \,,
\end{equation}
for both NO and IO. Light green bands for $\sin^2\theta_{23}$ and $\delta_{CP}$ indicate the anticipated $3\sigma$ sensitivities of DUNE and T2HK. Adopting representative uncertainties of $3\%$ for $\sin^2\theta_{23}$ and $12^\circ$ for $\delta_{CP}$, the projected $3\sigma$ ranges are 
\begin{equation}
		\label{eq:theta23_DUNE_and_T2HK}
	0.4559 \leq \sin^{2}\theta_{23} \leq 0.4841 \quad \text{for NO}, \qquad 0.5335 \leq \sin^{2}\theta_{23} \leq 0.5665 \quad \text{for IO}.
\end{equation}
and 
\begin{equation}
	\label{eq:deltaCP_DUNE_and_T2HK}
	0.978 \leq \delta_{CP}/\pi \leq 1.378 \quad \text{for NO}, \qquad 1.322 \leq \delta_{CP}/\pi \leq 1.722 \quad \text{for IO}.
\end{equation}
This concise graphical overview highlights how current constraints align with model predictions and demonstrates the discriminating power of upcoming experiments.
\begin{table}[t!]
\begin{center}
\small
\renewcommand{\tabcolsep}{0.1mm}
\renewcommand{\arraystretch}{1.15}
\begin{tabular}{|c||c|c|c||c|c|c|c|c|}\hline\hline
		
NO & $\eta /\pi $ & $\frac{m_{a}}{\text{meV}}$ & $r$ & $\sin ^2\theta _{12}$ & $\sin ^2\theta _{23}$ & $\delta_{CP}/\pi$ & $\beta/\pi$ & $\frac{m_{ee}}{\text{meV}}$ \\ \hline
$\mathcal{N}_{1}$	&[1.48,1.57]	&[3.37,3.66]	&[2.90,3.32]	&[0.3168,0.3193]	&[0.521,0.570]	&[1.53,1.60]	&[1.20,1.28]	&[3.18,3.45]\\ \hline
$\mathcal{N}_{2}$	&[0.431,0.509]	&[3.38,3.66]	&[2.90,3.30]	&[0.3168,0.3193]	&\cellcolor{red!20}[0.435,0.479]	&[1.40,1.47]	&[0.725,0.796]	&[3.18,3.45]\\ \hline
$\mathcal{N}_{3}$	&[1.83,1.88]	&[11.8,12.7]	&[1.17,1.31]	&[0.3168,0.3193]	&\cellcolor{red!20}[0.435,0.475]	&[1.41,1.46]	&[1.59,1.65]	&[2.85,3.08]\\ \hline
$\mathcal{N}_{4}$	&[1.62,1.69]	&[11.7,12.7]	&[1.17,1.34]	&[0.3168,0.3193]	&[0.525,0.583]	&[1.54,1.61]	&[0.322,0.407]	&[2.85,3.10]\\ \hline
$\mathcal{N}_{5}$	&[0.628,0.645]	&[5.91,6.29]	&[1.13,1.22]	&[0.3168,0.3193]	&[0.524,0.560]	&[1.54,1.58]	&[0.376,0.421]	&[2.69,3.25]\\ \hline
$\mathcal{N}_{6}$	&[0.855,0.872]	&[5.91,6.29]	&[1.13,1.22]	&[0.3168,0.3193]	&\cellcolor{red!20}[0.440,0.476]	&[1.42,1.46]	&[1.58,1.62]	&[2.69,3.26]\\ \hline
$\mathcal{N}_{7}$	&[0.731,0.774]	&[3.35,3.60]	&[4.25,4.75]	&[0.3168,0.3193]	&\cellcolor{red!20}[0.435,0.478]	&[1.40,1.47]	&[0.416,0.503]	&[2.24,2.40]\\ \hline
$\mathcal{N}_{8}$	&[1.21,1.27]	&[3.32,3.60]	&[4.25,4.83]	&[0.3168,0.3193]	&[0.522,0.584]	&[1.53,1.63]	&[1.46,1.58]	&[2.21,2.40]\\ \hline
$\mathcal{N}_{9}$	&[0.452,0.522]	&[4.97,5.22]	&[0.207,0.230]	&[0.2988,0.3013]	&[0.518,0.537]	&[0.899,0.925]	&[1.92,1.97]	&[3.33,3.60]\\ \hline
$\mathcal{N}_{10}$	&[1.31,1.38]	&[4.97,5.22]	&[0.207,0.229]	&[0.2988,0.3013]	&[0.518,0.537]	&\cellcolor{red!20}[1.07,1.10]	&[0.0294,0.0803]	&[3.33,3.60]\\ \hline
$\mathcal{N}_{11}$	&[1.34,1.49]	&[5.20,5.49]	&[0.142,0.160]	&[0.3138,0.3163]	&[0.556,0.584]	&\cellcolor{red!20}[1.31,1.37]	&[0,2.00)	&[2.21,2.38]\\ \hline
$\mathcal{N}_{12}$	&[0.0373,0.172]	&[3.30,3.46]	&[2.02,2.21]	&\cellcolor{red!20}[0.3109,0.3134]	&[0.536,0.546]	&[0.857,0.909]	&[0.555,0.696]	&[3.15,3.27]\\ \hline
$\mathcal{N}_{13}$	&[1.33,1.46]	&[3.30,3.46]	&[2.02,2.21]	&\cellcolor{red!20}[0.3109,0.3134]	&[0.536,0.546]	&\cellcolor{red!20}[1.09,1.14]	&[1.31,1.45]	&[3.15,3.27]\\ \hline
$\mathcal{N}_{14}$	&[1.05,1.08]	&[8.33,8.77]	&[0.259,0.271]	&[0.3229,0.3253]	&[0.524,0.557]	&[1.41,1.46]	&[0.559,0.624]	&[2.79,2.96]\\ \hline
$\mathcal{N}_{15}$	&[1.42,1.45]	&[8.32,8.77]	&[0.259,0.271]	&[0.3229,0.3253]	&\cellcolor{red!20}[0.443,0.476]	&[1.54,1.59]	&[1.38,1.44]	&[2.79,2.96]\\ \hline
$\mathcal{N}_{16}$	&[1.27,1.30]	&[3.01,3.19]	&[1.74,1.88]	&[0.3141,0.3166]	&\cellcolor{red!20}[0.484,0.503]	&\cellcolor{red!20}[1.03,1.05]	&[0.0350,0.0616]	&[3.62,3.85]\\ \hline
$\mathcal{N}_{17}$	&[0.701,0.729]	&[3.01,3.19]	&[1.74,1.88]	&[0.3141,0.3166]	&\cellcolor{red!20}[0.484,0.503]	&[0.953,0.969]	&[1.94,1.97]	&[3.62,3.85]\\ \hline
$\mathcal{N}_{18}$	&[0.404,0.493]	&[4.10,4.37]	&[1.63,1.78]	&\cellcolor{red!20}[0.3052,0.3077]	&\cellcolor{red!20}[0.435,0.466]	&[1.62,1.68]	&[1.20,1.30]	&[2.63,2.75]\\ \hline
$\mathcal{N}_{19}$	&[1.48,1.60]	&[4.10,4.37]	&[1.63,1.78]	&\cellcolor{red!20}[0.3052,0.3077]	&[0.534,0.575]	&\cellcolor{red!20}[1.32,1.39]	&[0.670,0.804]	&[2.63,2.75]\\ \hline
$\mathcal{N}_{20}$	&[1.17,1.21]	&[3.17,3.40]	&[1.30,1.45]	&[0.2912,0.2937]	&\cellcolor{red!20}[0.450,0.469]	&[1.35,1.38]	&[1.01,1.04]	&[3.27,3.50]\\ \hline
$\mathcal{N}_{21}$	&[0.289,0.325]	&[3.17,3.40]	&[1.30,1.45]	&[0.2912,0.2937]	&[0.531,0.550]	&[1.62,1.65]	&[0.961,0.989]	&[3.27,3.50]\\ \hline\hline
\end{tabular}
\caption{\label{tab:viable_regions_NO}The predicted ranges of the input parameters and lepton observables for all the 21 viable breaking patterns for the NO case. Note that for all these models, the reactor angle $\sin^2\theta_{13}$ is predicted to lie in $[0.02064,0.02418]$, the light neutrino mass $m_2$ in $[8.51\text{ meV}, 8.84\text{ meV}]$, and $m_3$ in $[49.5\text{ meV}, 50.8\text{ meV}]$. These ranges are not tabulated here for brevity. The highlighted regions in red color indicate predictions that overlap with the expected $3\sigma$ ranges of future experiments.}
\end{center}
\end{table}

\begin{table}[t!]
\begin{center}
\renewcommand{\tabcolsep}{1.5mm}
\renewcommand{\arraystretch}{1.15}
\begin{tabular}{|c||c|c|c||c|c|c|} \hline\hline

IO	& $\eta $/$\pi $ & $\frac{m_{a}}{\text{meV}}$ & $r$ & $\delta_{CP}/\pi$ & $\beta/\pi$ & $\frac{m_{ee}}{\text{meV}}$ \\ \hline 
			
$\mathcal{I}_{1}$	&[1.864,1.865]	&[4.10,4.20]	&[0.987,0.989]	&\cellcolor{red!20}[1.564,1.567]	&[1.925,1.928]	&[14.8,17.2]\\ \hline
$\mathcal{I}_{2}$	&[0.1350,0.1356]	&[4.10,4.20]	&[0.987,0.989]	&\cellcolor{red!20}[1.433,1.436]	&[0.07178,0.07535]	&[14.8,17.2]\\ \hline
$\mathcal{I}_{3}$	&[0.1381,0.1391]	&[4.09,4.20]	&[0.990,0.992]	&\cellcolor{red!20}[1.595,1.600]	&[0.07430,0.07822]	&[14.8,17.2]\\ \hline
$\mathcal{I}_{4}$	&[0.6151,0.6171]	&[3.79,3.89]	&[0.990,0.994]	&[1.186,1.275]	&[1.171,1.239]	&[12.6,14.9]\\ \hline
$\mathcal{I}_{5}$	&[1.383,1.385]	&[3.79,3.89]	&[0.990,0.994]	&[1.725,1.814]	&[0.7613,0.8286]	&[12.6,14.9]\\ \hline
$\mathcal{I}_{6}$	&[1.140,1.141]	&[2.59,2.66]	&[3.04,3.05]	&\cellcolor{red!20}[1.357,1.366]	&[1.807,1.814]	&[14.1,16.4]\\ \hline
$\mathcal{I}_{7}$	&[1.683,1.684]	&[2.57,2.64]	&[3.09,3.09]	&\cellcolor{red!20}[1.567,1.576]	&[1.812,1.819]	&[14.3,16.6]\\ \hline
$\mathcal{I}_{8}$	&[0.3589,0.3598]	&[2.59,2.66]	&[3.05,3.05]	&\cellcolor{red!20}[1.634,1.643]	&[0.1860,0.1925]	&[13.4,15.8]\\ \hline
$\mathcal{I}_{9}$	&[1.740,1.741]	&[3.85,3.96]	&[3.06,3.07]	&[1.145,1.174]	&[0.4772,0.5000]	&[14.1,16.5]\\ \hline
$\mathcal{I}_{10}$	&[1.093,1.093]	&[3.85,3.96]	&[3.06,3.07]	&[1.826,1.855]	&[1.500,1.523]	&[14.8,17.2]\\ \hline
$\mathcal{I}_{11}$	&[1.150,1.151]	&[2.57,2.64]	&[3.09,3.09]	&\cellcolor{red!20}[1.424,1.433]	&[0.1806,0.1877]	&[13.2,15.6]\\ \hline
$\mathcal{I}_{12}$	&[0.2130,0.2144]	&[2.32,2.38]	&[3.07,3.08]	&[1.128,1.153]	&[1.009,1.030]	&[16.6,17.1]\\ \hline
$\mathcal{I}_{13}$	&[0.3082,0.3104]	&[5.29,5.44]	&[0.414,0.415]	&[1.814,1.834]	&[1.619,1.633]	&[14.5,16.8]\\ \hline
$\mathcal{I}_{14}$	&[0.1896,0.1918]	&[5.29,5.44]	&[0.414,0.415]	&[1.166,1.186]	&[0.3666,0.3812]	&[14.5,16.9]\\ \hline\hline
\end{tabular}
\caption{\label{tab:viable_regions_IO}The predicted ranges of the input parameters and lepton observables for all the 14 viable breaking patterns for the IO case. Note that for all these models, $\sin^2\theta_{12}$ is predicted to lie in $[0.2893, 0.3295]$, the light neutrino masses $m_1$ and $m_2$ lie in $[48.4\text{ meV}, 49.7\text{ meV}]$ and $[49.2\text{ meV}, 50.5\text{ meV}]$, respectively. Moreover, as follows from Eq.~\eqref{eq:sum_rules}, within each pattern $\sin^2\theta_{13}$ and $\sin^2\theta_{23}$ are fixed exactly to single values, given by the corresponding entries in table~\ref{tab:bf_value}. 
}
\end{center}
\end{table}

To further explore the phenomenological implications of the 35 viable breaking patterns, we performed a comprehensive scan of their parameter spaces, imposing that all physical observables lie within the experimentally allowed $3\sigma$ ranges listed in table~\ref{tab:bf_13sigma_data}. Under these constraints, the three leptonic mixing angles, the two CP-violating phases, the neutrino masses, and the effective Majorana mass relevant to $0\nu\beta\beta$ decay are all confined to narrow regions. 
The resulting predictions are summarized in table~\ref{tab:viable_regions_NO} (NO) and table~\ref{tab:viable_regions_IO} (IO), demonstrating the strong predictive power of the Modular Littlest Seesaw framework.

\begin{figure}[t!]
\centering
\begin{tabular}{c}
\includegraphics[width=0.95\linewidth]{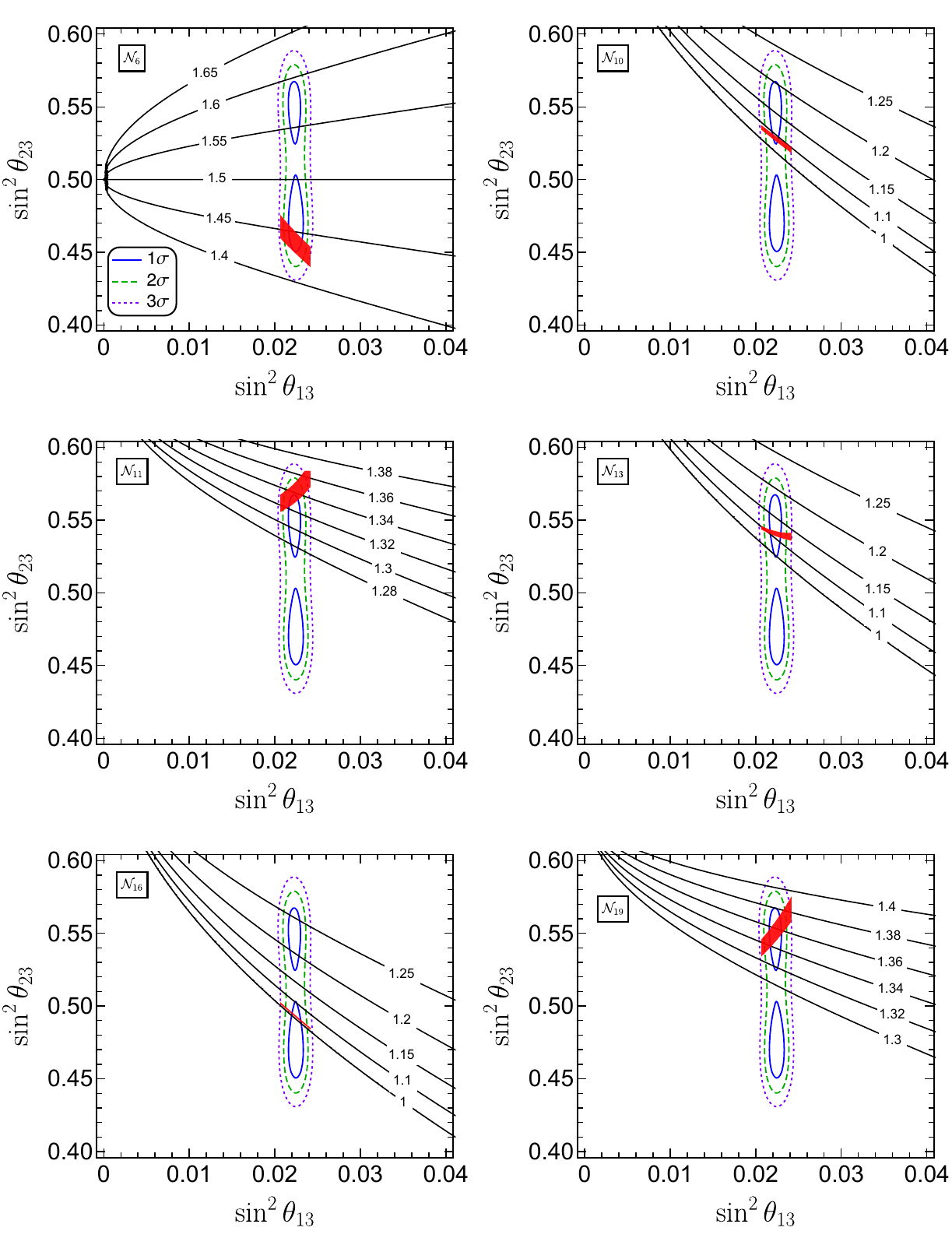}
\end{tabular}
\caption{\label{fig:sum_rules}
Contours of $\delta_{CP}/\pi$ in the $\sin^2\theta_{13}$--$\sin^2\theta_{23}$ plane for the representative TM$_1$ mixing pattern and five non-TM$_1$ viable patterns with normal neutrino mass ordering and $\chi^2_{\rm min}<5$. The red regions show the model predictions obtained from a scan over the input parameter space, subject to the requirement that all lepton mixing parameters and neutrino mass squared differences remain within their experimentally allowed $3\sigma$ ranges.}
\end{figure}

Using the general sum rules in Eq.~\eqref{eq:sum_rules} together with the fixed PMNS columns listed in table~\ref{tab:viable_BPs}, one can derive the  correlations among the lepton mixing angles and the Dirac CP phase for all 35 phenomenologically viable symmetry breaking patterns. These sum rules provide characteristic predictions for each pattern and establish strong correlations among the mixing parameters, thereby offering powerful tests of the models in future precision neutrino oscillation experiments such as JUNO, DUNE and T2HK. To illustrate these predictions, figure~\ref{fig:sum_rules} shows the contours of $\delta_{CP}/\pi$ in the $\sin^2\theta_{13}$--$\sin^2\theta_{23}$ plane for the representative TM$_1$ mixing pattern $\mathcal{N}_{6}$ together with five non-TM$_1$ patterns ($\mathcal{N}_{10}$, $\mathcal{N}_{11}$, $\mathcal{N}_{13}$, $\mathcal{N}_{16}$ and $\mathcal{N}_{19}$) in the NO spectrum satisfying $\chi^2_{\rm min}<5$\footnote{Note that the non-TM$_1$ patterns $\mathcal{N}_{12}$ and $\mathcal{N}_{17}$ also satisfy $\chi^{2}_{\text{min}}<5$. We do not plot their sum rules here, as they yield the same fixed columns as $\mathcal{N}_{13}$ and $\mathcal{N}_{16}$, respectively. }. The general analytic expressions are rather lengthy; for compactness, the explicit sum rules for the representative patterns displayed in figure~\ref{fig:sum_rules} are collected in appendix~\ref{sec:sum_rules}. The red regions denote the model predictions obtained by scanning the three independent input parameters $m_a$, $r$ and $\eta$ over their allowed ranges while requiring all neutrino mass squared differences and mixing observables to lie within their experimentally allowed $3\sigma$ intervals. The resulting contours clearly demonstrate that each viable symmetry breaking pattern occupies a distinct region in parameter space, highlighting the strong predictive power of the framework and its excellent prospects for discrimination by future high-precision measurements.

\begin{figure}[t!]
\centering
\begin{tabular}{c}
\includegraphics[width=0.92\linewidth]{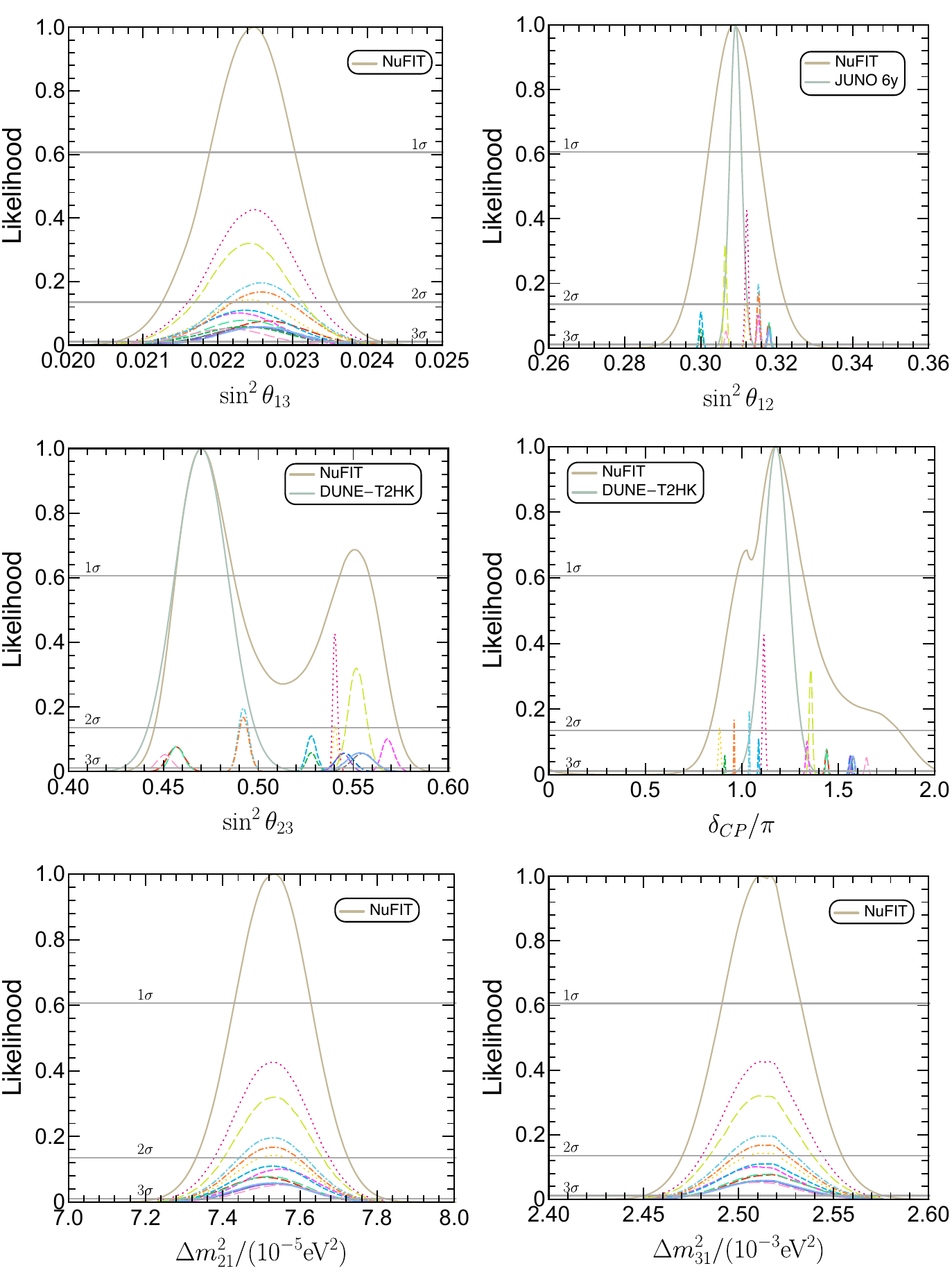}\\
\includegraphics[width=0.7\linewidth]{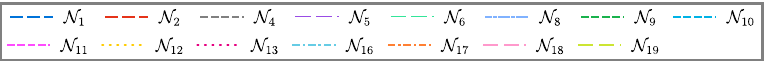}
\end{tabular}
\caption{\label{fig:likelihood_NO}Profile likelihoods for the six oscillation observables in the 15 NO lepton mixing patterns with $\chi^2_{\text{min}}\le 6$. The tan dotted curves denote the likelihoods obtained from \texttt{NuFIT}~\cite{Esteban:2024eli}. In the $\sin^{2}\theta_{12}$ panel, the green solid line shows the expected likelihood after six years of JUNO operation~\cite{JUNO:2022mxj,JUNO:2025gmd}. The green solid line in the $\sin^{2}\theta_{23}$ and $\delta_{CP}$ panels correspond to the anticipated sensitivities of DUNE~\cite{DUNE:2020ypp} and T2HK~\cite{Hyper-Kamiokande:2018ofw}. 
}
\end{figure}

\begin{figure}[t!]
\centering
\begin{tabular}{c}
\includegraphics[width=0.95\linewidth]{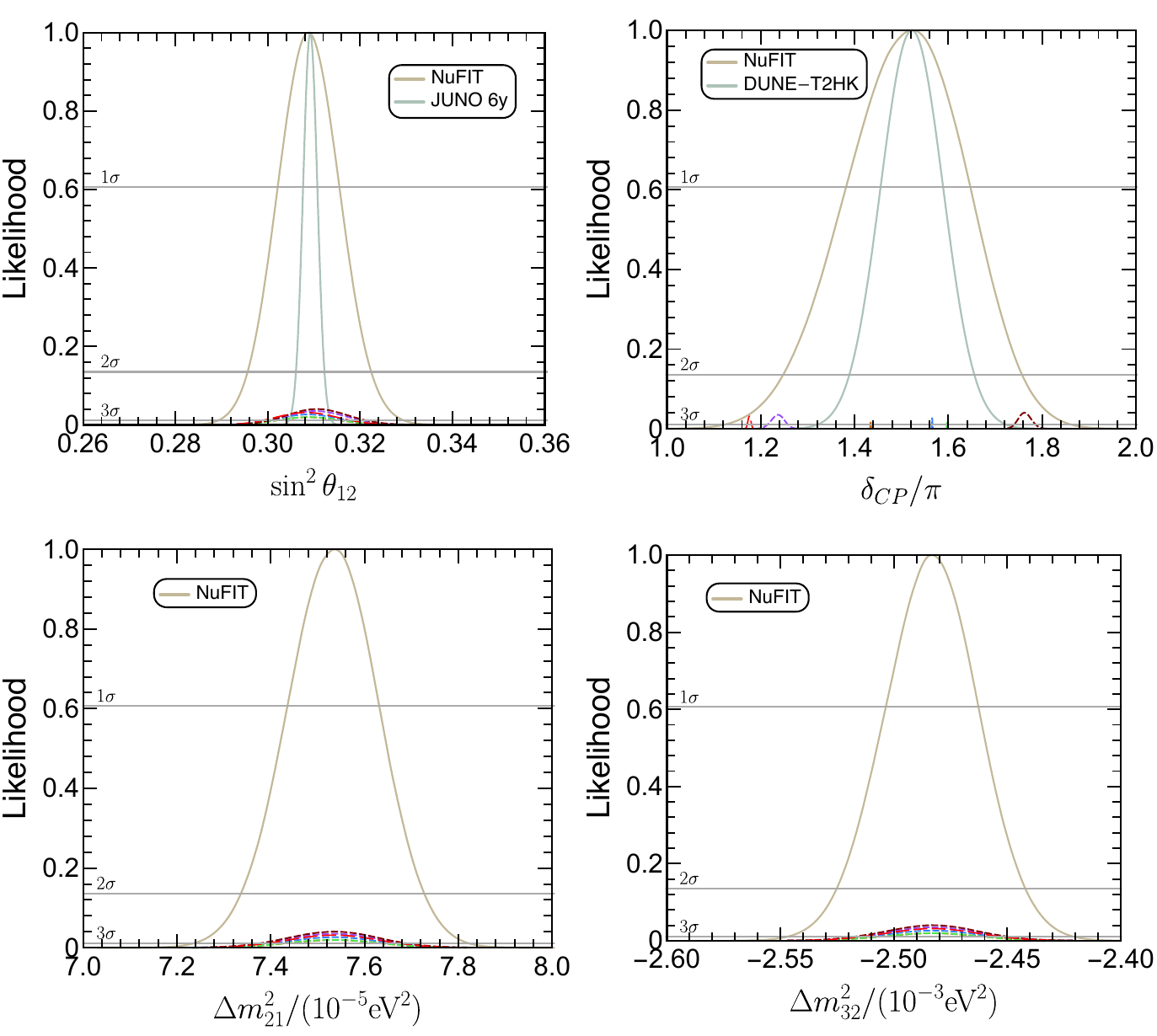}\\
\includegraphics[width=0.7\linewidth]{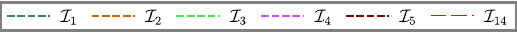}
\end{tabular}
\caption{\label{fig:likelihood_IO}Profile likelihoods for the six oscillation observables in the 6 IO lepton mixing patterns with $\chi^2_{\text{min}}\le 9$. The remaining conventions follow those used in figure~\ref{fig:likelihood_NO}.}
\end{figure}

To provide a more comprehensive view of the model predictions beyond the best fit points of viable breaking patterns, we construct the one-dimensional likelihood profile for an observable $o_i$,
\begin{equation}
	\label{eq:likelihood}
	L(o_{i})=\exp\!\left[-\frac{\chi^2(o_i)}{2}\right]\,,
\end{equation}
where the profile $\chi^{2}$ function is defined as
\begin{equation}\label{eq:chi2alpha}
\chi^2(o_{i}) =
\min\left[\chi^2(\vec{o})\Big|_{o_{i}}\right].
\end{equation}
That is, $\chi^2(o_i)$ is obtained by minimizing the global $\chi^{2}$ with respect to all model parameters while keeping the observable $o_i$ fixed. For the 15 NO patterns with $\chi^2_{\text{min}}\le 6$ and the 6 IO patterns with $\chi^2_{\text{min}}\le 9$, we use Eq.~\eqref{eq:likelihood} to construct one‑dimensional profile likelihoods for  $\sin^2\theta_{13}$, $\sin^2\theta_{12}$, $\sin^2\theta_{23}$, $\delta_{CP}$, $\Delta m^{2}_{21}/(10^{-5}\,\text{eV}^{2})$ and $\Delta m^{2}_{3\ell}/(10^{-3}\,\text{eV}^{2})$. The resulting likelihood curves are shown in figure~\ref{fig:likelihood_NO} for NO and figure~\ref{fig:likelihood_IO} for IO. 
By displaying the full correlated ranges predicted by each model, these figures extend the best-fit study and allow a quantitative assessment of both the predictive power of each pattern and its prospects for future experimental tests.

For the NO case, the predicted $\sin^2\theta_{12}$ ranges of 4 patterns overlap with the future sensitivity region defined in Eq.~\eqref{eq:s12sq_JUNO_6years}, namely $\mathcal{N}_{12}$, $\mathcal{N}_{13}$, $\mathcal{N}_{18}$ and $\mathcal{N}_{19}$, and therefore remain compatible with upcoming JUNO measurements. The remaining patterns predict $\sin^2\theta_{12}$ values that lie outside the future $3\sigma$ sensitivity reach of JUNO. In particular, all breaking patterns that yield the TM$_1$ mixing matrix may be ruled out by JUNO data. Nine of the viable patterns exhibit predicted ranges for $\sin^2\theta_{23}$ that intersect with the region in Eq.~\eqref{eq:theta23_DUNE_and_T2HK} which is the resolution after 15 years of
DUNE and T2HK running. These patterns are labelled as $\mathcal{N}_{2}$, $\mathcal{N}_{3}$, $\mathcal{N}_{6}$, $\mathcal{N}_{7}$, $\mathcal{N}_{15}$, $\mathcal{N}_{16}$, $\mathcal{N}_{17}$, $\mathcal{N}_{18}$ and $\mathcal{N}_{20}$. The remaining patterns yield predictions that fall outside the region of upcoming long‑baseline experiments. This renders them highly testable and, under the assumption that current best fit values persist, likely to be excluded by future data. Six patterns exhibit predicted ranges for $\delta_{CP}$ that overlap with the future $3\sigma$ region in Eq.~\eqref{eq:deltaCP_DUNE_and_T2HK}, namely $\mathcal{N}_{10}$, $\mathcal{N}_{11}$, $\mathcal{N}_{13}$, $\mathcal{N}_{16}$, $\mathcal{N}_{19}$ and $\mathcal{N}_{20}$, and therefore remain compatible with upcoming long‑baseline measurements. By contrast, the remaining patterns predict $\delta_{CP}$ values that fall outside the future $3\sigma$ sensitivity range of DUNE and T2HK. This renders them potentially testable and, if the current best fit values persist, disfavoured by forthcoming long‑baseline data. 

In contrast to the NO case, all 14 viable IO patterns predict values of $\sin^2\theta_{12}$ that overlap with the future $3\sigma$ sensitivity region of JUNO. Consequently, the forthcoming high precision JUNO measurement is unlikely to exclude any of these patterns.  For the atmospheric mixing angle, only one pattern, $\mathcal{I}_{14}$, predicts a value of $\sin^2\theta_{23}$ within the projected $3\sigma$ region of DUNE and T2HK given in Eq.~\eqref{eq:theta23_DUNE_and_T2HK}. The predictions of all remaining IO patterns lie outside this interval and can therefore be stringently tested by future long-baseline oscillation experiments. If the true value of $\sin^2\theta_{23}$ remains close to the current best fit point, these patterns would be strongly disfavoured. Regarding the Dirac CP phase, seven patterns, namely $\mathcal{I}_{1}$, $\mathcal{I}_{2}$, $\mathcal{I}_{3}$, $\mathcal{I}_{6}$, $\mathcal{I}_{7}$, $\mathcal{I}_{8}$ and $\mathcal{I}_{11}$ predict ranges of $\delta_{CP}$ that overlap with the projected $3\sigma$ sensitivity region of DUNE and T2HK in Eq.~\eqref{eq:deltaCP_DUNE_and_T2HK}. These patterns therefore remain compatible with the expected reach of future long-baseline experiments. The remaining patterns predict $\delta_{CP}$ outside this region and could be decisively tested by forthcoming measurements. Should future data confirm the current best-fit value, they would be strongly disfavoured.

It should be emphasized, however, that these conclusions rely on the assumption that the future central values of $\sin^2\theta_{23}$ and $\delta_{CP}$ remain close to their present best-fit values. Accordingly, the above assessment should be regarded as indicative rather than definitive. Given the current uncertainties in the determination of $\theta_{23}$ and $\delta_{CP}$, their best fit values may evolve with future data. Therefore, a more robust approach is to consider the combined constraining power of the next‑generation experiments. When the future sensitivities of JUNO, DUNE and T2HK are taken together, all viable patterns identified in our scan (for both NO and IO) will be subject to decisive and complementary tests. In this sense, irrespective of possible shifts in the central values of $\theta_{23}$ and $\delta_{CP}$, the suite of upcoming measurements ensures that essentially all currently viable constructions can be conclusively probed in the coming decades.

Furthermore, for the NO spectrum, the predicted value of $m_{ee}$ from all viable symmetry breaking patterns lies in the range $[2\,\text{meV},4\,\text{meV}]$, well below the sensitivity of current and future $0\nu\beta\beta$-decay experiments. This range is substantially lower than the latest KamLAND-Zen limit $m_{ee} < (28\text{--}122)\,\text{meV}$~\cite{KamLAND-Zen:2024eml} and also falls below the projected reach of next-generation experiments such as LEGEND-1000 ($m_{ee} < (9\text{--}21)\,\text{meV}$)~\cite{LEGEND:2021bnm} and nEXO ($m_{ee} < (4.7\text{--}20.3)\,\text{meV}$)~\cite{nEXO:2021ujk}. In contrast, for the IO spectrum, all viable patterns yield $m_{ee}$ values that lie within the expected sensitivity reach of forthcoming $0\nu\beta\beta$-decay searches, making them experimentally testable in the near future.

\FloatBarrier

\section{\label{sec:conclusion}Conclusion}

Modular symmetry provides an elegant and highly predictive framework for addressing the flavor puzzle. The Modular Littlest Seesaw implements this idea within the minimal seesaw model with two right-handed neutrinos, assuming that the three sectors (charged lepton, atmospheric neutrino and solar neutrino) are associated with distinct modular fixed points. 

Although the $\Delta(96)$ group has been extensively studied as a traditional flavor group~\cite{deAdelhartToorop:2011re,Ding:2012xx,King:2012in,Ding:2014ssa,Bernigaud:2020wvn,Gautam:2020bnx,Alvarado:2022lzx,Yan:2025itm}, its role as a finite modular group has received little attention.  It is particularly well-motivated in this role, because its complex triplet representations listed in table~\ref{tab:Delta96_Reps} offer new model-building possibilities that are absent in the more extensively studied modular $S_3$, $A_4$, $S_4$ and $A_5$ frameworks.

This study provides a comprehensive, model independent investigation of lepton mixing 
emerging from the finite modular group $\Delta(96)$ within the Modular Littlest Seesaw framework. The three left-handed lepton doublets are assigned to a triplet representation of $\Delta(96)$, while the two right-handed neutrinos, $N^{c}_{\text{atm}}$ and $N^{c}_{\text{sol}}$, transform as modular singlets. The modulus associated with the charged lepton sector is assumed to be fixed at a symmetry preserving point $\langle\tau_{\ell}\rangle$, thereby breaking the modular symmetry to the corresponding stabilizer subgroup. The modular forms $Y_{\text{atm}}$ and $Y_{\text{sol}}$ associated with the operators $LN^{c}_{\text{atm}}$ and $LN^{c}_{\text{sol}}$ are required to be triplet VVMFs of $\Delta(96)$, evaluated at the fixed points $\langle\tau_{\rm atm}\rangle$ and $\langle\tau_{\rm sol}\rangle$, respectively. For each assignment of $L$ and residual symmetry in the charged lepton sector, the corresponding charged lepton diagonalization matrices are given in table~\ref{tab:Ul}. We constructed the lowest- and next-to-lowest-weight triplet VVMFs of $\Delta(96)$ and determined the associated vacuum alignments at all independent fixed points in section~\ref{secVVMFs_Delta96}. Using these alignments as neutrino Yukawa couplings, the resulting light neutrino mass matrix is controlled by only three real parameters $|m_a|$, $r$ and $\eta$. Consequently, the framework predicts a massless lightest neutrino together with strong correlations among neutrino masses, mixing angles, and CP phases.

A central result of this work is summarized in table~\ref{tab:viable_BPs}. Through an exhaustive scan over the allowed assignments of $L$, $Y_{\rm atm}$ and $Y_{\rm sol}$, we found 35 phenomenologically viable and inequivalent symmetry breaking patterns, including 21 NO and 14 IO cases. 
Table~\ref{tab:viable_BPs} displays the residual symmetries, fixed PMNS columns,  and the corresponding 
$Y_{\rm atm}$ and $Y_{\rm sol}$ alignments for all viable patterns. 
It also makes transparent the key structural feature of the construction: once the fixed points and VVMFs are specified, the relevant VVMF directions are fixed by modular covariance up to an overall normalization. 
Consequently, the direction of the null vector $\hat v_{\rm fix}$, and hence one column of the PMNS matrix, is fixed algebraically. 
This algebraic rigidity is stronger than the constraint imposed by residual 
symmetry alone, especially when the residual stabilizer admits degenerate 
eigenspaces. It also sharply distinguishes the present modular construction from traditional residual-symmetry or tri-direct CP approaches, where residual constraints may leave additional alignment freedom. 
In the modular framework, the viable alignments form a finite set of exact 
algebraic directions, making the resulting flavor structures highly constrained.

A subsequent numerical scan of the parameter space gives the best-fit values of the input parameters and observables listed in table~\ref{tab:bf_value}, while the allowed ranges of the model parameters and observables are given in 
tables~\ref{tab:viable_regions_NO} and~\ref{tab:viable_regions_IO}. 
The profile likelihood analysis for the 15 NO patterns with 
$\chi^2_{\min}\leq 6$ and the 6 IO patterns with $\chi^2_{\min}\leq 9$, shown in figures~\ref{fig:likelihood_NO} and~\ref{fig:likelihood_IO}, reveals that the viable patterns predict highly restricted ranges for neutrino masses, mixing angles and CP-violating phases, highlighting the strong predictive power of the framework. In addition to the patterns reproducing the familiar TM$_1$ structure associated with conventional CSD$(n)$, new breaking patterns beyond the CSD$(n)$ paradigm also emerge, leading to novel sum rules among the lepton mixing angles and $\delta_{CP}$.

The testability of these constructions with forthcoming experimental facilities is a particularly compelling aspect of our findings. Future measurements of $\sin^2\theta_{12}$ at JUNO, and of $\sin^2\theta_{23}$ and $\delta_{CP}$ at DUNE and T2HK are expected to exclude a large fraction of the currently allowed viable breaking patterns, with all TM$_1$ patterns in the NO case potentially being ruled out by JUNO. Although the precise impact depends on the true values of $\theta_{23}$ and $\delta_{CP}$, the combined sensitivities of next-generation neutrino experiments will provide decisive and complementary probes of essentially all viable constructions. Furthermore, while the predicted effective Majorana mass $m_{ee}$ for NO lies below the reach of foreseeable $0\nu\beta\beta$-decay experiments, all viable IO patterns predict $m_{ee}$ values within the sensitivity of upcoming searches, offering an additional powerful test of the framework.

Finally, this bottom-up construction also has suggestive implications for 
top-down model building. In string compactifications, modular symmetries often 
arise as target-space dualities, and modular fixed points correspond to 
self-dual loci in moduli space, where residual symmetries may be enhanced. 
It is therefore natural to consider scenarios in which several moduli are 
stabilized at such special points. Moreover, low-order effective couplings in the four-dimensional theory may be organized in terms of low-weight VVMFs, in close analogy with the lowest- and next-to-lowest-weight 
VVMFs used in this work. In a factorized or multi-sector compactification, 
different effective operators can depend on different moduli, providing a 
possible top-down rationale for the multi-modulus structure assumed here. 
A complete realization of the finite modular group $\Delta(96)$ in string theory, 
including moduli stabilization and the control of cross-sector couplings, remains 
an important open direction for future work.

\section*{Acknowledgements}
CCL, HZH and LNY are supported by the National Natural Science Foundation of China under Grant Nos. 12547106, 12247103. XGL was supported by Universidad Nacional Autónoma de México Postdoctoral Program (POSDOC).

\section*{Appendix}
	
\begin{appendix}
		
\section{\label{sec:Delta96_group_theory}Group theory of the finite modular group $\Delta(96)$}

The finite modular group $\Delta(96)$ can be generated by the modular generators $S$ and $T$ which satisfy the following multiplication relations:
\begin{equation}
	S^2=(ST)^3=(ST^{-1} ST)^3=T^8=1,
\end{equation}	
The $\Delta(96)$ group can also be presented in terms of the four generators $a$, $b$, $c$ and $d$, which obey~\cite{Ding:2012xx,Ding:2014ssa}

\begin{eqnarray}
\nonumber && a^{3} = b^{2} = (ab)^{2} = c^{4} = d^{4} = 1,\quad cd = dc \\
\nonumber && aca^{-1}=c^{-1}d^{-1},\quad ada^{-1}=c,  \quad  bcb^{-1}=d^{-1},\quad bdb^{-1}=c^{-1} \,.
\end{eqnarray}
The two sets of generators are linked by
\begin{equation}
S=ba, \qquad \qquad T=bd\,,
\end{equation}
which is readily verified to satisfy the defining relations in both pictures. Inverting these gives
\begin{equation}
a=T^{3}ST^{6}, \qquad b=ST^{2}ST^{5}, \qquad c=ST^{2}ST^{4}, \qquad d=ST^{2}ST^{6}\,.
\end{equation}
When expressed in the $S$ and $T$ basis,  the 96 elements of $\Delta(96)$ belong to the following ten conjugacy classes:
\begin{eqnarray}
	\nonumber\hskip-0.15in1C_1&:& \{1\},\\
	\nonumber\hskip-0.15in3C_2&:& \{ (ST^4)^2, ST^4S, T^4 \},\\
	\nonumber\hskip-0.15in12C_2&:& \{ S, T^2ST^6, T^6ST^2, T^4ST^4, ST^6ST^3, ST^2ST^5, T^3ST^5,T^5ST^3, TST^7,\\
	\nonumber&&\quad ST^2ST, T^7ST, ST^6ST^7  \},\\
	\nonumber\hskip-0.15in 32C_3&:& \{  T^6ST^3, ST, T^4ST^5, T^2ST^7, T^2ST^5, T^4ST^3, ST^7, T^6ST, T^5ST^2, TS, TST^4, \\
	\nonumber&&~ T^5ST^6, T^7S, T^3ST^6, T^3ST^2, T^7ST^4, T^3ST^4, T^7ST^2, T^7ST^6, T^3S, T^2ST, T^4ST^7,\\
	\nonumber&&~  ST^3, T^6ST^5, TST^6, T^5ST^4, T^5S, TST^2, T^6ST^7, 	ST^5, T^4ST, T^2ST^3 \},\\
	\nonumber\hskip-0.15in3C_4^{(1)}&:& \{ T^2, ST^2S, (ST^6)^2 \},\\
    \nonumber\hskip-0.15in3C_4^{(2)}&:& \{  T^6, ST^6S, (ST^2)^2 \},\\
    \nonumber\hskip-0.15in6C_4&:& \{ ST^6ST^2, ST^2ST^6, ST^6ST^4, ST^4ST^6, ST^2ST^4,
    ST^4ST^2 
    \},\\
    \nonumber\hskip-0.15in12C_4&:& \{ ST^2ST^7, ST^6ST, T^3ST, T^5ST^7, TST^3, ST^6ST^5, T^7ST^5, ST^2ST^3, ST^4,\\
	\nonumber&&~T^2ST^2, T^6ST^6, T^4S  \},\\
	\nonumber\hskip-0.15in12C_8^{(1)}&:& \{ T, ST^4ST^3, T^7ST^7, TST^5, T^4ST^2, T^6S, T^2ST^4, ST^6, T^5ST, ST^4ST^7, T^3ST^3, T^5 \},\\
	\nonumber\hskip-0.15in12C^{(2)}_{8}&:& \{ ST^4ST^5, T^7, T^7ST^3, ST^2ST^7, T^6ST^4,T^2S, ST^2, \\
	\label{eq:Delta_96_CC}&&~TST, T^5ST^5, T^3, T^3ST^7, ST^4ST \}\,,
\end{eqnarray}	
where $kC_{n}$ denotes a conjugacy class which contains $k$ elements with order $n$. The group $\Delta(96)$ contains a total of fifty-six nontrivial Abelian subgroups: fifteen $Z_{2}$ subgroups, sixteen $Z_{3}$ subgroups, seven $K_{4}$ subgroups, twelve $Z_{4}$ subgroups and six $Z_{8}$ subgroups. In terms of the generators $S$ and $T$,  the explicit forms of these Abelian subgroups are given as follows:
	
\begin{itemize}[leftmargin=1.5em]
	\item{$Z_{2}$ subgroups}
	\begin{equation}
		\begin{aligned}
			&Z_{2}^{(ST^4)^2}=\left\{1,(ST^4)^2\right\},\quad &&Z_{2}^{ST^4S}=\left\{1,ST^4S\right\},\quad
		    &&Z_{2}^{T^4}=\left\{1,T^4\right\}, \\
		    &Z_{2}^{S}=\left\{1,S\right\}, \quad
		    &&Z_{2}^{T^2ST^6}=\left\{1,T^2ST^6\right\},\quad
			&&Z_{2}^{T^6ST^2}=\left\{1,T^6ST^2\right\},\\
			&Z_{2}^{T^4ST^4}=\left\{1,T^4ST^4\right\}, \quad
			&&Z_{2}^{ ST^6ST^3 }=\left\{1, ST^6ST^3 \right\},\quad
			&&Z_{2}^{ST^2ST^5}=\left\{1,ST^2ST^5\right\},\\
			&Z_{2}^{T^3ST^5 }=\left\{1,T^3ST^5\right\},\quad
			&&Z_{2}^{T^5ST^3}=\left\{1,T^5ST^3\right\},
			&&Z_{2}^{TST^7}=\left\{1, TST^7\right\},\\
		    &Z_{2}^{ST^2ST}=\left\{1,ST^2ST\right\}, \quad
			&&Z_{2}^{T^7ST }=\left\{1,T^7ST \right\},\quad
			&&Z_{2}^{ST^6ST^7}=\left\{1,ST^6ST^7\right\}
			\,.
		\end{aligned}
	\end{equation}
	The generators of the first three $Z_{2}$ subgroups form a single conjugacy class, and the generators of other twelve $Z_{2}$ subgroups constitute another conjugacy class.
	
	\item{$Z_{3}$ subgroups}
	\begin{equation}
		\begin{aligned}
			&Z_{3}^{T^6ST^3 } =\{1,T^6ST^3, T^5ST^2\}, \qquad\quad
			&& Z_{3}^{ST}=\{1, ST, T^7S \},\\ 
			& Z_{3}^{T^4ST^5}=\{1,T^4ST^5, T^3ST^4\}, 
			&&Z_{3}^{T^2ST^7}= \{1,T^2ST^7, TST^6\}, \\
			& Z_{3}^{T^3ST^6}=\{1, T^3ST^6,T^2ST^5\}, 
			&& {Z}_{3}^{T^5ST^4}=\{1, T^5ST^4,T^4ST^3\}\,, \\
			&Z_{3}^{TS} =\{1, TS,ST^7\}, 
			&& Z_{3}^{ T^6ST}=\{1,T^6ST, T^7ST^2\}, \\
			& Z_{3}^{ TST^4}=\{1, TST^4, T^4ST^7 \}, 
			&&Z_{3}^{T^2ST^3} =\{1, T^2ST^3,T^5ST^6\}, \\
			& Z_{3}^{T^3ST^2}=\{1,T^3ST^2, T^6ST^5 \}, 
			&& Z_{3}^{T^4ST}=\{1, T^4ST, T^7ST^4\}, \\
			&Z_{3}^{T^7ST^6}=\{1, T^7ST^6, T^2ST\},
			&& Z_{3}^{ST^5}=\{1, ST^5,T^3S\}, \\
			& Z_{3}^{T^5S}=\{1, T^5S,ST^3\}, 
			&&Z_{3}^{T^6ST^7} =\{1, T^6ST^7,TST^2\}\,.
		\end{aligned}
	\end{equation}
	All the sixteen $Z_{3}$ subgroups are conjugate to each other.

	\item{$Z_{4}$ subgroups}
	\begin{equation}
		\begin{aligned}
			&Z_{4}^{ST^6ST^2}=\{1, ST^2ST^6, (ST^4)^2,ST^6ST^2\}, 
			&&Z_{4}^{ST^2ST^4}=\{1, ST^2ST^4, ST^4S,ST^6ST^4\},\\  
			&Z_{4}^{ST^4ST^2}=\{1, ST^4ST^2, T^4,ST^4ST^6\},
			&&Z_{4}^{(ST^2)^2}=\{1, (ST^2)^2,(ST^4)^2, (ST^6)^2\},\\ 
			&Z_{4}^{T^2}=\{1,T^2, T^4, T^6\},  
			&&Z_{4}^{ST^2S}=\{1,ST^2S, ST^4S, ST^6S \},\\
			&Z_{4}^{ST^4 }=\{1, ST^4,(ST^4)^2, T^4S\}, 
			&&Z_{4}^{T^2ST^2}=\{1, T^2ST^2,(ST^4)^2, T^6ST^6\},\\
			&Z_{4}^{ST^2ST^3}=\{1, ST^2ST^3, T^4 ,ST^2ST^7\},
			&&Z_{4}^{ST^6ST }=\{1,ST^6ST, T^4, ST^6ST^5 \},\\ 
			&Z_{4}^{T^3ST}=\{1,T^3ST, ST^4S, T^7ST^5 \}, 
			&&Z_{4}^{TST^3}=\{1, TST^3, ST^4S, T^5ST^7\}\,.
		\end{aligned}
	\end{equation}
	The twelve $Z_{4}$ subgroups can be classified into three categories under similarity (conjugation) transformations within  $\Delta(96)$:  the first three $Z_{4}$, the fourth to sixth $Z_{4}$, and the remaining six subgroups.

	\item{$K_{4}$ subgroups}
	\begin{equation}
		\begin{aligned}
			&K_{4}^{(ST^4S, T^{4}) }=\{1,(ST^4)^2, ST^4S, T^4\}, \\
			&K_{4}^{(S,T^4ST^4)}=\{1,S, T^4ST^4, (ST^4)^2\},
			&&K_{4}^{(T^2ST^6,T^6ST^2)}=\{1,T^2ST^6, T^6ST^2, (ST^4)^2\}, \\
			&K_{4}^{(ST^6ST^3, ST^6ST^7)}=\{1,ST^6ST^3, ST^6ST^7, T^4\},
			&& K_{4}^{(ST^2ST^5, ST^2ST)}=\{1, ST^2ST^5, ST^2ST, T^4\},\\
			& K_{4}^{(T^3ST^5, T^7ST)}=\{1,T^3ST^5, T^7ST, ST^4S\},
		&&K_{4}^{(T^5ST^3, TST^7)}=\{1,T^5ST^3, TST^7, ST^4S\} \,.\end{aligned}
		\end{equation}
The subgroup $K ^{(ST^4S, T^{4} )} _{4}$ is a normal subgroup of $\Delta(96)$, whereas all other $K_{4}$ subgroups are related by conjugation.

	\item{$Z_{8}$ subgroups}
	\begin{equation}
		\begin{aligned}
			&Z_{8}^{ST^2 } =\{1, ST^2,(ST^2)^2,T^2ST^4,(ST^4)^2, ST^2ST^7, (ST^6)^2, T^6S\}, \\
			&Z_{8}^{ ST^6 } =\{1,ST^6,(ST^6)^2, T^6ST^4,(ST^4)^2, T^4ST^2, (ST^2)^2,T^2S \}, \\
			&Z_{8}^{T} =\{1,T, T^2, T^3, T^4, T^5, T^6, T^7 \}, \\
			&Z_{8}^{ST^4ST^3} =\{1, ST^4ST^3, T^2,ST^4ST^5, T^4, ST^4ST^7, T^6, ST^4ST\}, \\
			&Z_{8}^{TST} =\{1, TST, ST^6S, T^3ST^3, ST^4S, T^5ST^5, ST^2S,T^7ST^7\}, \\
			&Z_{8}^{TST^5} =\left\{1,TST^5,ST^2S,T^7ST^3,ST^4S,T^5ST, ST^6S,T^3ST^7\right\}\,.
		\end{aligned}
	\end{equation}
All the six $Z_{8}$ subgroups are conjugate to each other.
\end{itemize}

\begin{table}[t!]
\begin{center}
\renewcommand{\tabcolsep}{1.2mm}
\begin{tabular}{|c|c|c|}\hline\hline
~~  &  $S$  &   $T$  \\ \hline
&   &   \\ [-0.16in]
$\bm{1_{m}}$ & $(-1)^{m}$   &  $(-1)^{m}$  \\[0.03in] \hline
&   &    \\ [-0.15in]
$\bm{2}$ & $-\dfrac{1}{2}\begin{pmatrix}
				1 &\sqrt{3}\\
				\sqrt{3}&-1\\
\end{pmatrix}$
& $\begin{pmatrix}
				1 & 0 \\
				0 & -1 \\
\end{pmatrix}$
\\[0.1in] \hline
&   &   \\ [-0.15in]
			$\bm{3_{m}}$ &  $\frac{(-1)^m}{2}
			\begin{pmatrix}
				-1 & \sqrt{2} &1\\
				\sqrt{2}& 0 &\sqrt{2}\\
				1&\sqrt{2} & -1\\
			\end{pmatrix}$
			& $(-1)^{m}\begin{pmatrix}
				\omega^3_{8} & 0 & 0 \\
				0 & \omega^6_{8} & 0 \\
				0 & 0 & \omega^7_{8} \\
			\end{pmatrix}$
			\\ [0.1in]\hline
			&   &   \\ [-0.15in]
			$\bm{\bar{3}_{m}}$ &  $\frac{(-1)^m}{2}
			\begin{pmatrix}
				-1& \sqrt{2} & 1\\
				\sqrt{2} & 0 &\sqrt{2} \\
				1&\sqrt{2}& -1 \\
			\end{pmatrix}$
			& $(-1)^{m}\begin{pmatrix}
				\omega^5_{8} & 0 & 0 \\
				0 &\omega^2_{8}& 0 \\
				0 & 0 &\omega_{8} \\
			\end{pmatrix}$
			\\ [0.1in]\hline
			&   &   \\ [-0.15in]
			$\bm{\hat{3}_{m}}$ &  $\frac{(-1)^m}{2}
			\begin{pmatrix}
				0 & \sqrt{2}&\sqrt{2}\\
				\sqrt{2}& -1& 1\\
				\sqrt{2}&1& -1\\
			\end{pmatrix}$
			& $(-1)^{m}
			\begin{pmatrix}
				1 & 0 & 0 \\
				0 &\omega^2_{8}& 0 \\
				0 & 0 &\omega^6_{8}\\
			\end{pmatrix}$
			\\ [0.1in]\hline
			&   &      \\ [-0.16in]					
			$\bm{6}$ & 
			$\frac{1}{2}\begin{pmatrix}
				0 & 0 & 1 &1 &1&1\\
				0 & 0 & -1&1& -1&1\\
				1& -1& 0 & -1& 0 &1\\
				1&1& -1& 0 &1& 0 \\
				1& -1 & 0 & 1 & 0 & -1\\
				1&1 &1 & 0 & -1& 0 \\
			\end{pmatrix}$
			& $\begin{pmatrix}
				1 & 0 & 0 & 0 & 0 & 0 \\
				0 & -1 & 0 & 0 & 0 & 0 \\
				0 & 0 & \omega_{8} & 0 & 0 & 0 \\
				0 & 0 & 0 & \omega^3_{8} & 0 & 0 \\
				0 & 0 & 0 & 0 & \omega^5_{8} & 0 \\
				0 & 0 & 0 & 0 & 0 &\omega^7_{8} \\
			\end{pmatrix}$
			\\[0.12in] \hline\hline
		\end{tabular}
		\caption{\label{tab:Delta96_Reps}The representation matrices for the $\Delta(96)$  generators $S$ and $T$ in our chosen basis, where $m=0,1$ and $\omega_{8}\equiv e^{i\frac{\pi}{4}}$. 
        }
	\end{center}
\end{table}
	
The $\Delta(96)$ group has ten irreducible representations. These consist of two one-dimensional representations $\bm{1_m}$, one two-dimensional representation $\bm{2}$, six three-dimensional representations $\bm{3_m}$, $\bm{\bar{3}_m}$ and $\bm{\hat{3}_m}$, and one six-dimensional representation $\bm{6}$, where $m=0,1$. 
In this work, we adopt the $T$-diagonal basis. The explicit representation matrices for the generators $S$ and $T$ are listed in table~\ref{tab:Delta96_Reps}. Note that $\bm{\bar{3}_m}$ are the complex conjugates of $\bm{3_m}$. The representations $\bm{3_m}$, $\bm{\bar{3}_m}$ and $\bm{6}$ form faithful representations of $\Delta(96)$, while $\bm{\hat{3}_m}$ do not.  For the singlet and doublet representations, our basis coincides with that of Ref.~\cite{Yan:2025itm}. For the triplet and sextet representations, the representation matrices in our basis are related to those of Ref.~\cite{Yan:2025itm} by the following unitary transformations:
\begin{eqnarray}
\nonumber && \rho_{\bm{3_{m}}}(S)=U^{\dagger}_{3}\rho_{\bm{3_{m}}}(ab)U_{3}, \qquad \quad ~ \rho_{\bm{3_{m}}}(T)=U^{\dagger}_{3}\rho_{\bm{3_{m}}}(bd)U_{3}\,, \\ \nonumber &&\rho_{\bm{\bar{3}_{m}}}(S)=U^{T}_{3}\rho_{\bm{\bar{3}_{m}}}(ab)U^{*}_{3}, \qquad \quad  \rho_{\bm{\bar{3}_{m}}}(T)=U^{T}_{3}\rho_{\bm{\bar{3}_{m}}}(bd)U^{*}_{3}\,, \\
\nonumber && \rho_{\bm{\hat{3}_{m}}}(S)=V^{\dagger}_{3}\rho_{\bm{\hat{3}_{m}}}(ab)V_{3}, \qquad \quad ~ \rho_{\bm{\hat{3}_{m}}}(T)=V^{\dagger}_{3}\rho_{\bm{\hat{3}_{m}}}(bd)V_{3}, \\
 && \rho_{\bm{6}}(S)=U^{\dagger}_{6}\rho_{\bm{6}}(ab)U_{6}, \qquad \qquad ~~ \rho_{\bm{6}}(T)=U^{\dagger}_{6}\rho_{\bm{6}}(bd)U_{6}\,,
\end{eqnarray}
where the unitary matrices $U_{3}$, $V_{3}$ and $U_{6}$ are given by
\begin{eqnarray}
\nonumber && U_{3}=\frac{1}{\sqrt{2}}\left(
\begin{array}{ccc}
 -\omega^{3}_{8} & 0 & \omega^{3}_{8} \\
 0 & -\sqrt{2} & 0 \\
 1 & 0 & 1 \\
\end{array}
\right), \qquad \qquad  V_{3}=\frac{1}{\sqrt{2}}\left(
\begin{array}{ccc}
 0 & -i & i \\
 -\sqrt{2} & 0 & 0 \\
 0 & 1 & 1 \\
\end{array}
\right), \\
&& U_{6}=\frac{1}{\sqrt{2}}\left(
\begin{array}{cccccc}
 -i & i & 0 & 0 & 0 & 0 \\
 0 & 0 & 0 & 1 & 0 & 1 \\
 0 & 0 & \omega^{3}_{8} & 0 & -\omega^{3}_{8} & 0 \\
 1 & 1 & 0 & 0 & 0 & 0 \\
 0 & 0 & 0 & -\omega^{3}_{8} & 0 & \omega^{3}_{8} \\
 0 & 0 & -i & 0 & -i & 0 \\
\end{array}
\right)\,.
\end{eqnarray}
The Kronecker products between various irreducible representations follow immediately
\begin{eqnarray}
\nonumber &&\hskip-0.3in \bm{1_{m}}\otimes \bm{1_{n}}= \bm{1_{[m+n]}},\quad \bm{1_{m}}\otimes \bm{2}= \bm{2},\quad \bm{1_{m}}\otimes \bm{3_{n}}= \bm{3_{[m+n]}},\quad \bm{1_{m}}\otimes \bm{\bar{3}_{n}}= \bm{\bar{3}_{[m+n]}},  \\
\nonumber &&\hskip-0.3in \bm{1_{m}}\otimes \bm{\hat{3}_{n}}= \bm{\hat{3}_{[m+n]}}, \quad \bm{1_{m}}\otimes \bm{6}= \bm{6}, \quad \bm{2}\otimes\bm{2}=\bm{1_{0}}\oplus\bm{1_{1}}\oplus\bm{2}, \quad \bm{2}\otimes\bm{3_{m}}=\bm{3_{0}}\oplus\bm{3_{1}}, \\
\nonumber &&\hskip-0.3in \bm{2}\otimes\bm{\bar{3}_{m}}=\bm{\bar{3}_{0}}\oplus\bm{\bar{3}_{1}}, \quad \bm{2}\otimes\bm{\hat{3}_{m}}=\bm{\hat{3}_{0}}\oplus\bm{\hat{3}_{1}}, \quad \bm{2}\otimes\bm{6}=\bm{6_{i}}\oplus\bm{6_{ii}}, \quad \bm{3_{m}}\otimes \bm{3_{n}}= \bm{\bar{3}_{0}}\oplus\bm{\bar{3}_{1}}\oplus\bm{\hat{3}_{[m+n+1]}}, \\
\nonumber &&\hskip-0.3in \bm{3_{m}}\otimes \bm{\bar{3}_{n}}= \bm{1_{[m+n]}}\oplus\bm{2}\oplus\bm{6}, \quad \bm{3_{m}}\otimes \bm{\hat{3}_{n}}= \bm{\bar{3}_{[m+n+1]}}\oplus\bm{6}, \quad  \bm{3_{m}}\otimes \bm{6}= \bm{3_{0}}\oplus\bm{3_{1}}\oplus\bm{\hat{3}_{0}}\oplus\bm{\hat{3}_{1}}\oplus\bm{6}, \\
\nonumber &&\hskip-0.3in \bm{\bar{3}_{m}}\otimes \bm{\bar{3}_{n}}= \bm{3_{0}}\oplus\bm{3_{1}}\oplus\bm{\hat{3}_{[m+n+1]}}, ~~ \bm{\bar{3}_{m}}\otimes \bm{\hat{3}_{n}}= \bm{3_{[m+n+1]}}\oplus\bm{6}, ~~  \bm{\bar{3}_{m}}\otimes \bm{6}= \bm{\bar{3}_{0}}\oplus\bm{\bar{3}_{1}}\oplus\bm{\hat{3}_{0}}\oplus\bm{\hat{3}_{1}}\oplus\bm{6}, \\
\nonumber &&\hskip-0.3in \bm{\hat{3}_{m}}\otimes \bm{\hat{3}_{n}}= \bm{1_{[m+n]}}\oplus\bm{2}\oplus\bm{\hat{3}_{0}}\oplus\bm{\hat{3}_{1}}, \quad  \bm{\hat{3}_{m}}\otimes \bm{6}= \bm{3_{0}}\oplus\bm{3_{1}}\oplus\bm{\bar{3}_{0}}\oplus\bm{\bar{3}_{1}}\oplus\bm{6}, \\
\label{eq:kronecker_product}&&\hskip-0.3in \bm{6}\otimes \bm{6}= \bm{1_{0}}\oplus\bm{1_{1}}\oplus\bm{2_{S}}\oplus\bm{2_{A}}\oplus\bm{3_{0}}\oplus\bm{3_{1}}\oplus\bm{\bar{3}_{0}}\oplus\bm{\bar{3}_{1}}\oplus\bm{\hat{3}_{0}}\oplus\bm{\hat{3}_{1}}\oplus\bm{6_{S}}\oplus\bm{6_{A}}\,,
\end{eqnarray}
where $m,n=0,1$, and we have defined $[m+n]\equiv m+n$ (mod 2). The symbol $\bm{6_{i}}$ and $\bm{6_{ii}}$ denote two independent copies of the six-dimensional representation $\bm{6}$ appearing in the tensor products. The subscript ``$S$'' (``$A$'')  indicate symmetric  (antisymmetric) parts.  
In what follows, we present the Clebsch–Gordan (CG) coefficients in our working basis. The components of the first and second representations are denoted by $\alpha_{i}$ and $\beta_{i}$, respectively. For convenience, we also introduce the following notation to simplify the expressions of the CG coefficients
\begin{equation}\label{eq:P2P6_def}
	P_{2}=\begin{pmatrix}
		0&1\\
		-1&0
	\end{pmatrix},\qquad  
	P_{6}=\begin{pmatrix}
		0&1&0&0&0&0\\
		1&0&0&0&0&0\\
		0&0&0&0&1&0\\
		0&0&0&0&0&-1\\
		0&0&1&0&0&0\\
		0&0&0&-1&0&0
	\end{pmatrix}.
\end{equation}
Then the results are summarized in table~\ref{tab:Delta_96_CG}.

\begin{small}
	\renewcommand{\arraystretch}{1.13}
	\renewcommand{\tabcolsep}{0.9mm}
	\setlength\LTcapwidth{\textwidth}
	\setlength\LTleft{-0.0in}
	\setlength\LTright{0pt}
	\begin{longtable}{|c|c|c|c|c|c|c|c|c|c|c|c|}
		\caption{\label{tab:Delta_96_CG} Tensor products and the corresponding CG coefficients for the $\Delta(96)$, where the matrices $P_{2}$ and $P_{6}$ are defined in Eq.~\eqref{eq:P2P6_def}. }\\
		\midrule
		\specialrule{0em}{1.0pt}{1.0pt}
		\endfirsthead
		\multicolumn{12}{c}
		{{\bfseries \tablename\ \thetable{} -- continued from previous page}} \\
		\hline
		\endhead
		\caption[]{continues on next page}\\
		\endfoot
		\endlastfoot
		\hline
		
		\multicolumn{4}{|c}{$\bm{1_{m}}\otimes\bm{1_{n}}=\bm{1_{[m+n]}} $}& \multicolumn{4}{|c}{$\bm{1_{m}}\otimes\bm{2}=\bm{2}$} & \multicolumn{4}{|c|}{$\bm{1_{m}}\otimes\bm{3_{n}}=\bm{3_{[m+n]}} $} \\ \hline
		
		\multicolumn{4}{|c}{   $ \begin{array}{l}\bm{1_{[m+n]}}:~ \alpha_{1}\beta _{1}\end{array} $ } &
		\multicolumn{4}{|c}{$ \begin{array}{l}\bm{2}:P_{2}^{m} \begin{pmatrix} \alpha_{1}\beta _{1} \\ \alpha_{1}\beta _{2}\end{pmatrix}\end{array} $}&
		\multicolumn{4}{|c|}{ $ \begin{array}{l}\bm{3_{[m+n]}}:~\begin{pmatrix}\alpha_{1} \beta_{1}\\
					\alpha_{1} \beta_{2}\\\alpha_{1} \beta_{3}\end{pmatrix}\end{array} $  } \\ \hline 
		
		\multicolumn{4}{|c}{$\bm{1_{m}}\otimes\bm{\bar{3}_{n}}=\bm{\bar{3}_{[m+n]}}$}&  
		\multicolumn{4}{|c}{$\bm{1_{m}}\otimes\bm{\hat{3}_{n}}=\bm{\hat{3}_{[m+n]}}$}& \multicolumn{4}{|c|}{$\bm{1_{m}}\otimes\bm{6}=\bm{6}$}\\ \hline
		
		\multicolumn{4}{|c}{ $ \begin{array}{l}\bm{\bar{3}_{[m+n]}}:~\begin{pmatrix}\alpha_{1} \beta_{1}\\\alpha_{1} \beta_{2}\\\alpha_{1} \beta_{3}\end{pmatrix}\end{array} $  }&
		\multicolumn{4}{|c}{ $ \begin{array}{l}\bm{\hat{3}_{[m+n]}}:~\begin{pmatrix}\alpha_{1} \beta_{1}\\\alpha_{1} \beta_{2}\\\alpha_{1} \beta_{3}\end{pmatrix}\end{array} $ } &
		\multicolumn{4}{|c|}{  $ \begin{array}{l} \bm{6}:~P_{6}^{m}\begin{pmatrix} \alpha_{1} \beta_{1}\\ \alpha_{1} \beta_{2}\\ \alpha_{1} \beta_{3}\\ \alpha_{1} \beta_{4}\\ \alpha_{1} \beta_{5}\\ \alpha_{1} \beta_{6}\end{pmatrix}\\\end{array} $ }\\ \hline
		
		\multicolumn{4}{|c}{$\bm{2}\otimes\bm{2}=\bm{1_{0}}\oplus\bm{1_{1}}\oplus\bm{2}$}&
		\multicolumn{4}{|c}{$\bm{2}\otimes\bm{3_{m}}=\bm{3_{0}}\oplus\bm{3_{1}} $}&  		
		\multicolumn{4}{|c|}{$\bm{2}\otimes \bm{\bar{3}_{m}}=\bm{\bar{3}_{0}}\oplus\bm{\bar{3}_{1}}$}\\ \hline
		
		\multicolumn{4}{|c}{  $ \begin{array}{l}
				\bm{1_{0}}: \alpha_{1} \beta_{1}+\alpha_{2} \beta_{2}\\
				\bm{1_{1}}:	\alpha_{1} \beta_{2}-\alpha_{2} \beta_{1}\\
				\bm{2}:\begin{pmatrix} -\alpha_{1} \beta_{1}+\alpha_{2} \beta_{2}\\ \alpha_{1}\beta_{2}+\alpha_{2} \beta_{1}\\	\end{pmatrix}\end{array} $ } &
		\multicolumn{4}{|c}{  $ \begin{array}{l}
				\bm{3_{m}}:\quad~~ 
				\begin{pmatrix} \alpha _1 \beta _1+\sqrt{3} \alpha _2 \beta _3\\
					-2 \alpha _1 \beta _2\\
					\sqrt{3} \alpha _2 \beta _1+\alpha _1 \beta _3\end{pmatrix}\\
				\bm{3_{[m+1]}}:
				\begin{pmatrix} 3 \alpha _1 \beta _3-\sqrt{3} \alpha _2 \beta _1\\
					2\sqrt{3} \alpha _2 \beta _2\\
					3 \alpha _1 \beta _1-\sqrt{3} \alpha _2 \beta _3\end{pmatrix}	
			\end{array} $ }&
		\multicolumn{4}{|c|}{  $ \begin{array}{l}
				\bm{\bar{3}_{m}}:\quad~~ 
				\begin{pmatrix} 
					\alpha _1 \beta _1+\sqrt{3} \alpha _2 \beta _3\\
					-2 \alpha _1 \beta _2\\
					\sqrt{3} \alpha _2 \beta _1+\alpha _1 \beta _3
				\end{pmatrix}\\
				\bm{\bar{3}_{[m+1]}}:
				\begin{pmatrix} 3 \alpha _1 \beta _3-\sqrt{3} \alpha _2 \beta _1\\
					2\sqrt{3} \alpha _2 \beta _2\\
					3 \alpha _1 \beta _1-\sqrt{3} \alpha _2 \beta _3\end{pmatrix}\end{array} $ }\\ \hline		
		
		\multicolumn{4}{|c}{$\bm{2}\otimes\bm{\hat{3}_{m}}=\bm{\hat{3}_{0}}\oplus\bm{\hat{3}_{1}}$} & \multicolumn{4}{|c}{$\bm{3_{m}}\otimes\bm{3_{n}}=\bm{\bar{3}_{0}}\oplus\bm{\bar{3}_{1}}\oplus\bm{\hat{3}_{[m+n+1]}}$}& 
		\multicolumn{4}{|c|}{$\bm{\bar{3}_{m}}\otimes\bm{\bar{3}_{n}}=\bm{3_{0}}\oplus\bm{3_{1}}\oplus\bm{\hat{3}_{[m+n-1]}}$}\\ \hline
		
		\multicolumn{4}{|c}{ $ \begin{array}{l}
				\bm{\hat{3}_{m}}:\quad~~ 
				\begin{pmatrix} 
					2\alpha _1 \beta _1\\
					-\alpha _1 \beta _2-\sqrt{3} \alpha _2 \beta _3\\
					-\sqrt{3} \alpha _2 \beta _2-\alpha _1 \beta _3
				\end{pmatrix}\\
				\bm{\hat{3}_{[m+1]}}:
				\begin{pmatrix}2 \alpha _2 \beta _1\\
					\sqrt{3} \alpha _1 \beta _3-\alpha _2 \beta _2\\
					\sqrt{3} \alpha _1 \beta _2-\alpha _2 \beta _3\end{pmatrix}\end{array} $ } &
		\multicolumn{4}{|c}{ $ \begin{array}{l} 
				\bm{\bar{3}_{[m+n]}}:\quad~ 
				\begin{pmatrix} 
					\alpha _2 \beta _3-\alpha _3 \beta _2\\
					\alpha _3 \beta _1-\alpha _1 \beta _3\\
					\alpha _1 \beta _2-\alpha _2 \beta _1\end{pmatrix}\\
				\bm{\bar{3}_{[m+n+1]}}:
				\begin{pmatrix} \alpha _2 \beta _1+\alpha _1 \beta _2\\
					\alpha _1 \beta _1-\alpha _3 \beta _3\\
					-\alpha _3 \beta _2-\alpha _2 \beta _3\end{pmatrix}\\
				\bm{\hat{3}_{[m+n+1]}}:
				\begin{pmatrix} \sqrt{2}\alpha _2 \beta _2\\
					-\alpha _1 \beta _1-\alpha _3 \beta _3\\
					-\alpha _3 \beta _1-\alpha _1 \beta _3\end{pmatrix}	
			\end{array} $  } &
		\multicolumn{4}{|c|}{ $ \begin{array}{l}
				\bm{3_{[n+m]}}:\quad~ 
				\begin{pmatrix} 
					\alpha _2 \beta _3-\alpha _3 \beta _2\\
					\alpha _3 \beta _1-\alpha _1 \beta _3\\
					\alpha _1 \beta _2-\alpha _2 \beta _1\end{pmatrix}\\
				\bm{3_{[n+m+1]}}:
				\begin{pmatrix} \alpha _2 \beta _1+\alpha _1 \beta _2\\
					\alpha _1 \beta _1-\alpha _3 \beta _3\\
					-\alpha _3 \beta _2-\alpha _2 \beta _3\end{pmatrix}\\
				\bm{\hat{3}_{[m+n+1]}}:
				\begin{pmatrix} 
					\sqrt{2}\alpha _2 \beta _2\\
					-\alpha _3 \beta _1-\alpha _1 \beta _3\\
					-\alpha _1 \beta _1-\alpha _3 \beta _3\end{pmatrix}\end{array} $}\\ \hline	
		
		\multicolumn{6}{|c}{$\bm{3_{m}}\otimes\bm{\bar{3}_{n}}=\bm{1_{[m+n]}}\oplus\bm{2}\oplus\bm{6}$}& \multicolumn{6}{|c|}{$\bm{3_{m}}\otimes\bm{\hat{3}_{n}}=\bm{\bar{3}_{[m+n+1]}}\oplus\bm{6}$}\\ \hline
		
		\multicolumn{6}{|c}{  $ \begin{array}{l} \bm{1_{[m+n]}}:~~\alpha _1 \beta _1+\alpha _2 \beta _2+\alpha _3 \beta _3\\
				\bm{2}:P_{2}^{m+n}
				\begin{pmatrix}\alpha _1 \beta _1-2 \alpha _2 \beta _2+\alpha _3 \beta _3\\
					\sqrt{3} \left(\alpha _3 \beta _1+\alpha _1 \beta _3\right)\\\end{pmatrix}\\
				\bm{\bm{6}}:P_{6}^{m+n}
				\begin{pmatrix} \alpha _1 \beta _1-\alpha _3 \beta _3\\
					\alpha _3 \beta _1-\alpha _1 \beta _3\\
					\sqrt{2} \alpha _3 \beta _2\\
					-\sqrt{2} \alpha _2 \beta _1\\
					-\sqrt{2} \alpha _1 \beta _2\\
					\sqrt{2} \alpha _2 \beta _3\\\end{pmatrix} 	\end{array} $ } &
		\multicolumn{6}{|c|}{  $ \begin{array}{l}\bm{\bar{3}_{[m+n+1]}}:
				\begin{pmatrix}\alpha _3 \beta _2+\alpha _1 \beta _3\\
					-\sqrt{2} \alpha _2 \beta _1\\
					\alpha _1 \beta _2+\alpha _3 \beta _3\end{pmatrix} \\
				\bm{6}:P_{6}^{m+n}	
				\begin{pmatrix} 
					\sqrt{2}\alpha _2 \beta _2\\
					\sqrt{2}\alpha _2 \beta _3\\
					\alpha _1 \beta _3-\alpha _3 \beta _2\\
					\sqrt{2}\alpha _1 \beta _1\\
					\alpha _3 \beta _3-\alpha _1 \beta _2\\
					\sqrt{2}\alpha _3 \beta _1\end{pmatrix}\end{array} $ } \\ \hline		

\newpage		
		
		\multicolumn{6}{|c}{$\bm{\bar{3}_{m}}\otimes \bm{\hat{3}_{n}}=\bm{3_{[m+n+1]}}\oplus\bm{6}$}& \multicolumn{6}{|c|}{$\bm{\hat{3}_{m}}\otimes\bm{\hat{3}_{n}}=\bm{1_{[m+n]}}\oplus\bm{2}\oplus \bm{\hat{3}_{0}}\oplus\bm{\hat{3}_{1}}$}\\ \hline
		
		\multicolumn{6}{|c}{  $ \begin{array}{l}
				\bm{3_{[m+n+1]}}:
				\begin{pmatrix} \alpha _1 \beta _2+\alpha _3 \beta _3\\
					-\sqrt{2} \alpha _2 \beta _1\\
					\alpha _3 \beta _2+\alpha _1 \beta _3\end{pmatrix} \\
				\bm{6}:P_{6}^{m+n}	
				\begin{pmatrix}\sqrt{2}\alpha _2 \beta _3\\
					-\sqrt{2}\alpha _2 \beta _2\\
					\sqrt{2}\alpha _3 \beta _1\\
					\alpha _3 \beta _2-\alpha _1 \beta _3\\
					\sqrt{2}\alpha _1 \beta _1\\
					\alpha _1 \beta _2-\alpha _3 \beta _3\end{pmatrix}\end{array} $ } &
		\multicolumn{6}{|c|}{  $ \begin{array}{l} \bm{1_{[m+n]}}:\alpha _1 \beta _1+\alpha _3 \beta _2+\alpha _2 \beta _3\\
				\bm{2}:P_{2}^{m+n}	
				\begin{pmatrix}2 \alpha _1 \beta _1-\alpha _3 \beta _2-\alpha _2 \beta _3\\
					-\sqrt{3}(\alpha _2 \beta _2+\alpha _3 \beta _3)\end{pmatrix}\\
				\bm{\hat{3}_{[m+n]}}: 
				\begin{pmatrix}\alpha _2 \beta _3-\alpha _3 \beta _2\\
					\alpha _1 \beta _2-\alpha _2 \beta _1\\
					\alpha _3 \beta _1-\alpha _1 \beta _3\end{pmatrix}\\
				\bm{\hat{3}_{[m+n+1]}}:
				\begin{pmatrix} 
					\alpha _2 \beta _2-\alpha _3 \beta _3\\
					-\alpha _3 \beta _1-\alpha _1 \beta _3\\
					\alpha _2 \beta _1+\alpha _1 \beta _2\end{pmatrix}\end{array} $ }\\ \hline		
		\multicolumn{6}{|c}{$\bm{2}\otimes\bm{6}=\bm{6_{i}}\oplus\bm{6_{ii}}$}&  
		\multicolumn{6}{|c|}{$\bm{3_{m}}\otimes\bm{6}=\bm{3_{0}}\oplus\bm{3_{1}}\oplus\bm{\hat{3}_{0}}\oplus\bm{\hat{3}_{1}}\oplus\bm{6}$}\\ \hline  		
	
		\multicolumn{6}{|c}{  $ \begin{array}{l}
				\bm{6_{i}}:~~
				\begin{pmatrix}2\alpha _1 \beta _1 \\
					2\alpha _1 \beta _2 \\
					-\alpha _1 \beta _3-\sqrt{3}\alpha _2 \beta _5\\
					-\sqrt{3}\alpha _2 \beta _6 -\alpha _1 \beta _4 \\
					-\sqrt{3}\alpha _2 \beta _3 -\alpha _1 \beta _5 \\
					-\sqrt{3}\alpha _2 \beta _4 -\alpha _1 \beta _6 
					\end{pmatrix} \\
				\bm{6_{ii}}:
				\begin{pmatrix}2\alpha _2 \beta _2\\
					2\alpha _2 \beta _1\\
					\sqrt{3}\alpha _1 \beta _3-\alpha _2 \beta _5\\
					\alpha _2 \beta _6-\sqrt{3}\alpha _1 \beta _4\\
					-\alpha _2 \beta _3+\sqrt{3}\alpha _1 \beta _5\\
					\alpha _2 \beta _4-\sqrt{3}\alpha _1 \beta _6
					\end{pmatrix}\end{array} $ }& 		
		\multicolumn{6}{|c|}{ $ \begin{array}{l}
				\bm{3_{m}}:\quad~~ 
				\begin{pmatrix}\alpha _1 \beta _1-\alpha _3 \beta _2-\sqrt{2} \alpha _2 \beta _5\\
					\sqrt{2} \left(\alpha _3 \beta _6-\alpha _1 \beta _4\right)\\
					-\alpha _3 \beta _1+\alpha _1 \beta _2+\sqrt{2} \alpha _2 \beta _3\end{pmatrix}\\
				\bm{3_{[m+1]}}:
				\begin{pmatrix}-\alpha _3 \beta _1+\alpha _1 \beta _2-\sqrt{2} \alpha _2 \beta _3\\
					\sqrt{2} \left(\alpha _1 \beta _6-\alpha _3 \beta _4\right)\\
					\alpha _1 \beta _1-\alpha _3 \beta _2+\sqrt{2} \alpha _2 \beta _5\end{pmatrix}\\
				\bm{\hat{3}_{m}}:~\quad~ 
				\begin{pmatrix}2(\alpha _3 \beta _3+\alpha _1 \beta _5)\\
				\sqrt{2} \left(\alpha _3 \beta _4+\alpha _1 \beta _6\right)-2 \alpha _2 \beta _2\\
				2 \alpha _2 \beta _1-\sqrt{2} \left(\alpha _1 \beta _4+\alpha _3 \beta _6\right)\end{pmatrix}\\
				\bm{\hat{3}_{[m+1]}}:~ 
				\begin{pmatrix}2(\alpha _1 \beta _3+\alpha _3 \beta _5)\\
				-2 \alpha _2 \beta _1-\sqrt{2} \left(\alpha _1 \beta _4+\alpha _3 \beta _6\right)\\
				2 \alpha _2 \beta _2+\sqrt{2} \left(\alpha _3 \beta _4+\alpha _1 \beta _6\right)\end{pmatrix}\\
				\bm{6}:P_{6}^{m}
				\begin{pmatrix}\alpha _1 \beta _5-\alpha _3 \beta _3\\
					\alpha _1 \beta _3-\alpha _3 \beta _5\\
					\sqrt{2} \alpha _2 \beta _4\\
					-\alpha _1 \beta _1-\alpha _3 \beta _2\\
					-\sqrt{2} \alpha _2 \beta _6\\
					\alpha _3 \beta _1+\alpha _1 \beta _2\end{pmatrix}\end{array} $}
		\\ \hline  		
		
\newpage
		
		\multicolumn{6}{|c}{$\bm{\bar{3}_{m}}\otimes\bm{6}=\bm{\bar{3}_{0}}\oplus\bm{\bar{3}_{1}}\oplus\bm{\hat{3}_{0}}\oplus\bm{\hat{3}_{1}}\oplus\bm{6}$}& \multicolumn{6}{|c|}{$\bm{\hat{3}_{m}}\otimes\bm{6}=\bm{3_{0}}\oplus\bm{3_{1}}\oplus\bm{\bar{3}_{0}}\oplus\bm{\bar{3}_{1}}\oplus\bm{6}$}\\ \hline
		
		\multicolumn{6}{|c}{  $ \begin{array}{l} 	
				\bm{\bar{3}_{m}}:~\quad~ 
				\begin{pmatrix}\alpha _1 \beta _1+\alpha _3 \beta _2-\sqrt{2} \alpha _2 \beta _4\\
					\sqrt{2} \left(\alpha _3 \beta _3-\alpha _1 \beta _5\right)\\
					-\alpha _3 \beta _1-\alpha _1 \beta _2+\sqrt{2} \alpha _2 \beta _6\end{pmatrix}\\
				\bm{\bar{3}_{[m+1]}}:
				\begin{pmatrix}\alpha _3 \beta _1+\alpha _1 \beta _2+\sqrt{2} \alpha _2 \beta _6\\
					\sqrt{2} \left(\alpha _3 \beta _5-\alpha _1 \beta _3\right)\\
					-\alpha _1 \beta _1-\alpha _3 \beta _2-\sqrt{2} \alpha _2 \beta _4\end{pmatrix}\\
				\bm{\hat{3}_{m}}:~\quad~ 
				\begin{pmatrix}2(\alpha _1 \beta _4+\alpha _3 \beta _6)\\
					2 \alpha _2 \beta _1-\sqrt{2} \left(\alpha _3 \beta _3+\alpha _1 \beta _5\right)\\
					2 \alpha _2 \beta _2+\sqrt{2} \left(\alpha _1 \beta _3+\alpha _3 \beta _5\right)\end{pmatrix}\\
				\bm{\hat{3}_{[m+1]}}:~ 
				\begin{pmatrix}2(\alpha _3 \beta _4+\alpha _1 \beta _6)\\
				\sqrt{2} \left(\alpha _1 \beta _3+\alpha _3 \beta _5\right)-2 \alpha _2 \beta _2\\
				-2 \alpha _2 \beta _1-\sqrt{2} \left(\alpha _3 \beta _3+\alpha _1 \beta _5\right)\end{pmatrix}\\
				\bm{6}:P_{6}^{m}
				\begin{pmatrix}\alpha _1 \beta _4-\alpha _3 \beta _6\\
					\alpha _3 \beta _4-\alpha _1 \beta _6\\
					\alpha _3 \beta _1-\alpha _1 \beta _2\\
					-\sqrt{2} \alpha _2 \beta _3\\
					\alpha _3 \beta _2-\alpha _1 \beta _1\\
					\sqrt{2} \alpha _2 \beta _5\end{pmatrix}\end{array} $ } &
		\multicolumn{6}{|c|}{  $ \begin{array}{l}
				\bm{3_{m}}:~\quad~ 
				\begin{pmatrix}\sqrt{2} \alpha _2 \beta _3+2 \alpha _1 \beta _4-\sqrt{2} \alpha _3 \beta _5\\
					2(\alpha _3 \beta _1+\alpha _2 \beta _2)\\
					-\sqrt{2} \alpha _3 \beta _3+\sqrt{2} \alpha _2 \beta _5+2 \alpha _1 \beta _6\end{pmatrix}\\
				\bm{3_{[m+1]}}:
				\begin{pmatrix}\sqrt{2} \alpha _3 \beta _3-\sqrt{2} \alpha _2 \beta _5+2 \alpha _1 \beta _6\\
					2(-\alpha _2 \beta _1-\alpha _3 \beta _2)\\
					-\sqrt{2} \alpha _2 \beta _3+2 \alpha _1 \beta _4+\sqrt{2} \alpha _3 \beta _5\end{pmatrix}\\
				\bm{\bar{3}_{m}}:~\quad~ 
				\begin{pmatrix}-\sqrt{2} \alpha _2 \beta _4+2 \alpha _1 \beta _5+\sqrt{2} \alpha _3 \beta _6\\
					2(\alpha _2 \beta _1-\alpha _3 \beta _2)\\
				2 \alpha _1 \beta _3+\sqrt{2} \left(\alpha _3 \beta _4-\alpha _2 \beta _6\right)\end{pmatrix}\\
				\bm{\bar{3}_{[m+1]}}:~ 
				\begin{pmatrix}2 \alpha _1 \beta _3+\sqrt{2} \left(\alpha _2 \beta _6-\alpha _3 \beta _4\right)\\
					2(\alpha _2 \beta _2-\alpha _3 \beta _1)\\
				\sqrt{2} \alpha _2 \beta _4+2 \alpha _1 \beta _5-\sqrt{2} \alpha _3 \beta _6\end{pmatrix}\\
				\bm{6}:P_{6}^{m}
				\begin{pmatrix}\sqrt{2}\alpha _1 \beta _1\\
					-\sqrt{2}\alpha _1 \beta _2\\
					\alpha _3 \beta _4+\alpha _2 \beta _6\\
					\alpha _2 \beta _3+\alpha _3 \beta _5\\
					\alpha _2 \beta _4+\alpha _3 \beta _6\\
					\alpha _3 \beta _3+\alpha _2 \beta _5\end{pmatrix}
			\end{array} $  }\\ \hline
	
		\multicolumn{12}{|c|}{$\bm{6}\otimes \bm{6}=\bm{1_{0}}\oplus \bm{1_{1}}\oplus\bm{2_{S}}\oplus\bm{2_{A}}\oplus\bm{3_{0}}\oplus \bm{3_{1}}\oplus \bm{\bar{3}_{0}}\oplus\bm{\bar{3}_{1}}\oplus\bm{\hat{3}_{0}}\oplus\bm{\hat{3}_{1}}\oplus\bm6_{S}\oplus\bm{6_{A}}$}\\ \hline
		
		\multicolumn{12}{|c|}{ $ \begin{array}{l}
				\bm{1_{0}}:~\alpha _1 \beta _1-\alpha _2 \beta _2+\alpha _6 \beta _3+\alpha _5 \beta _4+\alpha _4 \beta _5+\alpha _3 \beta _6\\
				\bm{1_{1}}:~-\alpha _2 \beta _1+\alpha _1 \beta _2+\alpha _4 \beta _3-\alpha _3 \beta _4+\alpha _6 \beta _5-\alpha _5 \beta _6\\
				\bm{2_{S}}:~
				\begin{pmatrix}
					-2\alpha _1 \beta _1+2\alpha _2 \beta _2 +\alpha _4 \beta _5+\alpha _5 \beta _4+\alpha _3 \beta _6+\alpha _6 \beta _3\\
					\sqrt{3}(\alpha _4 \beta _3+\alpha _3 \beta _4+\alpha _6 \beta _5+\alpha _5 \beta _6) 
					\end{pmatrix} \\
				\bm{2_{A}}:~
				\begin{pmatrix}
					\sqrt{3}(-\alpha _6 \beta _3-\alpha _4 \beta _5+\alpha _5 \beta _4 +\alpha _3 \beta _6
					)\\
					2\alpha _2 \beta _1-2\alpha _1 \beta _2+\alpha _4 \beta _3-\alpha _3 \beta _4+\alpha _6 \beta _5-\alpha _5 \beta _6 
					\end{pmatrix}\end{array} $ } \\[0.55in]
		\multicolumn{3}{|c}{ } &	
		\multicolumn{3}{c}{ $ \begin{array}{l}
				\bm{3_{0}}:~
				\begin{pmatrix}-\alpha _4 \beta _1-\alpha _6 \beta _2+\alpha _1 \beta _4+\alpha _2 \beta _6\\
					\sqrt{2} \left(\alpha _3 \beta _5-\alpha _5 \beta _3\right)\\
					\alpha _6 \beta _1+\alpha _4 \beta _2-\alpha _2 \beta _4-\alpha _1 \beta _6\end{pmatrix} \\
				\bm{3_{1}}:~
				\begin{pmatrix}\alpha _6 \beta _1+\alpha _4 \beta _2+\alpha _2 \beta _4+\alpha _1 \beta _6\\
					\sqrt{2} \left(\alpha _5 \beta _5-\alpha _3 \beta _3\right)\\
					-\alpha _4 \beta _1-\alpha _6 \beta _2-\alpha _1 \beta _4-\alpha _2 \beta _6\end{pmatrix}\\
				\bm{\bar{3}_{0}}:~
				\begin{pmatrix}-\alpha _5 \beta _1+\alpha _3 \beta _2-\alpha _2 \beta _3+\alpha _1 \beta _5\\
					\sqrt{2} \left(\alpha _6 \beta _4-\alpha _4 \beta _6\right)\\
					\alpha _3 \beta _1-\alpha _5 \beta _2-\alpha _1 \beta _3+\alpha _2 \beta _5\end{pmatrix}\\
				\bm{\bar{3}_{1}}:~
				\begin{pmatrix}\alpha _3 \beta _1-\alpha _5 \beta _2+\alpha _1 \beta _3-\alpha _2 \beta _5\\
					\sqrt{2} \left(\alpha _4 \beta _4-\alpha _6 \beta _6\right)\\
					-\alpha _5 \beta _1+\alpha _3 \beta _2+\alpha _2 \beta _3-\alpha _1 \beta _5\end{pmatrix}\\
				\bm{\hat{3}_{0}}:~
				\begin{pmatrix}\sqrt{2}(\alpha _1 \beta _1+\alpha _2 \beta _2)\\
					\alpha _3 \beta _3+\alpha _6 \beta _4+\alpha _5 \beta _5+\alpha _4 \beta _6\\
					\alpha _5 \beta _3+\alpha _4 \beta _4+\alpha _3 \beta _5+\alpha _6 \beta _6\end{pmatrix}\end{array} $	} &
		\multicolumn{6}{c|}{ $ \begin{array}{l}
				\bm{\hat{3}_{1}}:~
				\begin{pmatrix}\sqrt{2}(\alpha _2 \beta _1+\alpha _1 \beta _2)\\
					\alpha _5 \beta _3-\alpha _4 \beta _4+\alpha _3 \beta _5-\alpha _6 \beta _6\\
					\alpha _3 \beta _3-\alpha _6 \beta _4+\alpha _5 \beta _5-\alpha _4 \beta _6\end{pmatrix}\\	
				\bm{6_{S}}:~
				\begin{pmatrix}\alpha _6 \beta _3 -\alpha _4 \beta _5-\alpha _5 \beta _4 +\alpha _3 \beta _6\\
				\alpha _4 \beta _3-\alpha _6 \beta _5+\alpha _3 \beta _4-\alpha _5 \beta _6 \\
				\alpha _3 \beta _1+\alpha _5 \beta _2+\alpha _1 \beta _3+\alpha _2 \beta _5\\
				-\alpha _1 \beta _4+\alpha _2 \beta _6-\alpha _4 \beta _1 +\alpha _6 \beta _2\\
				-\alpha _5 \beta _1-\alpha _3 \beta _2 -\alpha _2 \beta _3-\alpha _1 \beta _5\\
				-\alpha _2 \beta _4+\alpha _1 \beta _6+\alpha _6 \beta _1 -\alpha _4 \beta _2\end{pmatrix}\\
				\bm{6_{A}}:~
				\begin{pmatrix}-\alpha _6 \beta _3+\alpha _4 \beta _5-\alpha _5 \beta _4 +\alpha _3 \beta _6\\
					-\alpha _4 \beta _3 +\alpha _6 \beta _5+\alpha _3 \beta _4-\alpha _5 \beta _6\\
					-\alpha _3 \beta _1-\alpha _5 \beta _2+\alpha _1 \beta _3+\alpha _2 \beta _5\\
					\alpha _1 \beta _4-\alpha _2 \beta _6-\alpha _4 \beta _1 +\alpha _6 \beta _2\\
					\alpha _5 \beta _1+\alpha _3 \beta _2 -\alpha _2 \beta _3-\alpha _1 \beta _5\\
					\alpha _2 \beta _4-\alpha _1 \beta _6+\alpha _6 \beta _1 -\alpha _4 \beta _2\end{pmatrix}\\ \end{array} $ } \\	\hline
		\specialrule{0em}{1.0pt}{1.0pt}
		\midrule
	\end{longtable}
\end{small}

\section{\label{sec:qes_VVMFs}The $q$-expressions of lowest- and next-to-lowest-weight VVMFs}

In section~\ref{secVVMFs_Delta96}, a construction of the lowest-weight doublet and triplet VVMFs is presented in terms of generalized hypergeometric series. Their $q$-expressions are provided by
\begin{eqnarray}
\nonumber  Y_{\bm{2}}^{(2)}(\tau)&=&
\begin{pmatrix}
	1+24q+24q^{2}+96q^{3}+24q^{4}+144q^{5}+96q^{6}+\dots\\[4pt]
	8\sqrt{3}\,q^{1/2}\bigl(1+4q+6q^{2}+8q^{3}+13q^{4}+12q^{5}+14q^{6}+\dots\bigr)
\end{pmatrix}\,, \\
\nonumber 	Y_{\bm{\bar{3}_{0}}}^{(2)}(\tau) &=&
	\begin{pmatrix}
		q^{5/8}\bigl(1-3q+5q^{3}+q^{4}-3q^{5}-7q^{6}+\dots\bigr)\\[4pt]
		\frac{\sqrt{2}}{2}\,q^{1/4}\bigl(1-2q-3q^{2}+6q^{3}+2q^{4}-q^{6}+\dots\bigr)\\[4pt]
		-\frac{1}{2}\,q^{1/8}\bigl(1-3q+2q^{2}-q^{3}+10q^{5}-7q^{6}+\dots\bigr)
	\end{pmatrix},
	\\[4pt]
\nonumber 	Y_{\bm{\hat{3}_{0}}}^{(2)}(\tau) &=&
	\begin{pmatrix}
		1-8q+24q^{2}-32q^{3}+24q^{4}-48q^{5}+96q^{6}+\dots\\[4pt]
		-4\sqrt{2}\,q^{1/4}\bigl(1+6q+13q^{2}+14q^{3}+18q^{4}+32q^{5}+31q^{6}+\dots\bigr)\\[4pt]
		-16\sqrt{2}\,q^{3/4}\bigl(1+2q+3q^{2}+6q^{3}+5q^{4}+6q^{5}+10q^{6}+\dots\bigr)
	\end{pmatrix}\,, \\
\nonumber 	Y_{\bm{3_{1}}}^{(4)}(\tau) &=&
	\begin{pmatrix}
		q^{7/8}\bigl(1-5q+3q^{2}+20q^{3}-25q^{4}-26q^{5}+25q^{6}+\dots\bigr)\\[4pt]
		-\frac{\sqrt{2}}{8}\,q^{1/4}\bigl(1-10q+37q^{2}-50q^{3}-30q^{4}+128q^{5}-25q^{6}+\dots\bigr)\\[4pt]
		\frac{1}{2}\,q^{3/8}\bigl(1-5q+5q^{2}+10q^{3}-20q^{4}+19q^{5}-30q^{6}+\dots\bigr)
	\end{pmatrix}\,,
	\\[4pt]
\nonumber 	Y_{\bm{\bar{3}_{1}}}^{(4)}(\tau) &=&
	\begin{pmatrix}
		q^{1/8}\bigl(1+37q-30q^{2}-25q^{3}-320q^{4}+410q^{5}-87q^{6}+\dots\bigr)\\[4pt]
		-16\sqrt{2}\,q^{3/4}\bigl(1+2q-5q^{2}-10q^{3}+5q^{4}+6q^{5}+10q^{6}+\dots\bigr)\\[4pt]
		-2\,q^{5/8}\bigl(5+25q-64q^{2}+17q^{3}-155q^{4}+185q^{5}+205q^{6}+\dots\bigr)
	\end{pmatrix}\,,
	\\[4pt]
\nonumber 	Y_{\bm{\hat{3}_{1}}}^{(4)}(\tau) &=&
	\begin{pmatrix}
		q^{1/2}\bigl(1-4q-2q^{2}+24q^{3}-11q^{4}-44q^{5}+22q^{6}+\dots\bigr)\\[4pt]
		\sqrt{2}\,q^{3/4}\bigl(1-6q+11q^{2}-2q^{3}-11q^{4}+14q^{5}-38q^{6}+\dots\bigr)\\[4pt]
		-\frac{\sqrt{2}}{4}\,q^{1/4}\bigl(1-2q-11q^{2}+22q^{3}+50q^{4}-96q^{5}-121q^{6}+\dots\bigr)
	\end{pmatrix}\,, \\
\label{eq:qes_LVVMFs}Y_{\bm{3_{0}}}^{(6)}(\tau)&=&
	\begin{pmatrix}
		q^{3/8}\bigl(1+3q-75q^{2}+282q^{3}-276q^{4}-453q^{5}+962q^{6}+\dots\bigr)\\[4pt]
		-4\sqrt{2}\,q^{3/4}\bigl(1-6q+3q^{2}+46q^{3}-75q^{4}-114q^{5}+282q^{6}+\dots\bigr)\\[4pt]
		-2\,q^{7/8}\bigl(3-23q+57q^{2}-36q^{3}+21q^{4}-222q^{5}-69q^{6}+\dots\bigr)
	\end{pmatrix}\,.
\end{eqnarray}

Given the lowest-weight VVMFs obtained via the MLDE approach, higher-weight VVMFs for the finite modular group can be systematically constructed in several complementary ways: by applying the modular derivative $D_k$\footnote{The modular derivative $D_k \equiv \theta - \frac{kE_2}{12}$ is defined using the quasi‑modular Eisenstein series \newline$E_2(\tau)=1-24\sum_{n\ge1}\sigma_1(n)q^n$ and $\theta=q\frac{d}{dq}$. It raises the modular weight of a (vector‑valued) modular form by two units while preserving its modular transformation. The operator satisfies a graded Leibniz rule, that means $D_{k+m}(f g) = (D_k f)\, g + f\, (D_m g)$, and its n-fold iteration $D_k^{\,n}$ is given by $D_k^{\,n} \equiv D_{k+2(n-1)} \circ D_{k+2(n-2)} \circ \cdots \circ D_k$. }, by multiplying with scalar modular forms, or by forming tensor products and projecting with the CG coefficients of the finite modular group. In this work, we adopt the last approach to obtain the next-to-lowest-weight VVMFs corresponding to each of the six 3-$d$ irreducible representations of $\Delta(96)$.

For the finite modular group $\Delta(96)$, the lowest-weight doublet VVMF is of weight $2$. Using the CG coefficients of $\Delta(96)$ given in table~\ref{tab:Delta_96_CG}, the tensor product of the representation $\bm{2}$ with any 3-$d$ representation contains that same 3-$d$ representation in its decomposition. Consequently, to obtain the next-to-lowest-weight (i.e., weight $k_0+2$) VVMF for a given 3-$d$ representation, one simply contracts the doublet VVMF $Y^{(2)}_{\bm{2}}$ with the lowest-weight triplet VVMF of that representation. 
Applying this procedure, we obtain the following next-to-lowest-weight VVMF for each of the six 3-$d$ irreducible representations of $\Delta(96)$.

At weight 4, we have other two modular multiplets:
\begin{eqnarray}
\nonumber 	Y_{\bm{\bar{3}_{0}}}^{(4)}(\tau) &=&
	\bigl(Y_{\bm{2}}^{(2)}(\tau)\,Y_{\bm{\bar{3}_{0}}}^{(2)}(\tau)\bigr)_{\bm{\bar{3}_{0}}}
	=
	\begin{pmatrix}
		Y_{\bm{2},1}^{(2)}(\tau)\,Y_{\bm{\bar{3}_{0}},1}^{(2)}(\tau)
		+\sqrt{3}\,Y_{\bm{2},2}^{(2)}(\tau)\,Y_{\bm{\bar{3}_{0}},3}^{(2)}(\tau)\\[4pt]
		-2\,Y_{\bm{2},1}^{(2)}(\tau)\,Y_{\bm{\bar{3}_{0}},2}^{(2)}(\tau)\\[4pt]
		\sqrt{3}\,Y_{\bm{2},2}^{(2)}(\tau)\,Y_{\bm{\bar{3}_{0}},1}^{(2)}(\tau)
		+Y_{\bm{2},1}^{(2)}(\tau)\,Y_{\bm{\bar{3}_{0}},3}^{(2)}(\tau)
	\end{pmatrix}\,,
	\\[4pt]
	Y_{\bm{\hat{3}_{0}}}^{(4)}(\tau) &=&
	\frac{1}{2}\bigl(Y_{\bm{2}}^{(2)}(\tau)\,Y_{\bm{\hat{3}_{0}}}^{(2)}(\tau)\bigr)_{\bm{\hat{3}_{0}}}
	=
	\frac{1}{2}\begin{pmatrix}
		2\,Y_{\bm{2},1}^{(2)}(\tau)\,Y_{\bm{\hat{3}_{0}},1}^{(2)}(\tau)\\[4pt]
		-Y_{\bm{2},1}^{(2)}(\tau)\,Y_{\bm{\hat{3}_{0}},2}^{(2)}(\tau)
		-\sqrt{3}\,Y_{\bm{2},2}^{(2)}(\tau)\,Y_{\bm{\hat{3}_{0}},3}^{(2)}(\tau)\\[4pt]
		-\sqrt{3}\,Y_{\bm{2},2}^{(2)}(\tau)\,Y_{\bm{\hat{3}_{0}},2}^{(2)}(\tau)
		-Y_{\bm{2},1}^{(2)}(\tau)\,Y_{\bm{\hat{3}_{0}},3}^{(2)}(\tau)
	\end{pmatrix}\,.
\end{eqnarray}
The $q$-expansions for these two weight 4 modular forms are given by:
\begin{eqnarray}
\nonumber Y_{\bm{\bar{3}_{0}}}^{(4)}(\tau)&=&
\begin{pmatrix}
	-\,q^{5/8}\bigl(11-9q-65q^{3}+107q^{4}-297q^{5}+259q^{6}+\dots\bigr)\\[4pt]
	-\sqrt{2}\,q^{1/4}\bigl(1+22q-27q^{2}-18q^{3}-94q^{4}+359q^{6}+\dots\bigr)\\[4pt]
	-\frac{1}{2}\,q^{1/8}\bigl(1-27q-94q^{2}+359q^{3}-230q^{5}-343q^{6}+\dots\bigr)
\end{pmatrix}\,,
\\[4pt]
Y_{\bm{\hat{3}_{0}}}^{(4)}(\tau) &=&
\begin{pmatrix}
	1+16q-144q^{2}+448q^{3}-912q^{4}+2016q^{5}-4032q^{6}+\dots\\[4pt]
	2\sqrt{2}\,q^{1/4}\bigl(1+126q+757q^{2}+2198q^{3}+4914q^{4}+9632q^{5}+15751q^{6}+\dots\bigr)\\[4pt]
	8\sqrt{2}\,q^{3/4}\bigl(7+86q+333q^{2}+882q^{3}+1715q^{4}+3042q^{5}+5110q^{6}+\dots\bigr)
\end{pmatrix}\,.
\end{eqnarray}
At weight 6, the three next-to-lowest-weight triplet VVMFs are given by:
\begin{eqnarray}
\nonumber Y_{\bm{3_{1}}}^{(6)}(\tau) &=&
\bigl(Y_{\bm{2}}^{(2)}(\tau)\,Y_{\bm{3_{1}}}^{(4)}(\tau)\bigr)_{\bm{3_{1}}}
=
\begin{pmatrix}
	Y_{\bm{2},1}^{(2)}(\tau)\,Y_{\bm{3_{1}},1}^{(4)}(\tau)
	+\sqrt{3}\,Y_{\bm{2},2}^{(2)}(\tau)\,Y_{\bm{3_{1}},3}^{(4)}(\tau)\\[4pt]
	-2\,Y_{\bm{2},1}^{(2)}(\tau)\,Y_{\bm{3_{1}},2}^{(4)}(\tau)\\[4pt]
	\sqrt{3}\,Y_{\bm{2},2}^{(2)}(\tau)\,Y_{\bm{3_{1}},1}^{(4)}(\tau)
	+Y_{\bm{2},1}^{(2)}(\tau)\,Y_{\bm{3_{1}},3}^{(4)}(\tau)
\end{pmatrix}\,,
\\[4pt]
\nonumber Y_{\bm{\bar{3}_{1}}}^{(6)}(\tau) &=&
\bigl(Y_{\bm{2}}^{(2)}(\tau)\,Y_{\bm{\bar{3}_{1}}}^{(4)}(\tau)\bigr)_{\bm{\bar{3}_{0}}}
=
\begin{pmatrix}
	Y_{\bm{2},1}^{(2)}(\tau)\,Y_{\bm{\bar{3}_{1}},1}^{(4)}(\tau)
	+\sqrt{3}\,Y_{\bm{2},2}^{(2)}(\tau)\,Y_{\bm{\bar{3}_{1}},3}^{(4)}(\tau)\\[4pt]
	-2\,Y_{\bm{2},1}^{(2)}(\tau)\,Y_{\bm{\bar{3}_{1}},2}^{(4)}(\tau)\\[4pt]
	\sqrt{3}\,Y_{\bm{2},2}^{(2)}(\tau)\,Y_{\bm{\bar{3}_{1}},1}^{(4)}(\tau)
	+Y_{\bm{2},1}^{(2)}(\tau)\,Y_{\bm{\bar{3}_{1}},3}^{(4)}(\tau)
\end{pmatrix}\,,
\\[4pt]
Y_{\bm{\hat{3}_{1}}}^{(6)}(\tau) &=&
\bigl(Y_{\bm{2}}^{(2)}(\tau)\,Y_{\bm{\hat{3}_{1}}}^{(4)}(\tau)\bigr)_{\bm{\hat{3}_{1}}}
=
\begin{pmatrix}
	2\,Y_{\bm{2},1}^{(2)}(\tau)\,Y_{\bm{\hat{3}_{1}},1}^{(4)}(\tau)\\[4pt]
	-Y_{\bm{2},1}^{(2)}(\tau)\,Y_{\bm{\hat{3}_{1}},2}^{(4)}(\tau)
	-\sqrt{3}Y_{\bm{2},2}^{(2)}(\tau)\,Y_{\bm{\hat{3}_{1}},3}^{(4)}(\tau)\\[4pt]
	-\sqrt{3}\,Y_{\bm{2},2}^{(2)}(\tau)\,Y_{\bm{\hat{3}_{1}},2}^{(4)}(\tau)
	-Y_{\bm{2},1}^{(2)}(\tau)\,Y_{\bm{\hat{3}_{1}},3}^{(4)}(\tau)
\end{pmatrix}\,,
\end{eqnarray}
which have the following $q$-expansions:
\begin{eqnarray}
\nonumber Y_{\bm{3_{1}}}^{(6)}(\tau) &=&
\begin{pmatrix}
	q^{7/8}\bigl(13+7q-201q^{2}+164q^{3}+347q^{4}-2q^{5}+469q^{6}+\dots\bigr)\\[4pt]
	\frac{\sqrt{2}}{4}\,q^{1/4}\bigl(1+14q-179q^{2}+694q^{3}-1278q^{4}+1664q^{5}-2929q^{6}+\dots\bigr)\\[4pt]
	\frac{1}{2}\,q^{3/8}\bigl(1+67q-139q^{2}-422q^{3}+364q^{4}+2491q^{5}-1278q^{6}+\dots\bigr)
\end{pmatrix}\,,
\\[4pt]
\nonumber Y_{\bm{\bar{3}_{1}}}^{(6)}(\tau) &=&
\begin{pmatrix}
	q^{1/8}\bigl(1-179q-1278q^{2}-2929q^{3}+4288q^{4}+170q^{5}+26457q^{6}+\dots\bigr)\\[4pt]
	32\sqrt{2}\,q^{3/4}\bigl(1+26q+67q^{2}+14q^{3}-139q^{4}-402q^{5}-422q^{6}+\dots\bigr)\\[4pt]
	2\,q^{5/8}\bigl(7+347q+832q^{2}+1459q^{3}-4729q^{4}-1253q^{5}-11089q^{6}+\dots\bigr)
\end{pmatrix}\,,
\\[4pt]
Y_{\bm{\hat{3}_{1}}}^{(6)}(\tau) &=&
\begin{pmatrix}
	2\,q^{1/2}\bigl(1+20q-74q^{2}-24q^{3}+157q^{4}+124q^{5}+478q^{6}+\dots\bigr)\\[4pt]
	\sqrt{2}\,q^{3/4}\bigl(5-6q+31q^{2}-370q^{3}+761q^{4}+46q^{5}-430q^{6}+\dots\bigr)\\[4pt]
	\frac{\sqrt{2}}{4}\,q^{1/4}\bigl(1-74q+157q^{2}+478q^{3}-1198q^{4}-480q^{5}+2351q^{6}+\dots\bigr)
\end{pmatrix}.
\end{eqnarray}
At weight 8, there is a next-to-lowest-weight triplet VVMF
\begin{equation}
Y_{\bm{3_{0}}}^{(8)}(\tau) =
\bigl(Y_{\bm{2}}^{(2)}(\tau)\,Y_{\bm{3_{0}}}^{(6)}(\tau)\bigr)_{\bm{3_{0}}}
=
\begin{pmatrix}
	Y_{\bm{2},1}^{(2)}(\tau)\,Y_{\bm{3_{0}},1}^{(6)}(\tau)
	+\sqrt{3}\,Y_{\bm{2},2}^{(2)}(\tau)\,Y_{\bm{3_{0}},3}^{(6)}(\tau)\\[4pt]
	-2\,Y_{\bm{2},1}^{(2)}(\tau)\,Y_{\bm{3_{0}},2}^{(6)}(\tau)\\[4pt]
	\sqrt{3}\,Y_{\bm{2},2}^{(2)}(\tau)\,Y_{\bm{3_{0}},1}^{(6)}(\tau)
	+Y_{\bm{2},1}^{(2)}(\tau)\,Y_{\bm{3_{0}},3}^{(6)}(\tau)
\end{pmatrix}\,,
\end{equation}
with the following $q$-expansion
\begin{equation}
Y_{\bm{3_{0}}}^{(8)}(\tau) =
\begin{pmatrix}
	q^{3/8}\bigl(1-117q+549q^{2}-534q^{3}+1260q^{4}-10845q^{5}+16994q^{6}+\dots\bigr)\\[4pt]
	8\sqrt{2}\,q^{3/4}\bigl(1+18q-117q^{2}+70q^{3}+549q^{4}-522q^{5}-534q^{6}+\dots\bigr)\\[4pt]
	2\,q^{7/8}\bigl(9+35q-261q^{2}-972q^{3}+6879q^{4}-9738q^{5}-4095q^{6}+\dots\bigr)
\end{pmatrix}\,.
\end{equation}

With this, we have obtained all 3-$d$ VVMFs required for model building. Their explicit expressions, organized by modular weight, are summarized in table~\ref{tab:All three-dimensional VVMFs}.
\begin{table}[t!]
	\begin{center}
		\renewcommand{\tabcolsep}{4.5mm}
		\renewcommand{\arraystretch}{1.5}
		\begin{tabular}{|c|c||c|c|}\hline\hline
		
			Weight $k_Y$&$Y_{\bm{r}}^{(k_Y)}$&Weight $k_Y$&$Y_{\bm{r}}^{(k_Y)}$\\
			\hline
			$k_Y=2$&$Y_{\bm{2}}^{(2)}$,$Y_{\bm{\bar{3}_{0}}}^{(2)}$,$Y_{\bm{\hat{3}_{0}}}^{(2)}$&$k_Y=6$&$Y_{\bm{3_{0}}}^{(6)}$,$Y_{\bm{3_{1}}}^{(6)}$,$Y_{\bm{\bar{3}_{1}}}^{(6)}$,$Y_{\bm{\hat{3}_{1}}}^{(6)}$\\ 
			\hline
			$k_Y=4$&$Y_{\bm{3_{1}}}^{(4)}$,$Y_{\bm{\bar{3}_{1}}}^{(4)}$,$Y_{\bm{\hat{3}_{1}}}^{(4)}$,$Y_{\bm{\bar{3}_{0}}}^{(4)}$,$Y_{\bm{\hat{3}_{0}}}^{(4)}$&$k_Y=8$&$Y_{\bm{3_{0}}}^{(8)}$\\ 
			\hline\hline
		\end{tabular}
		\caption{\label{tab:All three-dimensional VVMFs}The lowest- and next-to-lowest-weight doublet and triplet VVMFs of $\Delta(96)$ employed in the model construction, organized by modular weight $k_Y$.}
	\end{center}
\end{table}

\subsection{Sextet VVMFs }
For completeness, we present here the construction of VVFMs in the unique 6-$d$ representation of $\Delta(96)$. For such a high-dimensional VVMF the pure MLDE method alone is no longer practical, and it has to be supplemented by the algebraic constraints satisfied by the VVMF. Alternatively, and more simply, the sextet VVMF can be obtained via tensor products of VVMFs of lower (fractional) weight. In fact, the 6-$d$ VVMF of lowest weight 2 for $\Delta(96)$ can precisely be constructed using polynomials in the following (weight 1/2) theta functions on $\Gamma(8)$:
\begin{subequations}
\begin{align}
\theta_i (\tau) &\equiv \sum_{n=-\infty}^{\infty} e^{4\pi i \tau (n+\frac{i-1}{4})^2}   \quad \text{with}  \quad i=1,\dots,4\,.
 \end{align}
\end{subequations}
The specific expression for sextet modular forms $Y^{(2)}_{\bm{6}}$ is given by~\cite{CentellesChulia:2026bkr}
\begin{align}
Y^{(2)}_{\bm{6}}(\tau)&=
\left(
\begin{array}{c}
 \theta_3^4-\theta_1^4\\
 2\theta_1\theta_3\left(\theta_1^2-\theta_3^2\right)
   \\
 \theta_1\left(\theta_2+\theta_4\right) \left(\theta_1^2+3\theta_3^2\right) \\
 \theta_1\left(\theta_2+\theta_4\right)^3  \\
  \theta_3\left(\theta_2+\theta_4\right) \left(3\theta_1^2+\theta_3^2\right) \\
  \theta_3\left(\theta_2+\theta_4\right)^3 
\end{array}
\right)\,,
\end{align}
with the $q$-expansion as follows
\begin{equation}
 Y^{(2)}_{\bm{6}}(\tau)=\left(
\begin{array}{c}   
-1+8 q^2-24 q^4+32 q^6-24 q^8+\dots \\
-4 \sqrt{q} \left(-1+4 q-6 q^2+8 q^3-13 q^4+12 q^5-14 q^6+24 q^7 +\dots \right) \\
2 q^{1/8} \left(1+13 q+18 q^2+31 q^3+48 q^4+42 q^5+57 q^6+80 q^7+\dots\right)\\
8 q^{3/8} \left(1+3 q+5 q^2+10 q^3+12 q^4+11 q^5+18 q^6+15 q^7+\dots\right)\\
4 q^{5/8} \left(3+7 q+16 q^2+15 q^3+19 q^4+39 q^5+27 q^6+31 q^7+\dots\right)\\
16 q^{7/8} \left(1+3 q+3 q^2+4 q^3+7 q^4+6 q^5+9 q^6+13 q^7+\dots\right)
\end{array}
\right)\,.
\end{equation}
There is another fundamental VVMF generator $Y^{(4)}_{\bm{6} I}(\tau)$ at weight 4, which can be constructed by the tensor product of this $Y^{(2)}_{\bm{6}}(\tau)$, i.e.,
\begin{equation}
 Y^{(4)}_{\bm{6} I}(\tau)\equiv\frac{1}{4}\left(Y^{(2)}_{\bm{6}}(\tau)Y^{(2)}_{\bm{6}}(\tau)
\right)_{\bm{6}} = \frac{1}{2}\left(
\begin{array}{c}
Y^{(2)}_{\bm{6},3}(\tau)Y^{(2)}_{\bm{6},6}(\tau)-Y^{(2)}_{\bm{6},4}(\tau)Y^{(2)}_{\bm{6},5}(\tau)\\
 Y^{(2)}_{\bm{6},3}(\tau)Y^{(2)}_{\bm{6},4}(\tau)-Y^{(2)}_{\bm{6},5}(\tau)Y^{(2)}_{\bm{6},6}(\tau)\\
 Y^{(2)}_{\bm{6},1}(\tau)Y^{(2)}_{\bm{6},3}(\tau)+Y^{(2)}_{\bm{6},2}(\tau)Y^{(2)}_{\bm{6},5}(\tau)\\
  Y^{(2)}_{\bm{6},2}(\tau)Y^{(2)}_{\bm{6},6}(\tau)-Y^{(2)}_{\bm{6},1}(\tau)Y^{(2)}_{\bm{6},4}(\tau)\\
  -Y^{(2)}_{\bm{6},2}(\tau)Y^{(2)}_{\bm{6},3}(\tau)-Y^{(2)}_{\bm{6},1}(\tau)Y^{(2)}_{\bm{6},5}(\tau)\\
Y^{(2)}_{\bm{6},1}(\tau)Y^{(2)}_{\bm{6},6}(\tau)-Y^{(2)}_{\bm{6},2}(\tau)Y^{(2)}_{\bm{6},4}(\tau)\\
\end{array}
\right)\,.
\end{equation}
Its $q$-expansion is 
\begin{equation}
 Y^{(4)}_{\bm{6} I}(\tau)=\left(
\begin{array}{c}   
-32 q \left( 1-4 q^2-2 q^4+24 q^6 + \dots\right) \\
-8 \sqrt{q} \left(-1-4 q+2 q^2+24 q^3+11 q^4-44 q^5-22 q^6+8 q^7 + \dots\right)\\
q^{1/8} \left(-1+11 q-50 q^2+121 q^3-176 q^4+198 q^5-233 q^6+176 q^7 + \dots\right)\\
4 q^{3/8} \left(1+11 q-11 q^2-38 q^3+12 q^4-13 q^5+50 q^6+167 q^7 + \dots\right)\\
-2 q^{5/8} \left(-1+11 q-48 q^2+99 q^3-81 q^4+11 q^5-121 q^6+275 q^7 + \dots\right)\\
8 q^{7/8} \left(-3-q+7 q^2+20 q^3+11 q^4-66 q^5-11 q^6+33 q^7 + \dots\right)\\
\end{array}
\right)\,.
\end{equation}

Another linearly independent weight 4 VVMF $Y^{(4)}_{\bm{6} II}(\tau)$ can be obtained via the following tensor product
\begin{equation}
 Y^{(4)}_{\bm{6} II}(\tau)\equiv
 -\frac{1}{2}\left(
 Y^{(2)}_{\bm{2}}(\tau)Y^{(2)}_{\bm{6}}(\tau)
 \right)_{\bm{6}_{i}}
 =
-\frac{1}{2}\left(
\begin{array}{c}
2Y^{(2)}_{\bm{2},1}(\tau)Y^{(2)}_{\bm{6},1}(\tau)\\
2Y^{(2)}_{\bm{2},1}(\tau)Y^{(2)}_{\bm{6},2}(\tau)\\
-Y^{(2)}_{\bm{2},1}(\tau)Y^{(2)}_{\bm{6},3}(\tau)
-\sqrt{3}Y^{(2)}_{\bm{2},2}(\tau)Y^{(2)}_{\bm{6},5}(\tau)\\
-\sqrt{3}Y^{(2)}_{\bm{2},2}(\tau)Y^{(2)}_{\bm{6},6}(\tau)
-Y^{(2)}_{\bm{2},1}(\tau)Y^{(2)}_{\bm{6},4}(\tau)\\
-\sqrt{3}Y^{(2)}_{\bm{2},2}(\tau)Y^{(2)}_{\bm{6},3}(\tau)
-Y^{(2)}_{\bm{2},1}(\tau)Y^{(2)}_{\bm{6},5}(\tau)\\
-\sqrt{3}Y^{(2)}_{\bm{2},2}(\tau)Y^{(2)}_{\bm{6},4}(\tau)
-Y^{(2)}_{\bm{2},1}(\tau)Y^{(2)}_{\bm{6},6}(\tau)
\end{array}
\right)\,.
\end{equation}
and its $q$-expansion reads
{\small\begin{equation}
 Y^{(4)}_{\bm{6} II}(\tau)=\left(
\begin{array}{c}   
1+24 q+16 q^2-96 q^3-144 q^4-48 q^5+448 q^6+576 q^7+\dots \\[1mm]
-4 \sqrt{q} \left(1+20 q-66 q^2+136 q^3-395 q^4+732 q^5-1066 q^6+1752 q^7+\dots\right)\\[1mm]
q^{1/8} \left(1+181 q+1266 q^2+3847 q^3+9456 q^4+17082 q^5+29673 q^6+47888 q^7+\dots\right)\\[1mm]
4 q^{3/8} \left(1+75 q+437 q^2+1306 q^3+2700 q^4+4979 q^5+8562 q^6+12711 q^7+\dots\right)\\[1mm]
2 q^{5/8} \left(15+283 q+1168 q^2+3123 q^3+6271 q^4+11931 q^5+18519 q^6+28579 q^7+\dots\right)\\[1mm]
8 q^{7/8} \left(13+111 q+375 q^2+916 q^3+1915 q^4+3294 q^5+5253 q^6+8113 q^7+\dots\right)
\end{array}
\right)\,.
\end{equation}}
One can verify that the VVMF generator $D_2Y^{(2)}_{\bm{6}}$ satisfies $D_2Y^{(2)}_{\bm{6}}(\tau)=\frac{1}{4}Y^{(4)}_{\bm{6} I}(\tau)+\frac{1}{6}Y^{(4)}_{\bm{6} II}(\tau)$.

The entire free module $\mathcal{M}(\bm{6})$ can be generated by the following six generators over the ring $\mathcal{M}(\bm{1})=\mathds{C}[E_4,E_6]$:
\begin{equation}
\mathcal{M}(\bm{6}) = \left\langle Y^{(2)}_{\bm{6}},~~ Y^{(4)}_{\bm{6} I}, ~~ D_2Y^{(2)}_{\bm{6}},~~D^2_2Y^{(2)}_{\bm{6}}, ~~ D_4Y^{(4)}_{\bm{6} I}, ~~ D^3_2Y^{(2)}_{\bm{6}}\right\rangle\,.
\end{equation}
For example, at weight 6, in addition to the two generators $Y^{(6)}_{\bm{6} I}\equiv D^2_2Y^{(2)}_{\bm{6}}$ and $Y^{(6)}_{\bm{6} II}\equiv D_4Y^{(4)}_{\bm{6}I}$ mentioned above, there is another linearly independent 6-$d$ VVMF: $Y^{(6)}_{\bm{6} III}\equiv E_4 Y^{(2)}_{\bm{6}}$. We omit their explicit $q$-expansions and tensor-product constructions here.

\section{\label{sec:fixed_column}Analytical results for the fixed PMNS columns of the viable patterns beyond TM$_1$}

The viable patterns $\mathcal{N}_{1}$--$\mathcal{N}_{8}$ all predict the first column of the TM$_1$ mixing matrix. The fixed PMNS columns for the remaining 27 viable patterns are summarized in table~\ref{tab:fixed_columns}. For compactness, we define $\varrho\equiv 2^{1/4}$ in this appendix.

\newpage

\begin{footnotesize}
	\renewcommand{\arraystretch}{1.35}
	\renewcommand{\tabcolsep}{0.9mm}
	\setlength\LTcapwidth{\textwidth}
	\setlength\LTleft{-0.0in}
	\setlength\LTright{0pt}
	
	\begin{longtable}{|c|c|c|c|c|c|c|c|c|c|c|c|}
		\caption{\label{tab:fixed_columns}
			Analytical expressions for the fixed PMNS columns of the 27 viable patterns beyond the TM$_1$ mixing scheme.
		}\\
		\midrule
		\specialrule{0em}{1.0pt}{1.0pt}
		
		\endfirsthead
		
		\multicolumn{12}{c}
		{{\bfseries \tablename\ \thetable{} -- continued from previous page}}\\
		\hline
		
		\endhead
		
		\caption[]{Continued on next page.}\\
		\endfoot
		
		\endlastfoot
		
		\hline
		
		\multicolumn{12}{|c|}{Fixed columns for NO case}
		\\ \hline

		\multicolumn{4}{|c}{$\mathcal N_{9},\mathcal N_{10}$}
		&
		\multicolumn{4}{|c}{$\mathcal N_{11}$}
		&
		\multicolumn{4}{|c|}{$\mathcal N_{12},\mathcal N_{13}$}
		\\ \hline
		
		\multicolumn{4}{|c}{
			$
			\left(
			\begin{array}{c}
				\sqrt{\dfrac{
						6+4\sqrt3+7\sqrt2+2\varrho(1-\sqrt2-\sqrt3)}
					{10+16\sqrt2-4\varrho^3}}
				\\[2mm]
				\sqrt{\dfrac{
						-2-4\sqrt3+9\sqrt2+2\varrho(\sqrt3-\sqrt2-1)}
					{10+16\sqrt2-4\varrho^3}}
				\\[2mm]
				\sqrt{\dfrac{6}{10+16\sqrt2-4\varrho^3}}
			\end{array}
			\right)
			$
		}
		&
		\multicolumn{4}{|c}{
			$
			\left(
			\begin{array}{c}
				\sqrt{\dfrac{6+4\sqrt3+7\sqrt2}
					{8+16\sqrt2+2\sqrt3}}
				\\[2mm]
				\sqrt{\dfrac{-2-4\sqrt3+9\sqrt2}
					{8+16\sqrt2+2\sqrt3}}
				\\[2mm]
				\sqrt{\dfrac{4+2\sqrt3}
					{8+16\sqrt2+2\sqrt3}}
			\end{array}
			\right)
			$
		}
		&
		\multicolumn{4}{|c|}{
			$
			\left(
			\begin{array}{c}
				\dfrac{1}{2}
				\sqrt{\dfrac{
						4+12\sqrt2-\sqrt3+4\sqrt3\,\varrho^3}
					{3+6\sqrt2}}
				\\[2mm]
				\dfrac{1}{\sqrt{3+6\sqrt2}}
				\\[2mm]
				\dfrac{1}{2}
				\sqrt{\dfrac{
						4+12\sqrt2+\sqrt3-4\sqrt3\,\varrho^3}
					{3+6\sqrt2}}
			\end{array}
			\right)
			$
		}
		\\ \hline
		
		\multicolumn{6}{|c}{$\mathcal N_{14}$}
		&
		\multicolumn{6}{|c|}{$\mathcal N_{15}$}
		\\ \hline
		
		\multicolumn{6}{|c}{
			$
			\left(
			\begin{array}{c}
				\dfrac{2}{
					\sqrt{
						24+12\sqrt2-3\sqrt6-6(1+\sqrt3)\varrho^3
				}}
				\\[2mm]
				\sqrt{\dfrac{
						4+6\sqrt2-3\sqrt6+2(\sqrt3-3)\varrho^3}
					{
						24+12\sqrt2-3\sqrt6-6(1+\sqrt3)\varrho^3
				}}
				\\[2mm]
				\sqrt{\dfrac{
						16+6\sqrt2-8\sqrt3\,\varrho^3}
					{
						24+12\sqrt2-3\sqrt6-6(1+\sqrt3)\varrho^3
				}}
			\end{array}
			\right)
			$
		}
		&
		\multicolumn{6}{|c|}{
			$
			\left(
			\begin{array}{c}
				\dfrac{2}{
					\sqrt{
						24+12\sqrt2-3\sqrt6-6(1+\sqrt3)\varrho^3
				}}
				\\[2mm]
				\sqrt{\dfrac{
						16+6\sqrt2-8\sqrt3\,\varrho^3}
					{
						24+12\sqrt2-3\sqrt6-6(1+\sqrt3)\varrho^3
				}}
				\\[2mm]
				\sqrt{\dfrac{
						4+6\sqrt2-3\sqrt6+2(\sqrt3-3)\varrho^3}
					{
						24+12\sqrt2-3\sqrt6-6(1+\sqrt3)\varrho^3
				}}
			\end{array}
			\right)
			$
		}
		\\ \hline

		\multicolumn{4}{|c}{$\mathcal N_{16},\mathcal N_{17}$}
		&
		\multicolumn{4}{|c}{$\mathcal N_{18}$}
		&
		\multicolumn{4}{|c|}{$\mathcal N_{19}$}
		\\ \hline
		
		\multicolumn{4}{|c}{
			$
			\left(
			\begin{array}{c}
				\dfrac{1}{\sqrt2}
				\sqrt{\dfrac{6+6\varrho-\sqrt2-2\varrho^3}{11-4\varrho}}
				\\[2mm]
				\dfrac{1}{\sqrt2}
				\sqrt{\dfrac{14-10\varrho-3\sqrt2+2\varrho^3}{11-4\varrho}}
				\\[2mm]
				\sqrt{\dfrac{1+2\sqrt2-2\varrho}{11-4\varrho}}
			\end{array}
			\right)
			$
		}
		&
		\multicolumn{4}{|c}{
			$
			\left(
			\begin{array}{c}
				\sqrt{\dfrac{4+3\sqrt2}
					{8+7\sqrt2-2\varrho-2\varrho^3}}
				\\[2mm]
				\sqrt{\dfrac{
						4+3\sqrt2-2\varrho-2\varrho^3}
					{8+7\sqrt2-2\varrho-2\varrho^3}}
				\\[2mm]
				\sqrt{\dfrac{\sqrt2}
					{8+7\sqrt2-2\varrho-2\varrho^3}}
			\end{array}
			\right)
			$
		}
		&
		\multicolumn{4}{|c|}{
			$
			\left(
			\begin{array}{c}
				\sqrt{\dfrac{4+3\sqrt2}
					{8+7\sqrt2-2\varrho-2\varrho^3}}
				\\[2mm]
				\sqrt{\dfrac{\sqrt2}
					{8+7\sqrt2-2\varrho-2\varrho^3}}	
				\\[2mm]
				\sqrt{\dfrac{
						4+3\sqrt2-2\varrho-2\varrho^3}
					{8+7\sqrt2-2\varrho-2\varrho^3}}
			\end{array}
			\right)
			$
		}
		\\ \hline
		
		\multicolumn{6}{|c}{$\mathcal N_{20}$}
		&
		\multicolumn{6}{|c|}{$\mathcal N_{21}$}
		\\ \hline
		
		\multicolumn{6}{|c}{
			$
			\left(
			\begin{array}{c}
				\sqrt{\dfrac{8+7\sqrt2+2\varrho}
					{4(6+5\sqrt2-2\varrho-2\varrho^3)}}
				\\[2mm]
				\sqrt{\dfrac{8+7\sqrt2-6\varrho-4\varrho^3}
					{4(6+5\sqrt2-2\varrho-2\varrho^3)}}
				\\[2mm]
				\sqrt{\dfrac{8+6\sqrt2-4\varrho-4\varrho^3}
					{4(6+5\sqrt2-2\varrho-2\varrho^3)}}
			\end{array}
			\right)
			$
		}
		&
		\multicolumn{6}{|c|}{
			$
			\left(
			\begin{array}{c}
				\sqrt{\dfrac{8+7\sqrt2+2\varrho}
					{4(6+5\sqrt2-2\varrho-2\varrho^3)}}
				\\[2mm]
				\sqrt{\dfrac{8+6\sqrt2-4\varrho-4\varrho^3}
					{4(6+5\sqrt2-2\varrho-2\varrho^3)}}	
				\\[2mm]
				\sqrt{\dfrac{8+7\sqrt2-6\varrho-4\varrho^3}
					{4(6+5\sqrt2-2\varrho-2\varrho^3)}}
			\end{array}
			\right)
			$
		}
		\\ \hline
		
		\multicolumn{12}{|c|}{	Fixed columns for IO case}
		\\ \hline
		
		\multicolumn{6}{|c|}{$\mathcal I_{1}$}
		&
		\multicolumn{6}{c|}{$\mathcal I_{2},\mathcal I_{3}$}
		\\ \hline
		
		\multicolumn{6}{|c|}{
			$
			\left(
			\begin{array}{c}
				\sqrt{\dfrac{
						1+8\sqrt2-4\sqrt6+(1-\sqrt3)\varrho^3}
					{3+36\sqrt2}}
				\\[2mm]
				\sqrt{\dfrac{
						1+20\sqrt2-2\varrho^3}
					{3+36\sqrt2}}
				\\[2mm]
				\sqrt{\dfrac{
						1+8\sqrt2+4\sqrt6+(1+\sqrt3)\varrho^3}
					{3+36\sqrt2}}
			\end{array}
			\right)
			$
		}
		&
		\multicolumn{6}{c|}{
			$
			\left(
			\begin{array}{c}
				\sqrt{\dfrac{
						1+8\sqrt2-4\sqrt6+(1-\sqrt3)\varrho^3}
					{3+36\sqrt2}}
				\\[2mm]
				\sqrt{\dfrac{
						1+8\sqrt2+4\sqrt6+(1+\sqrt3)\varrho^3}
					{3+36\sqrt2}}	
				\\[2mm]
				\sqrt{\dfrac{
						1+20\sqrt2-2\varrho^3}
					{3+36\sqrt2}}
			\end{array}
			\right)
			$
		}
		\\ \hline
		\newpage
		
		\multicolumn{6}{|c|}{$\mathcal I_{4}$}
		&
		\multicolumn{6}{c|}{$\mathcal I_{5}$}
		\\ \hline
		
		\multicolumn{6}{|c|}{
			$
			\left(
			\begin{array}{c}
				\sqrt{\dfrac{
						36-4\sqrt6+2\sqrt3
						+8\varrho^3+4\sqrt3\,\varrho^3
						-16\varrho(1+\sqrt3)}
					{3(35+4\varrho^3)}}
				\\[2mm]
				\sqrt{\dfrac{
						33+32\varrho-4\varrho^3}
					{3(35+4\varrho^3)}}
				\\[2mm]
				\sqrt{\dfrac{
						36+4\sqrt6-2\sqrt3
						+8\varrho^3-4\sqrt3\,\varrho^3
						+16\varrho(\sqrt3-1)}
					{3(35+4\varrho^3)}}
			\end{array}
			\right)
			$
		}
		&
		\multicolumn{6}{c|}{
			$
			\left(
			\begin{array}{c}
				\sqrt{\dfrac{
						36-4\sqrt6+2\sqrt3
						+8\varrho^3+4\sqrt3\,\varrho^3
						-16\varrho(1+\sqrt3)}
					{3(35+4\varrho^3)}}
				\\[2mm]	
				\sqrt{\dfrac{
						36+4\sqrt6-2\sqrt3
						+8\varrho^3-4\sqrt3\,\varrho^3
						+16\varrho(\sqrt3-1)}
					{3(35+4\varrho^3)}}	
				\\[2mm]
				\sqrt{\dfrac{
						33+32\varrho-4\varrho^3}
					{3(35+4\varrho^3)}}
			\end{array}
			\right)
			$
		}
		\\ \hline

		\multicolumn{6}{|c|}{$\mathcal I_{6},\mathcal I_{7}$}
		&
		\multicolumn{6}{c|}{$\mathcal I_{8},\mathcal I_{11}$}
		\\ \hline
		
		\multicolumn{6}{|c|}{
			$
			\left(
			\begin{array}{c}
				\sqrt{\dfrac{
						2(\sqrt6+\sqrt2-\varrho^3-2)}
					{3(-\varrho\sqrt3+\varrho^3\sqrt3+2\sqrt6)}}
				\\[2mm]
				\sqrt{\dfrac{
						7\sqrt2+\sqrt3\,\varrho^3+1-2\varrho^3}
					{3(5\sqrt2+2-\varrho^3)}}
				\\[2mm]
				\sqrt{\dfrac{
						3\sqrt3+\sqrt3\,\varrho^3+(2+\sqrt3)\varrho
						+2-\varrho^3-\sqrt6}
					{3(-\sqrt6+2\sqrt3+2\sqrt3\,\varrho^3)}}
			\end{array}
			\right)
			$
		}
		&
		\multicolumn{6}{c|}{
			$
			\left(
			\begin{array}{c}
				\sqrt{\dfrac{
						2(\sqrt6+\sqrt2-\varrho^3-2)}
					{3(-\varrho\sqrt3+\varrho^3\sqrt3+2\sqrt6)}}
				\\[2mm]
				\sqrt{\dfrac{
						3\sqrt3+\sqrt3\,\varrho^3+(2+\sqrt3)\varrho
						+2-\varrho^3-\sqrt6}
					{3(-\sqrt6+2\sqrt3+2\sqrt3\,\varrho^3)}}	
				\\[2mm]
				\sqrt{\dfrac{
						7\sqrt2+\sqrt3\,\varrho^3+1-2\varrho^3}
					{3(5\sqrt2+2-\varrho^3)}}
			\end{array}
			\right)
			$
		}
		\\ \hline
		
		\multicolumn{6}{|c|}{$\mathcal I_{9},\mathcal I_{10}$}
		&
		\multicolumn{6}{c|}{$\mathcal I_{12}$}
		\\ \hline
		
		\multicolumn{6}{|c|}{
			$
			\left(
			\begin{array}{c}
				\sqrt{\dfrac{
						2\sqrt2+2-(1+\sqrt3)\varrho^3}
					{3(3\sqrt2+2-\sqrt3\,\varrho^3)}}
				\\[2mm]
				\sqrt{\dfrac{
						2\sqrt2+2+\sqrt3-\varrho^3}
					{3(3\sqrt2+2-\sqrt3\,\varrho^3)}}
				\\[2mm]
				\sqrt{\dfrac{
						5\sqrt2+2+2\varrho^3-\sqrt3-2\sqrt3\,\varrho^3}
					{3(3\sqrt2+2-\sqrt3\,\varrho^3)}}
			\end{array}
			\right)
			$
		}
		&
		\multicolumn{6}{c|}{
			$
			\left(
			\begin{array}{c}
				\sqrt{\dfrac{
						-5\varrho^3-4\sqrt3\,\varrho-\sqrt6
						+2\varrho+\sqrt3\,\varrho^3+4\sqrt3+7\sqrt2}
					{
						28+28\sqrt2+12\varrho+12\sqrt3+4\sqrt6
						+4\sqrt3\,\varrho+8\sqrt3\,\varrho^3}}
				\\[2mm]
				\sqrt{\dfrac{
						16+11\sqrt2+4\sqrt3+3\sqrt6
						+2\varrho+4\sqrt3\,\varrho+\varrho^3+3\sqrt3\,\varrho^3}
					{
						28+28\sqrt2+12\varrho+12\sqrt3+4\sqrt6
						+4\sqrt3\,\varrho+8\sqrt3\,\varrho^3}}
				\\[2mm]
				\sqrt{\dfrac{
						12+10\sqrt2+4\sqrt3+2\sqrt6
						+8\varrho+4\sqrt3\,\varrho+4\varrho^3+4\sqrt3\,\varrho^3}
					{
						28+28\sqrt2+12\varrho+12\sqrt3+4\sqrt6
						+4\sqrt3\,\varrho+8\sqrt3\,\varrho^3}}
			\end{array}
			\right)
			$
		}
		\\ \hline
		
		\multicolumn{6}{|c|}{$\mathcal I_{13}$}
		&
		\multicolumn{6}{c|}{$\mathcal I_{14}$}
		\\ \hline
		
		\multicolumn{6}{|c|}{
			$
			\left(
			\begin{array}{c}
				\sqrt{\dfrac{
						7+6\sqrt2+\sqrt3-\varrho-\varrho^3
						-3\sqrt3\,\varrho-2\sqrt3\,\varrho^3}
					{
						28+21\sqrt2+8\varrho+8\sqrt3+5\sqrt6
						+2\varrho^3+2\sqrt3\,\varrho^3}}
				\\[2mm]
				\sqrt{\dfrac{
						13+6\sqrt2+3\sqrt3+4\sqrt6
						+5\varrho-\varrho^3-\sqrt3\,\varrho+2\sqrt3\,\varrho^3}
					{
						28+21\sqrt2+8\varrho+8\sqrt3+5\sqrt6
						+2\varrho^3+2\sqrt3\,\varrho^3}}
				\\[2mm]
				\sqrt{\dfrac{
						8+9\sqrt2+4\sqrt3+\sqrt6
						+4\varrho+4\sqrt3\,\varrho+4\varrho^3+2\sqrt3\,\varrho^3}
					{
						28+21\sqrt2+8\varrho+8\sqrt3+5\sqrt6
						+2\varrho^3+2\sqrt3\,\varrho^3}}
			\end{array}
			\right)
			$
		}
		&
		\multicolumn{6}{c|}{
			$
			\left(
			\begin{array}{c}
				\sqrt{\dfrac{
						7+6\sqrt2+\sqrt3-\varrho-\varrho^3
						-3\sqrt3\,\varrho-2\sqrt3\,\varrho^3}
					{
						28+21\sqrt2+8\varrho+8\sqrt3+5\sqrt6
						+2\varrho^3+2\sqrt3\,\varrho^3}}
				\\[2mm]
				\sqrt{\dfrac{
						8+9\sqrt2+4\sqrt3+\sqrt6
						+4\varrho+4\sqrt3\,\varrho+4\varrho^3+2\sqrt3\,\varrho^3}
					{
						28+21\sqrt2+8\varrho+8\sqrt3+5\sqrt6
						+2\varrho^3+2\sqrt3\,\varrho^3}}	
				\\[2mm]
				\sqrt{\dfrac{
						13+6\sqrt2+3\sqrt3+4\sqrt6
						+5\varrho-\varrho^3-\sqrt3\,\varrho+2\sqrt3\,\varrho^3}
					{
						28+21\sqrt2+8\varrho+8\sqrt3+5\sqrt6
						+2\varrho^3+2\sqrt3\,\varrho^3}}
			\end{array}
			\right)
			$
		}
		\\ \hline
		
		\specialrule{0em}{1.0pt}{1.0pt}
		\midrule
		
	\end{longtable}
\end{footnotesize}

\section{\label{sec:sum_rules}Explicit sum rules for selected patterns}

For the six NO patterns shown in figure~\ref{fig:sum_rules}, namely $\mathcal{N}_{6}$, $\mathcal{N}_{10}$, $\mathcal{N}_{11}$, $\mathcal{N}_{13}$, $\mathcal{N}_{16}$ and $\mathcal{N}_{19}$, we substitute the corresponding fixed PMNS columns listed in table~\ref{tab:fixed_columns} into the general relations in Eq.~\eqref{eq:sum_rules}. This gives five distinct sets of explicit sum rules, which are presented below.

For the TM$_1$ mixing pattern $\mathcal{N}_{6}$, the sum rules among mixing angles and Dirac CP phase can be written as
\begin{equation}
\begin{cases}
	\sin^2\theta_{12}
	=
	1-
	\frac{2}{3\cos^2\theta_{13}
	}, \\[5pt]
\cos\delta_{CP}
	=\frac{(3-5\cos2\theta_{13})\cot2\theta_{23}\csc\theta_{13}}{4 \sqrt{3\cos2\theta_{13}-1}}.
\end{cases}
\end{equation}

For $\mathcal N_{10}$, one obtains the mixing angles relations
\begin{equation}
\begin{cases}
	\sin^2\theta_{12}
	=
	1-
	\frac{
		6+4\sqrt3+7\varrho^2+2\varrho(1-\varrho^2-\sqrt3)
	}{
		\left(10+16\varrho^2-4\varrho^3\right)
		\cos^2\theta_{13}
	}, \\[5pt]
\cos\delta_{CP}
	=
\scalebox{0.8}{$\frac{
			\left[-8-4\sqrt3+9\varrho^2
			+2\varrho(\sqrt3-1-\varrho^2)\right]\cos^2\theta_{13}
			+
			\left\{
			\left[16+4\sqrt3+23\varrho^2
			+2\varrho(1-3\varrho^2-\sqrt3)\right]\sin^2\theta_{13}
			-
			\left[4-4\sqrt3+9\varrho^2
			+2\varrho(\sqrt3-1-\varrho^2)\right]
			\right\}\cos 2\theta_{23}
		}{
			2\sqrt{
				\left[6+4\sqrt3+7\varrho^2
				+2\varrho(1-\varrho^2-\sqrt3)\right]
				\left[
				\left(10+16\varrho^2-4\varrho^3\right)\cos^2\theta_{13}
				-6-4\sqrt3-7\varrho^2
				-2\varrho(1-\varrho^2-\sqrt3)
				\right]
			}
			\,\sin 2\theta_{23}\sin\theta_{13}
		}$}.
\end{cases}
\end{equation}

The pattern $\mathcal N_{11}$ gives the following mixing angles relations
\begin{equation}
\begin{cases}
	\sin^2\theta_{12}
	=
	1-
	\frac{
		6+4\sqrt3+7\varrho^2
	}{
		\left(8+16\varrho^2+2\sqrt3\right)
		\cos^2\theta_{13}
	}\,,\\[5pt]
\cos\delta_{CP}
		=
		\frac{
			\left[9\varrho^2-6(1+\sqrt3)\right]\cos^2\theta_{13}
			+
			\left\{
			\left[14+6\sqrt3+23\varrho^2\right]\sin^2\theta_{13}
			-
			\left[2-2\sqrt3+9\varrho^2\right]
			\right\}\cos 2\theta_{23}
		}{
			2\sqrt{
				\left[6+4\sqrt3+7\varrho^2\right]
				\left[
				\left(8+16\varrho^2+2\sqrt3\right)\cos^2\theta_{13}
				-6-4\sqrt3-7\varrho^2
				\right]
			}
			\,\sin 2\theta_{23}\sin\theta_{13}
		}\,.
\end{cases}
\end{equation}

For $\mathcal N_{13}$, substitution yields the mixing angles relations
\begin{equation}
\begin{cases}
	\sin^2\theta_{12}
	=
	1-
	\frac{
		4(1+3\varrho^2)+\sqrt3(4\varrho^3-1)
	}{
		12(1+2\varrho^2)\cos^2\theta_{13}
	}\,, \\[5pt]
\cos\delta_{CP}
		=
		\frac{
			\left[\sqrt3(4\varrho^3-1)-12\varrho^2\right]\cos^2\theta_{13}
			+
			\left\{
			\left[16+36\varrho^2+\sqrt3(4\varrho^3-1)\right]\sin^2\theta_{13}
			-
			\left[8+12\varrho^2-\sqrt3(4\varrho^3-1)\right]
			\right\}\cos 2\theta_{23}
		}{
			2\sqrt{
				\left[4(1+3\varrho^2)+\sqrt3(4\varrho^3-1)\right]
				\left[
				12(1+2\varrho^2)\cos^2\theta_{13}
				-4(1+3\varrho^2)-\sqrt3(4\varrho^3-1)
				\right]
			}
			\,\sin 2\theta_{23}\sin\theta_{13}
		}\,.
\end{cases}
\end{equation}

For $\mathcal N_{16}$, the mixing angles relations are
\begin{equation}
\begin{cases}
	\sin^2\theta_{12}
	=
	1-
	\frac{
		6(1+\varrho)-\varrho^2(1+2\varrho)
	}{
		2(11-4\varrho)\cos^2\theta_{13}
	}\,, \\[5pt]
\cos\delta_{CP}
		=
		\frac{
			\left[6(2-\varrho)-\varrho^2(7-2\varrho)\right]\cos^2\theta_{13}
			+
			\left\{
			\left[28-2\varrho-\varrho^2-2\varrho^3\right]\sin^2\theta_{13}
			-
			\left[16-14\varrho+\varrho^2+2\varrho^3\right]
			\right\}\cos 2\theta_{23}
		}{
			2\sqrt{
				\left[6(1+\varrho)-\varrho^2(1+2\varrho)\right]
				\left[
				2(11-4\varrho)\cos^2\theta_{13}
				-6(1+\varrho)+\varrho^2(1+2\varrho)
				\right]
			}
			\,\sin 2\theta_{23}\sin\theta_{13}
		}\,.
\end{cases}
\end{equation}

Finally, $\mathcal N_{19}$ gives the mixing angles relations
\begin{equation}
\begin{cases}
	\sin^2\theta_{12}
	=
	1-
	\frac{
		4+3\varrho^2
	}{
		\left[8+7\varrho^2-2\varrho(1+\varrho^2)\right]
		\cos^2\theta_{13}
	}\,, \\[5pt]
	\cos\delta_{CP}
	=
	\frac{
		\left[\varrho(1+\varrho^2)-(2+\varrho^2)\right]\cos^2\theta_{13}
		+
		\left\{
		\left[6+5\varrho^2-\varrho-\varrho^3\right]\sin^2\theta_{13}
		-
		\left[2+2\varrho^2-\varrho-\varrho^3\right]
		\right\}\cos 2\theta_{23}
	}{
		\sqrt{
			(4+3\varrho^2)
			\left[
			\left(8+7\varrho^2-2\varrho(1+\varrho^2)\right)\cos^2\theta_{13}
			-(4+3\varrho^2)
			\right]
		}
		\,\sin 2\theta_{23}\sin\theta_{13}
	}\,.
\end{cases}
\end{equation}

\end{appendix}
	
\vskip 2cm

\bibliographystyle{utphys}
\bibliography{references}	
	
\end{document}